\documentclass[nofootinbib, aps,11pt,preprintnumbers]{revtex4-1}

\usepackage{graphicx}
\usepackage{amssymb}
\usepackage{textcomp}
\usepackage{amsmath}
\usepackage{bm}
\usepackage{times}
\usepackage{epsfig}
\usepackage{color}
\usepackage{array,multirow}

\newcommand{\todo}[1]{{\color{red}#1}}
\newcommand{\Tr}{\mbox{Tr$\;$}}
\newcommand{\susy}{{\mbox{\scriptsize SUSY}}}

\renewcommand{\i}{{\mbox{\scriptsize I}}}

\newcommand{\trix}[1]{\left(\begin{array}{#1}}
\newcommand{\notrix}{\end{array}\right)}
\newcommand{\comment}[1]{}

\def\beq{\begin{equation}}
\def\eeq{\end{equation}}
\def\bea{\begin{eqnarray}}
\def\eea{\end{eqnarray}}

\begin{document}
\title{\Large  {{\bf{The Minimal SUSY $B-L$ Model: From the Unification Scale to the LHC}}}}
\author{{Burt A.~Ovrut, Austin Purves and Sogee Spinner} \\[2mm]
    {\it  Department of Physics, University of Pennsylvania} \\
   {\it Philadelphia, PA 19104--6396}\\[4mm]
}
\date{\today}

\let\thefootnote\relax\footnotetext{\mbox{ovrut@elcapitan.hep.upenn.edu,  ~apurves@sas.upenn.edu, 
~sogee@sas.upenn.edu} }

\begin{abstract}
\ \\ [10mm]
{\bf ABSTRACT:} This paper introduces a random statistical scan over the high-energy initial parameter space of the minimal SUSY $B-L$ model--denoted as the $B-L$ MSSM. Each initial set of points is renormalization group evolved to the electroweak scale--being subjected, sequentially, to the requirement of radiative $B-L$ and electroweak symmetry breaking, the present experimental lower bounds on the $B-L$ vector boson and sparticle masses, as well as the lightest neutral Higgs mass of $\sim$125 GeV. The subspace of initial parameters that satisfies all such constraints is presented, shown to be robust and to contain a wide range of different configurations of soft supersymmetry breaking masses. The low-energy predictions of each such ``valid'' point--such as the sparticle mass spectrum and, in particular, the LSP--are computed and then statistically analyzed over the full subspace of valid points. Finally, the amount of fine-tuning required is quantified and compared to the MSSM computed using an identical random scan. The $B-L$ MSSM is shown to generically require less fine-tuninng.

\end{abstract}

\maketitle

\tableofcontents

%
\section{Introduction}
%
The minimal supersymmetric standard model (MSSM) is the simplest possible $N=1$ supersymmetric extension of the standard model of particle physics. It has the gauge group $SU(3)_{C} \times SU(2)_{L} \times U(1)_{Y}$,  replaces each gauge field by a vector supermultiplet, each matter fermion by a chiral supermultiplet, has a conjugate pair of Higgs chiral superfields and has no right-handed neutrino multiplets. The MSSM was introduced in various contexts in \cite{Dimopoulos:1981zb,Nappi:1982hm} and has been extensively reviewed in~\cite{Martin:1997ns}. However, without further modification, the most general superpotential for the MSSM contains cubic interactions that explicitly violate both lepton and baryon number and lead, among other things, to rapid proton decay--which is unobserved. The traditional solution to this problem is to demand that, in addition to the gauge group, the MSSM be invariant under an ad hoc ${\mathbb{Z}}_{2}$ finite symmetry--$R$-parity--which acts on individual component fields as $(-1)^{3(B-L)+2s}$.  Although there have been many attempts to explain how $R$-parity can arise in the MSSM~\cite{Aulakh:1999cd, Aulakh:2000sn, Babu:2008ep, Feldman:2011ms, FileviezPerez:2011dg} or be spontaneously broken~\cite{Aulakh:1982yn, Hayashi:1984rd, Masiero:1990uj}. These are all ``non-minimal'' in the sense, for example, that they require additional matter multiplets, have other ad hoc assumptions and so on. Is there a more natural and minimal approach to $R$-parity in the MSSM?

One begins by noting that, within the context of supersymmetry, $R$-parity is a finite subgroup of $U(1)_{B-L}$--see for example \cite{Mohapatra:1986su}. It follows that $R$-parity might  arise as a consequence of demanding that the MSSM be invariant under a global $U(1)_{B-L}$ symmetry. This would be consistent with current bounds on baryon and lepton number violation, and can also be imposed on the MSSM extended to include three families of right-handed neutrino chiral supermultiplets . However, one of the implications of the standard model is that a global continuous symmetry group is likely to appear in its local form--that is, as a gauge symmetry. With this in mind, one might ask if $R$-parity could arise as the consequence of extending the MSSM gauge group to include a new gauged
$U(1)_{B-L}$ symmetry. It has long been known that the standard MSSM is anomalous with respect to gauged $B-L$ symmetry, whereas the MSSM extended by three families of right-handed neutrino supermultiplets is anomaly free. Furthermore, this is the minimal such extension of the MSSM. We will call this anomaly free, minimal content theory the $B-L$ MSSM, and propose that this is a more ``natural'' way in which $R$-parity can arise in low energy supersymmetric particle physics.

The $B-L$ MSSM was identified from a low-energy ``bottom-up'' point of view in \cite{FileviezPerez:2008sx, Barger:2008wn, Everett:2009vy}$^1$\footnote{$^1$See~\cite{Mohapatra:1986aw} for a similar idea in the context of $E_6$.}. In several papers \cite{FileviezPerez:2012mj,Perez:2013kla}, these authors explored its structure and some phenomenological consequences. In addition, these ideas, and other directions, have recently been reviewed in~\cite{Perez:2015rza}. Interestingly, the $B-L$ MSSM was also discovered from a high-energy ``top-down'' viewpoint in \cite{Braun:2005ux,Braun:2005nv,Braun:2013wr}, where it was shown that  this model arises within the context of 
heterotic superstring theory \cite{Lukas:1998yy,Lukas:1998tt}. More specifically, the $B-L$ MSSM is the low-energy effective theory associated with compactifying the $E_{8} \times E_{8}$ heterotic string on a Schoen Calabi-Yau threefold \cite{Braun:2004xv} with a specific class of $SU(4)$ vector bundles \cite{Braun:2005zv}. An important aspect of this high-energy point of view is that the parameters of the theory are specified near the gauge coupling unification scale, and then run down to the electroweak scale using the renormalization group (RG). This allows one to explore fundamental aspects of the theory--such as $B-L$ and electroweak  symmetry breaking. First steps in this direction were taken in \cite{Ambroso:2009jd,Ambroso:2009sc,Ambroso:2010pe}, where it was shown--for a restrictive set of initial parameters--that radiative breaking of both of these symmetries  can indeed occur. A further study of the Wilson lines, spectra, mass scales and the unification of gauge couplings from the high-energy superstring point of view was presented in \cite{Ovrut:2012wg}. Combining the bottom-up and top-down approaches to the $B-L$ MSSM, various aspects of both LHC and neutrino phenomenology were studied in the special case where a stop or a sbottom sparticle is the lightest supersymmetric particle (LSP) \cite{Marshall:2014kea,Marshall:2014cwa}.  

However, all of the previous analyses involved specific assumptions--either about high-energy initial conditions or the low-energy structure of the theory. For the $B-L$ MSSM to be a realistic contender for the low-energy theory of particle physics, it is essential that its initial parameter space be explored in a generic way, and that its low-energy predictions be compared with all present experimental data. This will be carried out in detail in this paper. We perform a statistical scan over a well-defined and wide range of initial parameters and, for each fixed set of such parameters, scale the theory down to low energy using the RGEs with specified threshold conditions. The results will be examined to determine  the subset of the parameter space that, sequentially, 1) breaks $B-L$ symmetry at a scale consistent with experiment, 2) breaks electroweak symmetry, 3) has all sparticle masses above their present experimental lower bounds and 4) predicts the mass of the lightest neutral Higgs scalar to be within 2$\sigma$ standard deviation from the ATLAS measured value of 125.36 GeV. A small subset of important results from this statistical scan were given in \cite{Ovrut:2014rba}. Here, we give present the details of the method use, as well as a wide array of new results and experimental predictions.

This paper is organized as follows. In Section~\ref{sec:model}, the $B-L$ MSSM model at the TeV scale is described in detail. This includes a discussion of both $B-L$ and electroweak symmetry breaking and much of our notation. The connection between the UV picture at the scale of gauge coupling unification and the TeV scale is outlined in Section~\ref{sec:unif}.  Specifically, the important mass scales and energy regimes are fully described and the relationships between the different mass scales are presented. The content of these two sections is expanded upon in Section~\ref{sec:1115}. Here, the details of the renormalization group equation (RGE) evolution of the parameters of the theory between the UV and TeV scales are discussed. This includes the appropriate input values of the parameters and the relevant equations--with reference to Appendix~\ref{sec:719} when necessary. Special attention is given to the right-handed sneutrino RGE which drives radiative $B-L$ symmetry breaking.  The relationship between the different running parameters and scales--introduced in Section~\ref{sec:unif}--is further discussed. The latter part of this section outlines the experimental bounds used in our analysis. First, collider bounds are discussed. These correspond to lower bounds on the physical sparticle masses--which are closely related to the running mass parameters. The section finishes by describing--and implementing--the well-known bounds from flavor changing neutral currents and CP violation.

This paper approaches the connection between UV and TeV physics in a novel way. Specifically, instead of assuming some universal conditions or relationships between UV soft SUSY breaking parameters, we simply allow all 
such parameters to be within about an order of magnitude of some chosen SUSY breaking scale. Our approach will then be to scan all relevant SUSY breaking parameters, at the high scale, over this possible range. These parameters will then be RG evolved to the TeV scale, following the discussion in Section~\ref{sec:1115}. This very general approach, as well as other details of our scan, are described in Section~\ref{sec:1141}. Such an approach is especially applicable to the string realization of the $B-L$ MSSM model, since, in that case, each chiral supermultiplet  arises from a different $\mathbf{16}$ representation of $SO(10)$.  Therefore, their soft SUSY breaking masses do not obey boundary conditions at the high scale. Our analysis is also valid for a wide range of pure GUTs, which can impose disparate  boundary conditions at the unification scale--including none at all. Finally a ``meta'' scan is conducted to choose the optimal value for the range of SUSY breaking mass scales.

The optimal range, arrived at in Section~\ref{sec:1141}, is used to generate all subsequent results in this paper--the bulk of which are presented in Section~\ref{sec:1058}. Relating these results to the twenty-four phenomenologically relevant scanned parameters is daunting at best. Fortunately, a cohesive picture can be presented in terms of two
so-called $S$-parameters. These $S$-parameters, which are each the sum of the squares of SUSY breaking mass parameters, play an important role in radiative $B-L$ symmetry breaking. Their role in this capacity is presented in Fig.~\ref{fig:1204}. The results for all other experimental constraints can also be expressed in terms of these two parameters--see Fig.~\ref{fig:1205} and Fig.~\ref{fig:1148}. A central result of this work is given is Fig.~\ref{fig:1039}, which displays the frequency at which particles appear as LSPs in our scan. The section closes with histograms of the spectra,  as well as some spectrum plots to help characterize specific features of our results.

Fine-tuning in our approach is addressed in Section~\ref{sec:FT}. While fine-tuning in the $B-L$ MSSM is not drastically different than in the MSSM, there are several key issues to highlight. First, while one might expect that the scale associated with $B-L$ symmetry breaking  could introduce new contributions to fine-tuning, it is shown that this is not the case. Second, the MSSM, analyzed using the same methods as in this paper, typically yields equal or more fine-tuning than in the our model for similar initial points. Finally, an LSP analysis similar to Fig.~\ref{fig:1039}, but with fine-tuning constrained to be better than one part in a thousand, is presented in Fig.~\ref{fig:417}. We conclude in Section~\ref{sec:conclusion}.

In addition to the main sections of this paper, three Appendices are included to help elucidate various topics. Appendix~\ref{sec:719} contains all one-loop RGEs for this model in the different regimes. The first part of Appendix~\ref{sec:131} specifies how to relate the running soft mass parameters to the physical masses of the SUSY particles. The second part of Appendix~\ref{sec:131} describes the procedure used to calculate the SM-like Higgs mass. Those readers interested in the details of how the random scan in this paper was conducted, are directed to Appendix~\ref{sec:536}.

%
\section{The TeV Scale Model}
\label{sec:model}
%
Motivated by both phenomenological considerations and string theory, we analyze the minimal anomaly free extension of the MSSM with gauge group
\begin{eqnarray}
	SU(3)_C\otimes SU(2)_L\otimes U(1)_{3R}\otimes U(1)_{B-L} \ .
	\label{eq:458}
\end{eqnarray}
As discussed in~\cite{Ovrut:2012wg}, we prefer to work with the Abelian factors $U(1)_{3R}\otimes U(1)_{B-L}$ rather than $U(1)_{Y}\otimes U(1)_{B-L}$-- although they are physically equivalent. This is motivated by the fact that the former is the unique choice that does not introduce kinetic mixing between the associated field strengths at any scale in their renormalization group equation (RGE) evolution. 
The gauge covariant derivative can be written as
\beq
D=\partial-iI_{3R}g_RW_{R}-i\frac{B-L}{2}g_{BL}B^\prime \ ,
\eeq
where $I_{3R}$ is the $U(1)_{3R}$ charge and the factor of $\frac{1}{2}$ is introduced in the last term by a redefinition of the gauge coupling $g_{BL}$-- thus simplifying many equations. 
As discussed in~\cite{Ovrut:2012wg} and throughout this paper, a radiatively induced vacuum expectation value (VEV) for a right-handed sneutrino will spontaneously  break the Abelian factors $U(1)_{3R} \times U(1)_{B-L}$ to $U(1)_Y$, in analogy with the way that the Higgs fields break $SU(2)_L\otimes U(1)_Y$ to $U(1)_{EM}$ in the SM. 
For simplicity, we will refer to this as ``$B-L$'' symmetry breaking--even though it is technically the breaking of a linear combination of the $U(1)_{3R}$ and $U(1)_{B-L}$ generators, leaving the hypercharge group generated by
\beq
	Y = I_{3R} + \frac{B-L}{2} 
\eeq
invariant.
The particle content of the minimal model is simply that of the MSSM plus three right-handed neutrino chiral multiplets. That is, three generations of matter superfields
\begin{eqnarray}
	Q=\trix{c}u\\d\notrix\sim({\bf 3}, {\bf 2}, 0, \frac{1}{3}) & \begin{array}{rl}u^c\sim&(\bar{\bf 3}, {\bf 1}, -1/2, -\frac{1}{3}) \\
	d^c\sim&(\bar{\bf 3}, {\bf 1}, 1/2, -\frac{1}{3})\end{array} \ , \nonumber\\
	L=\trix{c}\nu\\e\notrix\sim({\bf 1}, {\bf 2}, 0, -1)&\begin{array}{rl}\nu^c\sim&({\bf 1}, {\bf 1}, -1/2, 1)\\
	e^c\sim&({\bf 1}, {\bf 1}, 1/2, 1)\end{array} \ ,
	\label{eq:246}
\end{eqnarray}
along with two Higgs supermultiplets
\begin{eqnarray}
	H_u=\trix{c}H_u^+\\H_u^0\notrix&\sim&({\bf 1}, {\bf 2}, 1/2, 0) \ ,\nonumber\\
	H_d=\trix{c}H_d^0\\H_d^-\notrix&\sim&({\bf 1}, {\bf 2}, -1/2, 0) \ .
	\label{eq:247}
\end{eqnarray}
We refer to this model throughout the remainder of this paper as the $B-L$ MSSM.

The superpotential of the $B-L$ MSSM is given by
\begin{eqnarray}
	W=Y_u Q H_u u^c - Y_d Q H_d d^c -Y_e L H_d e^c +Y_\nu L H_u \nu^c+\mu H_u H_d \ ,
\end{eqnarray}
where flavor and gauge indices have been suppressed and the Yukawa couplings are three-by-three matrices in flavor space. In principle, the Yukawa matrices are arbitrary complex matrices. However, the observed smallness of the three CKM mixing angles and the CP-violating phase dictate that the quark Yukawa matrices be taken to be nearly diagonal and real. The lepton Yukawa coupling matrix can also be chosen to be diagonal and real. This is accomplished  by moving the rotation angles and phases into the neutrino Yukawa couplings which, henceforth, must be complex matrices. Furthermore, the smallness of the first and second family fermion masses implies that all components of the up, down, and lepton Yukawa couplings--with the exception of the (3,3) components--can be neglected for the purposes of this paper. Similarly, the very light neutrino masses imply that the neutrino Yukawa couplings are sufficiently small so as to be neglected for the purposes of this paper. The $\mu$-parameter can be chosen to be real, but not necessarily positive, without loss of generality. The soft supersymmetry breaking Lagrangian is then given by
\begin{align}
\begin{split}
	-\mathcal L_{\mbox{\scriptsize soft}}  = &
	\left(
		\frac{1}{2} M_3 \tilde g^2+ \frac{1}{2} M_2 \tilde W^2+ \frac{1}{2} M_R \tilde W_R^2+\frac{1}{2} M_{BL} \tilde {B^\prime}^2
	\right.
		\\
	& \left.
		\hspace{0.4cm} +a_u \tilde Q H_u \tilde u^c - a_d \tilde Q H_d \tilde d^c - a_e \tilde L H_d \tilde e^c
		+ a_\nu \tilde L H_u \tilde \nu^c + b H_u H_d + h.c.
	\right)
	\\
	& + m_{\tilde Q}^2|\tilde Q|^2+m_{\tilde u^c}^2|\tilde u^c|^2+m_{\tilde d^c}^2|\tilde d^c|^2+m_{\tilde L}^2|\tilde L|^2
	+m_{\tilde \nu^c}^2|\tilde \nu^c|^2+m_{\tilde e^c}^2|\tilde e^c|^2 \\
	&+m_{H_u}^2|H_u|^2+m_{H_d}^2|H_d|^2 \ .
\end{split}
\end{align}
The $b$ parameter can be chosen to be real and positive without loss of generality. The gaugino soft masses can, in principle, be complex. This, however, could lead to CP-violating effects that are not observed. Therefore, we proceed by assuming they all are real. The $a$-parameters and scalar soft mass can, in general, be Hermitian matrices in family space. Again, however, this could lead to unobserved flavor and CP violation. Therefore, we will assume they all are diagonal and real. Furthermore, we assume that only the (3,3) components of the up, down, and lepton $a$-parameters are significant and that the neutrino $a$ parameters are negligible. For more explanation of these assumptions, see Section \ref{sec:857}.

Spontaneous breaking of $B-L$ symmetry results from a right-handed sneutrino developing a non-vanishing VEV, since it carries the appropriate $I_{3R}$ and $B-L$ charges. However, since sneutrinos are singlets under the $SU(3)_C\otimes SU(2)_L\otimes U(1)_{Y}$ gauge group, it does not break any of the SM symmetries.  To acquire a VEV, a right-handed sneutrino must develop a tachyonic mass\footnote{Here and throughout this paper we use the term ``tachyon'' to describe a scalar particle whose $m^{2}$ parameter is negative. Although all $m^{2}$ parameters at high scale will be chosen positive, one or more can be driven negative at lower energy by radiative corrections. This signals dynamical instability at the origin--although a stable VEV may, or may not, develop.}. As discussed in \cite{Mohapatra:1986aw, Ghosh:2010hy, Barger:2010iv}, a VEV can only be generated in one linear combination of the right-handed sneutrinos. Furthermore, beyond the fact that its VEV breaks $B-L$ symmetry, in which combination it occurs has no further observable effect. This is because there is no right-handed charged current to link the right-handed neutrinos to a corresponding right-handed charged lepton. Therefore, without loss of generality, one can assume that it is the third generation right-handed sneutrino that acquires a VEV. At a lower mass scale, electroweak symmetry is spontaneously broken by the neutral components of both the up and down Higgs multiplets acquiring non-zero VEV's.  In combination with the right-handed sneutrino VEV, this also induces a VEV in each of the three generations of left-handed sneutrinos. The notation for the relevant VEVs is
\beq
	\left< \tilde \nu^c_3 \right> \equiv \frac{1}{\sqrt 2} v_R, \ \ \left<\tilde \nu_i\right> \equiv \frac{1}{\sqrt 2} {v_L}_i, \ \
	\left< H_u^0\right> \equiv \frac{1}{\sqrt 2}v_u, \ \ \left< H_d^0\right> \equiv \frac{1}{\sqrt 2}v_d,
\eeq
where $i=1,2,3$ is the generation index.

The neutral gauge boson that becomes massive due to $B-L$ symmetry breaking, $Z_R$, has a mass at leading order, in the relevant limit that $v_R \gg v$, of
\beq
	M_{Z_R}^2 = \frac{1}{4}\left(g_R^2+g_{BL}^2 \right) v_R^2
				\left(1+\frac{g_R^4}{g_R^2+g_{BL}^2} \frac{v^2}{v_R^2}\right) \ ,
	\label{eq:237}
\eeq
where
\beq
	v^2 \equiv v_d^2 + v_u^2 \ .
\eeq
The second term in the parenthesis is a small effect due to mixing in the neutral gauge boson sector. The hypercharge gauge coupling is given by
\beq
	\label{eq:Y.3R.BL}
	g_Y = g_R \sin \theta_R = g_{BL}\cos \theta_R \ ,
\eeq
where
\beq
		\cos \theta_R = \frac{g_R}{\sqrt{g_R^2+g_{BL}^2}} \ .
\eeq

Since the neutrino masses are roughly proportional to the ${Y_\nu}_{ij}$ and ${v_L}_i$ parameters, it follows that ${Y_\nu}_{ij} \ll 1$ and ${v_L}_i\ll v_{u,d}, v_R$. In this phenomenologically relevant limit, the minimization conditions of the potential are simple and worthwhile to note. They are
\begin{align}
	\label{eq:MC.vR}
	v_R^2=&\frac{-8m^2_{\tilde \nu_{3}^c}  + g_R^2\left(v_u^2 - v_d^2 \right)}{g_R^2+g_{BL}^2} \ ,
	\\
	{v_L}_i=&\frac{\frac{v_R}{\sqrt 2}(Y_{\nu_{i3}}^* \mu v_d-a_{\nu_{i3}}^* v_u)}
			{m_{\tilde L_{i}}^2-\frac{g_2^2}{8}(v_u^2-v_d^2)-\frac{g_{BL}^2}{8}v_R^2} \ ,
	\\
	\label{eq:EW.mu}
	\frac{1}{2} M_Z^2 =&-\mu^2+\frac{m_{H_u}^2\tan^2\beta-m_{H_d}^2}{1-\tan^2\beta} \ ,
	\\
	\label{eq:EW.b}
	\frac{2b}{\sin2\beta}=&2\mu^2+m_{H_u}^2+m_{H_d}^2 \ .
\end{align}
Here, the first two equations correspond to the sneutrino VEVs. The third and fourth equations are of the same form as in the MSSM, but new $B-L$ scale contribution to $m_{H_u}$ and $m_{H_d}$ shift their values significantly compared to the MSSM. Eq.~(\ref{eq:MC.vR}) can be used to re-express the $Z_R$ mass as
\beq
	\label{eq:MZR.mnuc}
	M_{Z_R}^2 = -2 m_{\tilde \nu^c_3}^2
				\left(1+\frac{g_R^4}{g_R^2+g_{BL}^2} \frac{v^2}{v_R^2}\right) \ .
\eeq
This makes it clear that, to leading order, the $Z_R$ mass is determined by the soft SUSY breaking mass of the third family right-handed sneutrino. The term proportional to $v^2/v_R^2$ is insignificant in comparison and, henceforth, neglected  in our calculations.

A direct consequence of generating a VEV for the third family sneutrino is the spontaneous breaking of $R$-parity. The induced operators in the superpotential are
\begin{equation}
	\label{W.brpv}
	W \supset \epsilon_i \,  L_i \,  H_u - \frac{1}{\sqrt 2 }{Y_e}_i \, {v_L}_i \,  H_d^- \,   e^c_i \ ,
\end{equation}
where
\begin{equation}
	\epsilon_i  \equiv \frac{1}{\sqrt 2} {Y_\nu}_{i3} v_R \ .
\end{equation}
This general pattern of $R$-parity violation is referred to as bilinear $R$-parity breaking and has been discussed in many different contexts, especially in reference to neutrino masses-- see references~\cite{Mukhopadhyaya:1998xj,Chun:1998ub, Chun:1999bq, Hirsch:2000ef} for early works. In addition, the Lagrangian contains additional bilinear terms generated by ${v_L}_i$ and $v_R$ from the super-covariant derivative. These are
\begin{align}
\begin{split}
	\label{L.n}
	\mathcal{L} \supset &
	- \frac{1}{2}{v_L}_i^* \left[ g_2 \left(\sqrt 2 \, e_i \tilde W^+ 
	+  \nu_i \tilde W^0\right) - g_{BL} \nu_i \tilde B' \right]
	\\
	&
	-\frac{1}{2} v_R \left[-g_R \nu_3^c \tilde W_R + g_{BL} \nu_3^c \tilde B' \right]+ \text{h.c.}
\end{split}
\end{align}
The consequences of spontaneous $R$-parity violation are quite interesting, and have been discussed in a variety of papers. For LHC studies, see~\cite{FileviezPerez:2012mj, Perez:2013kla} as well as recent work on stop and sbottom LSP's in this context and the connection between their decays and the neutrino sector~\cite{Marshall:2014kea, Marshall:2014cwa}. Predictions for the neutrino sector were discussed in~\cite{Mohapatra:1986aw, Ghosh:2010hy, Barger:2010iv}. It was shown that the lightest left-handed, or active, neutrino is massless and that the model contains two right-handed neutrinos, referred to as sterile neutrinos, that are lighter than the remaining two active neutrinos. Sterile neutrinos can influence the cosmological evolution of the universe due to their role as dark radiation. This effect was studied in~\cite{Perez:2013kla}. 

In this section, we have focussed on the TeV scale manifestation of the $B-L$ MSSM. However, the main content of this paper will be to study the connection between this low energy theory and its possible origins in $E_8 \otimes E_8$ heterotic string theory, thereby linking some of the $B-L$ MSSM phenomenology to high scale physics. In this context, our model is the remnant of an $SO(10)$ unified symmetry broken by two Wilson lines. Although the details of the various physical regimes of this theory, and the renormalization group scaling between them, were given in~\cite{Ovrut:2012wg}, we review them in the next section for completeness.

%
\section{Journey From the Unification Scale}
\label{sec:unif}
%
The goal of this section is to review the physics associated with the string construction of the $B-L$ MSSM--from unification to the electroweak scale. After compactification to four-dimensions, the unified gauge group is $SO(10)$. This is then further broken to the  $B-L$ MSSM gauge group by the turning on of two Abelian Wilson lines, denoted by $\chi_{3R}$ and $\chi_{B-L}$ respectively. The energy scales associated with these Wilson lines need not be the same. In fact, exact gauge coupling unification at one-loop, which we will assume throughout this paper, requires that the scales be different-- implying there is a two-step symmetry breaking process from $SO(10)$ to the gauge group of the $B-L$ MSSM. This leads to an intermediate regime between the two scales associated with the Wilson lines. The particle content and gauge group in this regime depends on which Wilson line turns on first. Defining the mass scales of $\chi_{3R}$ and $\chi_{B-L}$ as $M_{\chi_{3R}}$ and $M_{\chi_{B-L}}$ respectively, we find the following two initial symmetry breaking patterns.
\begin{itemize}
	\item $M_{\chi_{B-L}} > M_{\chi_{3R}}$: ~~$SO(10) \to SU(3)_C \otimes SU(2)_L \otimes SU(2)_R \otimes U(1)_{B-L}$, ~the ``left-right'' model 
	\item  $M_{\chi_{3R}} > M_{\chi_{B-L}}$: ~~$SO(10) \to SU(4)_C \otimes SU(2)_L \otimes U(1)_{3R}$, ~a modified version of the ``Pati-Salam'' model
\end{itemize}
In each case, the subsequent turning on of the second Wilson line breaks the intermediate model to the $B-L$ MSSM.

Reference~\cite{Ovrut:2012wg} studied these two cases and found that gauge coupling unification dictates that the Wilson line scales should be separated by less than an order of magnitude; that is, the intermediate regime is not very large. It follows that the TeV scale physics has little dependence on which of the above two models inhabit the intermediate regime. For simplicity, we will carry out our analysis under the assumption that it is the first of these symmetry breaking patterns that occurs. Hence, the intermediate regime contains the left-right model. We then make the identifications
\begin{align}
	 M_U & \equiv M_{\chi_{B-L}},~ \text{ the scale of gauge coupling unification}
	 \\
	 M_\i & \equiv M_{\chi_{3R}}, ~ \ \text{ the intermediate scale}
\end{align}
These scales will be further discussed below.

In the intermediate regime, the particle content of the left-right model consists of nine copies of the matter family
\begin{eqnarray}
	\label{eq:LR.matter}
	Q\sim({\bf 3}, {\bf 2}, {\bf 1}, \frac{1}{3}), \quad &Q^c=
	\trix{c}
	d^c\\
	u^c
	\notrix
	\sim(\bar{\bf 3}, {\bf 1}, {\bf 2}, -\frac{1}{3})\\
	L\sim({\bf 1}, {\bf 2}, {\bf 1}, -1), \quad &L^c=\trix{c}e^c\\ \nu^c\notrix\sim({\bf 1}, {\bf 1}, {\bf 2}, 1),
\end{eqnarray}
two copies of a Higgs bi-doublet, which contains the MSSM Higgs fields,
\begin{eqnarray}
	\label{eq:LR.Higgs}
	\mathcal{H}_1, \ \mathcal{H}_2 \sim({\bf 1}, {\bf 2}, {\bf 2}, 0) \ ,
\end{eqnarray}
and a pair of color triplets
\begin{equation}
	\label{eq:LR.triplets}
	H_C\sim({\bf 3}, {\bf 1}, {\bf 1}, 2), \quad \bar H_C\sim(\bar{\bf 3}, {\bf 1}, {\bf 1}, -2) \ .
\end{equation}
Once the second Wilson line turns on, the extra particle content integrates out and one is left with exactly the spectrum of the $B-L$ MSSM.

At this point, it is important to make a quick note on notation for the $B-L$ gauge coupling. Thus far, we have discussed the gauge parameter $g_{BL}$, which couples to $\frac{1}{2}(B-L)$ charge. As is well known, this gauge coupling has to be properly normalized so as to unify with the other gauge parameters. We use $g_{BL}'$ defined by 
\beq
	g_{BL}' = \sqrt{\frac{2}{3}} g_{BL}
\eeq	
to denote the properly unifying coupling. The parameter $g_{BL}'$ couples to $\sqrt{\frac{3}{8}} (B-L)$ charge and will appear in the RGEs. For quantities of physical interest, such as physical masses, $g_{BL}$ will be used.

To fully understand the evolution of this model from unification to the electroweak scale, it should be noted that there are five relevant mass scales of interest, two of which were mentioned briefly above. All five are described in the following:
\begin{itemize}
	\item $M_U$: ~The unification mass and the scale of the first Wilson line. We assume that all gauge couplings unify at this scale. That is, ~$g_3 = g_2 = g_R = g_{BL}' = g_U$.
	\item $M_\i$: ~The intermediate scale associated with the second Wilson line and the symmetry breaking $SU(2)_R \to U(1)_{3R}$ -- that is, the right-handed isospin breaks into its third component. Since the gauge coupling of $SU(2)_R$ slightly above $M_\i$ is equal to the $U(1)_{3R}$ gauge coupling slightly below $M_\i$, we use $g_R$ for both $SU(2)_R$ and $U(1)_{3R}$. All gauge couplings have trivial thresholds at this scale.
	\item $M_{B-L}$: ~The $B-L$ scale is the mass at which the right-handed sneutrino VEV triggers $U(1)_{3R}\otimes U(1)_{B-L} \to U(1)_Y$. Physically, this corresponds to the mass of the neutral gauge boson $Z_R$ of the broken symmetry and, therefore, the scale of $Z_R$ decoupling. Specifically
	\beq
		M_{Z_R} = M_{B-L},
		\label{eq:249}
	\eeq
	where $M_{Z_R}$ depends on parameters evaluated at $M_{B-L}$--see Eq. (\ref{eq:237}). Substituting Eq. (\ref{eq:237}) into this relation yields a transcendental equation that must be solved using iterative numerical methods to obtain the correct value for $M_{B-L}$.
	
	At this scale, we also evaluate the hypercharge gauge coupling using its relationship to the gauge parameters of $B-L$ and the third component of right-handed isospin. This is given by
	\begin{equation}
		\label{eq:1.3R.BL}
		g_1 = \sqrt{\frac{5}{3}} g_R \sin \theta_R = \sqrt{\frac{5}{2}} g_{BL}'\cos \theta_R \ ,
	\end{equation}
	where
	\beq
		\cos \theta_R = \frac{g_R}{\sqrt{g_R^2+\frac{3}{2} g_{BL}'^2}} \ .
	\eeq
	Note that Eq.~(\ref{eq:1.3R.BL}) is just a restatement of Eq.~(\ref{eq:Y.3R.BL}) with gauge couplings properly normalized for unification, including a rescaled hypercharge gauge coupling $g_1$ defined by
	\beq
		g_1 = \sqrt{\frac{5}{3}} g_Y \ .
	\eeq
	\item $M_\text{SUSY}$: The soft SUSY breaking scale. This is the scale at which all sparticles are integrated out with the exception of the right-handed sneutrinos. The right-handed sneutrinos are associated with $B-L$ breaking and, therefore, are integrated out at the $B-L$ scale. While there is obviously no single scale associated with the masses of all the SUSY partners, we use the scale of stop decoupling given by
\beq
	M_\text{SUSY} = \sqrt{m_{\tilde t_1}\ m_{\tilde t_2}}.
	\label{eq:358}
\eeq	
This scale is useful because when the stops decouple, the parameter that controls electroweak symmetry breaking, that is, the soft $H_u$ mass parameter, effectively stops running-- see~\cite{Gamberini:1989jw} for more details. Like the $B-L$ scale, the SUSY scale must be determined using iterative numerical methods because the physical stop masses in Eq. (\ref{eq:358}) depend implicitly on the SUSY scale. 
	\item $M_{\text{EW}}$:~The electroweak scale. This is the well-known scale associated with the $Z$ and $W$ gauge bosons of the SM. We will make the identification
	\beq
		M_{\text{EW}} = M_Z.
	\eeq
	For correct electroweak breaking, one must satisfy the conditions
	\begin{align}
		2b&<2\mu^2+m_{H_u}^2+m_{H_d}^2 \ ,
		\\
		b^2&> (\mu^2+M_{H_d}^2)(\mu^2+M_{H_u}^2)  \ .
	\end{align}
	The first constraint guarantees that the Higgs potential is bounded from below while the second indicates that the trivial vacuum is not stable.
\end{itemize}

With the relevant mass scales appropriately defined, we can now discuss the physical regimes that exist in between them. To begin with, we will be interested in the evolution of the gauge couplings--since our assumption that they unify will help relate these disparate scales to each other. We present below, for each regime, the slope factors $b_a$ appearing in the gauge RGE's 
\beq
	\frac{d}{d t} \alpha_a^{-1} = -\frac{b_a}{2 \pi} \ ,
\eeq
where $a$ indexes the associated  gauge groups. Note that while $M_U > M_\i \gg M_{B-L}, M_{\text{SUSY}}$, the hierarchy between the SUSY and $B-L$ scales depends on the point chosen in the initial parameter space. Each of the two possibilities will be addressed below.
\begin{itemize}
	\item $M_U \ - \ M_\i$: This regime is populated by the left-right model discussed above. In this interval, the $b_{a}$ factors are
	\begin{equation}
		b_3 = 10,\ b_2 = 14,\ b_R=14,\ b_{B-L}=19 \ .
		\label{eq:646}
	\end{equation}
We will refer to this scaling interval as the ``left-right regime'' and, when required, denote the associated $b$-coefficients by $b_{a}^{\rm LR}$.
	\item $M_\i - {\rm max}(M_{\text{SUSY}}, M_{B-L})$: This regime is populated by the $B-L$ MSSM model. The $b_{a}$ factors in this case are
	\begin{equation}
		b_3 = -3,\ b_2 = 1,\ b_{R}=7,\ b_{B-L}=6 \ .
	\end{equation}
We will refer to this scaling interval as the ``B-L MSSM regime'' and, when required, denote its $b$-coefficients by $b_{a}^{\rm BL}$.
\end{itemize}
The remaining two regimes depend on which of the following two cases occurs: $M_{B-L} > M_{\text{SUSY}}$--the ``right-side-up'' hierarchy--and $M_{\text{SUSY}} > M_{B-L}$--the ``upside-down'' hierarchy. \\

\noindent \underline{right-side-up hierarchy}:
\begin{itemize}
	\item $M_{B-L} - M_{\text{SUSY}}$: ~In this case $B-L$ has been broken but SUSY is still a good symmetry, thereby giving an MSSM-like theory--that is, the MSSM plus two light right-handed neutrino chiral multiplets. Another possible deviation might occur in the composition of the bino--more about this late. In general, however, this is the MSSM. Specifically, the gauge couplings in this regime evolve like the well-known MSSM gauge couplings with $b_{a}$ coefficients
	\begin{equation}
		b_3 = -3,\ b_2 = 1,\ b_{1}=\frac{33}{5} \ .
	\end{equation}
We refer to this interval as the ``MSSM'' regime and  denote the associated $b$-coefficients by $b_{a}^{\rm MSSM}$.
	\item $M_{\text{SUSY}} - M_{\text{EW}}$: ~In this regime, one simply has the SM with two sterile neutrinos. It has the well-known slope factors
	\begin{equation}
		b_3 = -7,\ b_2 = -\frac{19}{6},\ b_{1}=\frac{41}{10} \ .
		\label{eq:644}
	\end{equation}
We refer to this as the ``SM'' regime and  denote the $b$-coefficients by $b_{a}^{\rm SM}$.
\end{itemize}
\noindent \underline{upside-down hierarchy}:
\begin{itemize}
	\item $M_{\susy} - M_{B-L}$: ~Now $B-L$ remains a good symmetry below the average stop mass, where we effectively integrated out the SUSY partners. The resulting theory is simply a non-SUSY $SU(3)_C \otimes SU(2)_L \otimes U(1)_{3R} \otimes U(1)_{B-L}$ model, which also includes three generations of right-handed sneutrinos--the third of which acts as the $B-L$ Higgs. The slope factors are
	\begin{equation}
		b_3 = -7,\ b_2 = \frac{19}{6},\ b_{R}= \frac{53}{12}, \ b_{BL} = \frac{33}{8}.
	\end{equation}
	\item $M_{B-L} - M_{\text{EW}}$: ~Here, again, we have the SM with two sterile neutrinos and  the slope factors given in Eq.~(\ref{eq:644}).
\end{itemize}

Given the above information, and the demand that all gauge couplings unify, we can solve for a given mass scale in terms of the others. First consider the unification mass--corresponding to the scale at which the four gauge couplings  become equal to each other. Practically, it is derived as the energy-momenta at which $g_3 = g_2$. As is well-known, this will not be influenced by any scale that acts as a threshold for complete multiplets of a minimal group that unifies $SU(3)$ and $SU(2)$--for example, $SU(5)$. The $B-L$  and intermediate scales are both such thresholds. The $B-L$ scale is a threshold for singlets of $SU(5)$, that is, the right-handed neutrinos, while $M_\i$ is a threshold for six new matter generations, a pair of Higgs doublets and their $SU(5)$ color partners. All of these particles fit into the \textbf{1}, \textbf{5}, $\mathbf{\bar 5}$ and \textbf{10} of $SU(5)$-- see Eqs.~(\ref{eq:LR.matter})-(\ref{eq:LR.triplets}). Working through the algebra of setting $g_3(M_U) = g_2(M_U)$ yields
\beq
	M_U  = \left[
		e^{\frac{2 \pi \left(\alpha_3 - \alpha_2\right)}{\alpha_2 \alpha_3}}
		M_Z^{\left(b_2^\text{SM} - b_3^\text{SM}\right)}
		M_\text{SUSY}^{\left(b_2^\text{MSSM} - b_2^\text{SM}+b_3^\text{SM} - b_3^\text{MSSM}\right)}
	\right]^\frac{1}{b_2^\text{LR} - b_3^\text{LR}},
\eeq
where the superscripts on the slope factors indicate their regime of relevance and the $\alpha_{i}$ take their experimental values at $M_Z$~\cite{PDG}:
\begin{equation}
	\label{eq:alpha.ew}
	\alpha_3(M_Z)=0.118,\ \alpha_2(M_Z)=0.0337,\ \alpha_1(M_Z)=0.0170 \ .
\end{equation}
Inserting all the coefficients, the unification scale becomes
\beq
	M_U \simeq
	2.186 \times 10^{16} \left(\frac{M_\text{SUSY}}{\text{1~GeV}}\right)^{0.0417} \ (\text{GeV}) \ .
\label{b1}
\eeq

\begin{figure}[h]
\centering
	\includegraphics[scale=0.6]{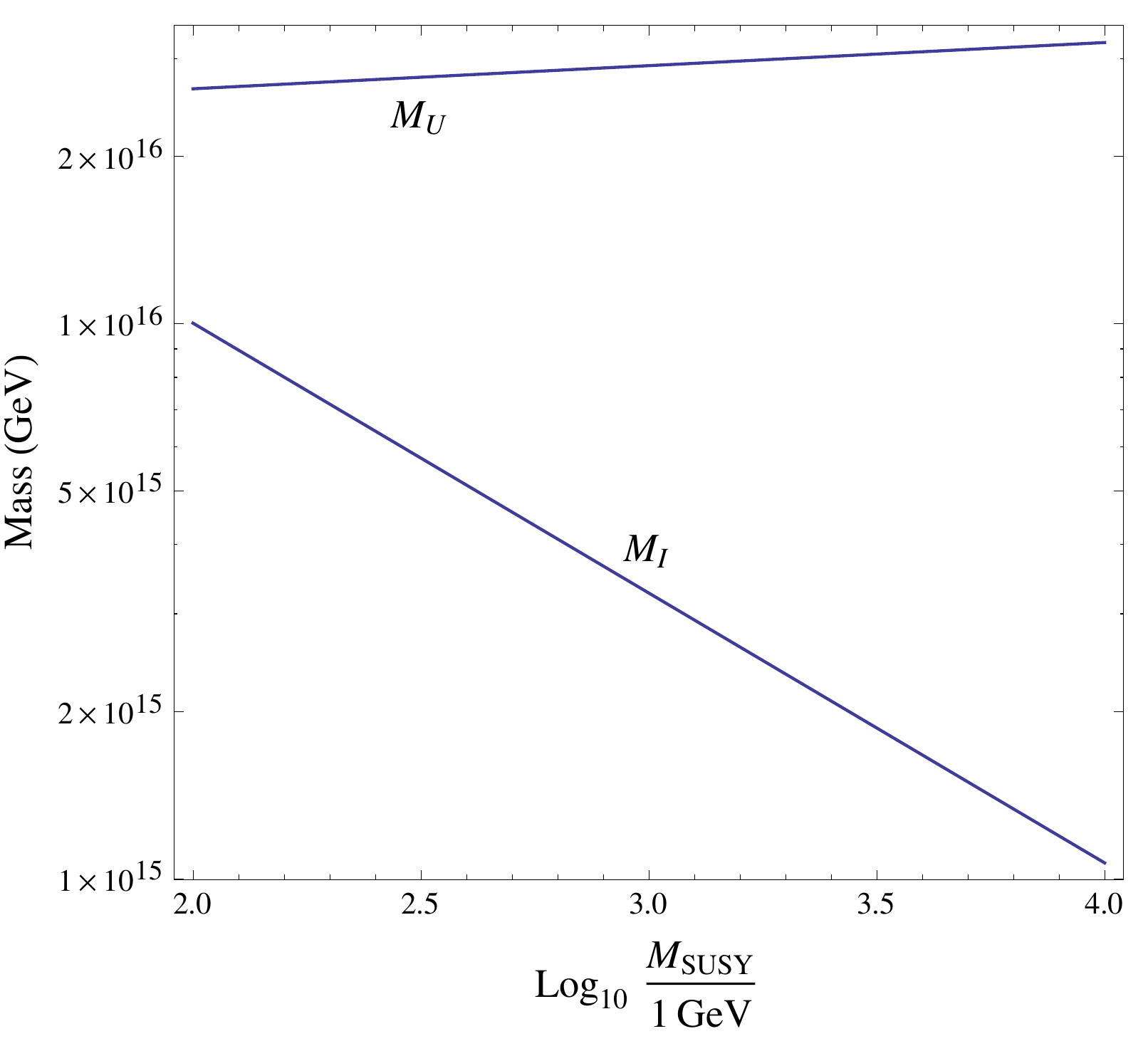}
	\caption{The unification and intermediate scales as a function of the SUSY mass scale. These two scales have no dependence on the $B-L$ scale.}
	\label{fig:MU.MI.MSUSY}
\end{figure}

Similarly, the intermediate scale can be solved for by setting $g_{R}(M_U)=g'_{BL}(M_U) = g_3(M_U)$ and using the relationship between the gauge couplings of hypercharge , $B-L$ and the third component of right-handed isospin given in Eq.~(\ref{eq:1.3R.BL}). The intermediate scale is found to be
\begin{align}
\begin{split}
	M_\i & =
	\left[
		e^{\frac{10 \pi \left(\alpha_1 - \alpha_2\right)}{\alpha_1 \alpha_2}}
		M_Z^{5\left(b_2^\text{SM} - b_1^\text{SM}\right)}
		M_\text{SUSY}^{5\left(b_2^\text{MSSM} - b_2^\text{SM}+b_1^\text{SM} - b_1^\text{MSSM}\right)}
	\right.
	\\
	& \hspace{1cm}  \left.
		M_U^{\left(3 b_R^\text{LR} + 2 b_{BL}^\text{LR} - 5 b_2^\text{LR}  \right)}
	\right]^\frac{1}
		{
			5 \left(b_2^\text{BL} -b_2^\text{LR}\right)
			+ 2\left(b_{BL}^{\text{LR}} - b_{BL}^{\text{BL}} \right)
			+ 3 \left(b_R^\text{LR} - b_R^\text{BL} \right)}.
\end{split}
\end{align}
Substituting for $M_U$ using Eq. (\ref{b1}) gives
\beq
	M_\i \simeq 1.835 \times 10^{17} \left(\frac{M_\text{SUSY}}{\text{1~GeV}}\right)^{-0.486} \ (\text{GeV}).
\eeq
These relationships are displayed in Fig.~\ref{fig:MU.MI.MSUSY}.

%
\section{The Framework}
\label{sec:1115}
%
The purpose of this paper is twofold: (i) to analyze how the high scale parameter space is associated with the TeV scale phenomenology of the $B-L$ MSSM and (ii) to introduce a new way of analyzing this question. More specifically, (i) is an attempt to understand the  kind of general UV conditions that yield valid, experimentally consistent points at the TeV scale--especially conditions that lead to $B-L$ breaking and a viable $B-L$/electroweak hierarchy, a topic which has not been widely studied before. Such a study requires a framework for evolving parameters through the relevant regimes. The skeleton of this framework was discussed in the previous section. This section will continue this discussion and review the well-known experimental constraints on the SUSY parameters. Utilizing these constraints, the next section presents the motivation and the strategy for our scan. This addresses (ii).

The approach to the RG evolution of the parameters is similar to other such work, with several deviations that will be highlighted below. The RGEs of interest are calculated using reference~\cite{Martin:1993zk} and are presented in Appendix~\ref{sec:719}. Gauge couplings and gaugino masses are evolved up to the unification scale. The remaining parameters, Yukawa couplings, sfermion mass parameters and $a$-terms, are only evaluated in the scaling regimes below the intermediate scale. This is because in the string construction considered here, the scaling regime between the unification scale and $M_\i$ contains six additional copies of matter fields as well as an additional copy of Higgs fields.We note that each component field of a given generation of matter originates from a different \textbf{16} of $SO(10)$. This is important and will be discussed later. Since these new Yukawa couplings are unknown, RG  running them through this regime would not contribute to the predictability of this study. In practice, we implement these calculations piecewise starting with the analytically tractable equations first. These are the gauge couplings, gaugino mass parameters and the first and second generation sfermion mass parameters, as well as all sneutrino mass parameters. We then numerically calculate the evolution of the remaining parameters.

As is traditional, we begin by inputting the experimentally determined parameters--that is, the gauge couplings and Yukawa couplings derived from fermion masses--at the electroweak scale. The initial values of the gauge couplings were given above in Eq.~(\ref{eq:alpha.ew}). For the purposes of this paper, the SM Yukawa couplings, which are three-by-three matrices in flavor space,  can all be approximated to be zero except for the three-three elements which give mass to the third generation SM fermions. We use the initial conditions
\beq
	y_t=0.955,\quad y_b=0.0174,\quad y_\tau=0.0102.
\eeq
For details on relating fermion masses to Yukawa couplings, see~\cite{Djouadi:2005gi}. Here the lower case $y$ represents Yukawa couplings in the non-SUSY regime. These can be evolved to the SUSY scale, both in the right-side-up hierarchy, Eqs.~(\ref{eq:SM:yt}) -~(\ref{eq:SM:ytau}), and up-side-down hierarchy, Eqs.~(\ref{eq:BLSM:yt}) -~(\ref{eq:BLSM:ytau}). At the SUSY scale, one has the non-trivial boundary conditions
\begin{eqnarray}
	y_{t}(M_\susy)&=&Y_{t}(M_\susy)\sin\beta\nonumber\\
	y_{b,\tau}(M_\susy)&=& Y_{b,\tau}(M_\susy)\cos\beta.
\end{eqnarray}
The boundary condition at the $B-L$ scale is trivial. Above the $B-L$ and SUSY scales, the Yukawa couplings are only evolved up to the intermediate scale utilizing the RGEs in Eqs.~(\ref{eq:$B-L$ MSSM.yt}) - (\ref{eq:$B-L$ MSSM.ytau}).

The gauge couplings in the various regimes were discussed in the previous section. With those solutions in hand, the RGE evolution of the gauginos can be easily derived. Gaugino masses are inputted at the unification scale and evolved down. Naively, one might expect gaugino mass unification. However, this is not always the case--as has been discussed in a number of contexts, see for example~\cite{Ellis:1985jn, Choi:2005ge}. Therefore, and to be as general as possible, we impose no relationship between the different gaugino masses at the unification scale. The general RGE for a gaugino mass parameter is
\begin{equation}
	\frac{d}{dt} M_a = \frac{b_a \alpha_a M_a}{2 \pi},
\end{equation}
where $a$ indexes the gauge groups. These equations can be solved analytically. For the gauginos associated with $SU(3)_{C}$, $SU(2)_{L}$, $U(1)_{3R}$ and $U(1)_{B-L}$ one has
\begin{equation}
	M_a(t) = \frac{M_a(M_U)}{\alpha_a(M_U)} \alpha_a(t).
\end{equation}

The bino, however, is a treated somewhat differently for each of the two possible hierarchies between the $B-L$ and SUSY scales. For the right-side-up hierarchy, at the $M_{B-L}$ scale, we have three neutral fermions that mix: the third generation right-handed neutrino $\nu^c_3$, the $B-L$ gaugino (blino) and the $I_{3R}$ gaugino (rino). This is a direct consequence of $R$-parity violation in the $B-L$ MSSM. As we will see, it is possible for a neutralino LSP mass eigenstate to have a significant $\nu^c_3$ component.
The mixing between the third-family right-handed neutrino and the $U(1)$ gauginos is described in the $(\nu^c_3, \tilde W_R, \tilde B^\prime)$ basis by the mass matrix\footnote
{
	This mass matrix neglects mixing with the Higgsinos through the electroweak breaking Higgs VEV. This is a safe approximation since the lower bound on the $Z_R$ mass implies that the electroweak Higgs VEV will be negligible compared to the third-family right-handed sneutrino VEV.
}
\begin{eqnarray}
\trix{ccc}
0				&-\cos\theta_RM_{Z_R}	&\sin\theta_RM_{Z_R}	\\
-\cos\theta_RM_{Z_R}	&M_R				&0	\\
\sin\theta_RM_{Z_R}	&0				&M_{BL}	\\
\notrix \ .
\label{eq:1059}
\end{eqnarray}
Due to the RGEs monotonically pushing the values of $M_R$ and $M_{BL}$ down, they will typically be significantly lighter than $M_{Z_R}$. It is, therefore, instructive to perturbatively diagonalize this mass matrix in the limit $M_R, M_{BL}\ll M_{Z_R}$. At zeroth order, the mass eigenstates are
\begin{eqnarray}
	\label{eq:322}
	\tilde B&=&\tilde W_R\sin\theta_R+\tilde B^\prime\cos\theta_R\\
	\label{eq:323}
	\nu^c_{3a}&=&\frac{1}{\sqrt 2}(\nu^c_3-\tilde W_R\cos\theta_R+\tilde B^\prime\sin\theta_R)\\
	\label{eq:324}
	\nu^c_{3b}&=&\frac{1}{\sqrt 2}(\nu^c_3+\tilde W_R\cos\theta_R-\tilde B^\prime\sin\theta_R),
\end{eqnarray}
with masses
\begin{equation}
	M_1=0, ~~~m_{\nu^c_{3a}}=M_{Z_R}, ~~~m_{\nu^c_{3b}}=M_{Z_R} \ .
\label{buddy1}
\end{equation}

At first order, the effect of adding $M_R$ and $M_{BL}$ back into the mass matrix is to give the bino a mass of
\begin{eqnarray}
	M_1=\sin^2\theta_R M_R+\cos^2\theta_R M_{BL}.
\end{eqnarray}
This shows that, in the right-side-up hierarchy, between the scales $M_{B-L}$ and $M_\susy$ we have the gauge group and particle content of the MSSM plus two right-handed neutrino supermultiplets--that is, the two sneutrino generations that do not acquire a VEV.\footnote{At some points in parameter space, it is possible that the required limit will not be satisfied and there will not be a mass eigenstate that can clearly be identified as the bino. However, since the scaling regime between $M_{B-L}$ and $M_\susy$ is always small, the errors introduced by assuming the existence of a bino are insignificant.} Below the $B-L$ scale, the bino mass is
\begin{equation}
	M_1(t) = \frac{M_1(M_{B-L})}{\alpha_1(M_{B-L})} \alpha_1(t).
\end{equation}
In the upside-down case, all neutralinos are diagonalized at the SUSY mass scale. 

The running of the tri-linear $a$-terms is straightforward. Their initial values are randomly generated at the intermediate scale. The $a$-term RGEs in the $B-L$ MSSM regime are given in Eqs.~(\ref{eq:$B-L$ MSSM.at}) - (\ref{eq:$B-L$ MSSM.atau}), while those for the MSSM are in Eqs.~(\ref{eq:MSSM.at}) - (\ref{eq:MSSM.atau}). All relevant threshold conditions are trivial. 

The RGEs for the square of the soft sfermion mass parameters can be broken into two categories: 1) those with simple analytic solutions--given in Eqs.~(\ref{eq:$B-L$ MSSM.mQ1}) -~(\ref{eq:$B-L$ MSSM.mec}) and Eqs.~(\ref{eq:MSSM.mQ1}) -~(\ref{eq:MSSM.mec}) for the $B-L$ MSSM and MSSM regimes respectively--and 2) those requiring numerical solutions--given in Eqs.~(\ref{eq:$B-L$ MSSM.Hu}) - (\ref{eq:$B-L$ MSSM.tauc}) and Eqs.~(\ref{eq:MSSM.Hu}) -~(\ref{eq:MSSM.tauc}) for the $B-L$ MSSM and MSSM regimes. These parameters are all inputted at the intermediate scale. The third generation right-handed sneutrino is then evolved to the $B-L$ scale--while all other sfermion mass squared parameters are RG evolved to the SUSY scale. The third generation right-handed sneutrino mass squared plays an important role here since, when it runs negative, it triggers $B-L$ breaking as was discussed in detail in~\cite{Ambroso:2009jd, Ambroso:2009sc}. The right-handed sneutrino mass RGE is
\beq
	\label{eq:RGE.snu}
	16\pi^2\frac{d}{dt}m_{\tilde \nu^c_{3}}^2=-3g_{BL}^2M_{BL}^2-2g_{R}^2M_{R}^2+\frac{3}{4}g_{BL}^2S_{BL}-g_{R}^2S_{R} \ ,
\eeq
where
\begin{eqnarray}
	\label{eq:S.BLA}
	S_{BL}&=&\Tr(2m_{\tilde Q}^2-m_{\tilde u^c}^2-m_{\tilde d^c}^2-2m_{\tilde L}^2+m_{\tilde \nu^c}^2+m_{\tilde e^c}^2) \ ,\\
	\label{eq:S.RA}
	S_{R}&=&m_{H_u}^2-m_{H_d}^2+\Tr\left(-\frac{3}{2}m_{\tilde u^c}^2+\frac{3}{2}m_{\tilde d^c}^2-\frac{1}{2} m_{\tilde \nu^c}^2+\frac{1}{2} 	m_{\tilde e^c}^2\right).
\end{eqnarray}

Despite the lack of a large Yukawa coupling, the right-handed sneutrino mass can still be driven tachyonic by appropriate signs and magnitudes of the $S$-terms defined in Eqns (\ref{eq:S.BLA}, \ref{eq:S.RA}). To emphasize this, the analytic solution to the sneutrino mass RGE is presented here. It is
\begin{align}
\begin{split}
	m_{\nu_3}^2&(M_{B-L})=m_{\nu_3}^2(M_\i)
	\\
	&+\frac{1}{14}\frac{g_R^4(M_\i) - g_R^4(M_{B-L})}{g_U^4}M_R(M_U)^2
	+\frac{1}{8}\frac{g_{BL}^4(M_\i) - g_{BL}^4(M_{B-L})}{g_U^4}M_{BL}(M_U)^2
	\\
	&+\frac{1}{14}\frac{g_R^2(M_\i) - g_R^2(M_{B-L})}{g_R^2(M_\i)}S_R(M_\i)
	-\frac{1}{16}\frac{g_{BL}^2(M_\i) - g_{BL}^2(M_{B-L})}{g_{BL}^2(M_\i)}S_{BL}({M_\i}) \ .
\label{eq:315}
\end{split}
\end{align}
Recall that the value of any Abelian gauge couplings grows larger at higher scale. Therefore, we see that a tachyonic sneutrino is only possible when $S_R(M_\i)$ is negative and/or $S_{BL}(M_\i)$ is positive. This demonstrates the central role played by the $S$-terms in the breaking of $B-L$ symmetry. 
Note that in typical unification scenarios all soft masses are ``universal'' and, hence, both $S$-terms vanish. However, it was mentioned earlier that, in this string construction, different elements of a given generation arise from different \textbf{16} representations of $SO(10)$. Therefore, the soft masses of a given generation are generically non-degenerate. Hence, the $S$-terms can be non-zero. 

As mentioned above, $M_{Z_R} \simeq \sqrt 2 |m_{\tilde \nu^c_{3}}|$ and the relationship
\beq
	M_{Z_R}(M_{B-L}) = M_{B-L}
\eeq
is used to iteratively solve for the $B-L$ scale. The SUSY mass scale must also be solved for iteratively using the equation
\beq
	\sqrt{m_{\tilde t_1}(M_\susy) m_{\tilde t_2}(M_\susy)} = M_\susy \ ,
\eeq
where $m_{\tilde t_1} < m_{\tilde t_2}$ are the physical stop masses. The relationships between the soft mass parameters and the physical masses are given in Appendix~\ref{sec:546}. The soft mass squared parameter for the up-type Higgs is driven tachyonic, as usual, by the large top Yukawa coupling. Furthermore, the decoupled values of the soft Higgs mass squared parameters are used to calculate the $\mu$- and $b$-terms using Eqs.~(\ref{eq:EW.mu}) and (\ref{eq:EW.b}).

The soft mass parameters have non-trivial boundary conditions at the $B-L$ scale due to the effects of the $B-L$ and $I_{3R}$ $D$-terms:
\begin{align}
\begin{split}
	\label{eq:BC.BL}
	m_\phi^2(M_{B-L}^-) - m_\phi^2(M_{B-L}^+) &= -\frac{1}{4}\left(g_R^2+g_{BL}^2\right) v_R^2 \left( I_{3R} - Y \sin^2 \theta_R \right)
	\\
	& \simeq  -M_{Z_R}^2 \left( I_{3R} - Y \sin^2 \theta_R \right) \ ,
\end{split}
\end{align}
where $M_{B-L}^-$ and $M_{B-L}^+$ indicate a scale slightly below and slightly above the $B-L$ scale respectively, and $I_{3R}$ and $Y$ are the third component of right-handed isospin and hypercharge of a generic scalar $\phi$.

Once the mass parameters have been properly evolved to their appropriate scales, the physical masses can be evaluated. For much of the spectrum, this has been discussed in the literature, see for example~\cite{Martin:1997ns}, and has been included in Appendix~\ref{sec:546}. The new element here is the mass of the scalar associated with the third generation right-handed sneutrino--the $B-L$ Higgs--degenerate with the $Z_R$ mass. In addition, the calculation for the SM-like Higgs mass is crucial since the experimentally measured value of $\sim$125 GeV requires substantial radiative corrections from the stop sector in the MSSM. 
 In this paper we follow the approach of references~\cite{ArkaniHamed:2004fb,Cabrera:2011bi,Giudice:2011cg}--taking into account the decoupling scales of the two stops, matching the quartic Higgs coupling at those scales and RGE evolving the quartic coupling to the electroweak scale to calculate the Higgs mass. Full details are given in Appendix~\ref{sec:915}.

Once a given physical mass is calculated, it is compared to current lower bounds or, in the case of the SM Higgs, the experimentally measured value. If a given initial set of parameters predicts a physical mass that is inconsistent with current bounds, it is rejected as being an invalid point. These lower bounds are discussed in the next subsection.

\subsection{Collider Constraints}
\label{sec:856}
The bounds placed by collider data on SUSY masses are, in general, model dependent. That is, they depend on the spectrum and decay modes. Despite the much larger energy of the LHC, LEP 2 still has competitive bounds on colorless particles that couple to the $Z$ and/or the photon--including sleptons in scenarios with both $R$-parity conservation~\cite{LEPSUSYWG/04-01.1,LEPSUSYWG/02-09.2} and violation~\cite{LEPSUSYWG/02-10.1}, bounds on charginos~\cite{LEPSUSYWG/01-03.1, LEPSUSYWG/02-04.1} and bounds on sneutrinos in the case of $R$-parity violation~\cite{LEPSUSYWG/02-10.1}. As one may expect, due to the relatively clean environment at LEP, these bounds are close to one half the center of mass energy of LEP 2. Therefore, for simplicity, we proceed with the bound that all colorless fields that couple to the photon must be heavier than 100 GeV. That is,
\beq
	m_{\tilde \ell}, m_{\tilde \chi_1^\pm} > 100 \text{ GeV},
\eeq
where $\tilde\ell$ is any charged slepton. Colorless states that couple to the $Z$, \textit{the left-handed sneutrino}, must be heavier than half the $Z$ mass:
\beq
	m_{\tilde \nu_L} > 45.6 \text{ GeV},
\eeq
Colorless states that do not couple to the Z, such as right-handed sneutrinos/neutrinos and the bino, have such small collider production cross-sections that they do not have collider-based lower bounds. Wino and Higgsino neutralinos are degenerate with their chargino partner, thereby effectively putting a lower bound of 100 GeV on those states as well.

The bounds from the LHC are much more dependent on the parameters. For example, if one investigated the bound on, for example, degenerate squarks in this model with a neutralino LSP, those bounds could be significantly different than in the case of a sneutrino, or some other,  LSP. Allowing the squark masses to split would further alter the lower bounds. In fact, a full treatment would involve calculating the signatures of a given point in parameter space, comparing the number of events to the most recent LHC bounds on such events, and determining if the parameter point is valid. We do not expect the details of these lower bounds to heavily affect our results. We will, therefore, simply use the naive bounds
\beq
	m_{\tilde q} > 1000 \text{ GeV}, \quad m_{\tilde g} > 1300 \text{ GeV} \ ,
\eeq
which are based on recent CMS~\cite{CMS:2014ksa} and ATLAS~\cite{Aad:2014wea} studies of the $R$-parity conserving MSSM. In these studies, the colored states decay into jets and missing energy--possible final states in our model whenever the LSP decays into neutrinos. In this paper, we impose these bounds except in the case of a stop or sbottom LSP. These two cases were explicitly studied in~\cite{Marshall:2014kea, Marshall:2014cwa} and yielded the following lower bounds:
\beq
	\text{admixture (right-handed) stop LSP: } m_{\tilde t_1} > 450 \ (400) \text{ GeV}, \quad m_{\tilde b_1} > 500 \text{ GeV},
\eeq
where $\tilde t_1$ ($\tilde b_1$) denotes the lightest stop (sbottom). Here, right-handed refers to a stop that is almost completely right-handed--that is, a stop mixing angle, $\theta_{\tilde t} > 85^\circ$ or, equivalently, a state composed of 99$\%$ right-handed stop--while admixture stop refers to all other stops. This distinction is based on the phenomenology of the stops; right-handed stops have significant decays into a top quark and neutrinos while admixture stops decay almost exclusively to a bottom quark and a charged lepton.

The lower bound on the $Z_R$ mass from LHC searches is 2.5 TeV~\cite{ATLAS:2013jma,CMS:2013qca}. Finally, we require that the Higgs mass be within the $2\sigma$ allowed range from the value measured at the ATLAS experiment at the LHC. We naively obtain the two sigma range by adding in quadrature the systematic and statistical uncertainties from \cite{Aad:2014aba}, and multiplying the result by two:
\begin{eqnarray}
	m_{h^0}=125.36\pm 0.82\mbox{ GeV}.
\end{eqnarray}
See \cite{Chatrchyan:2013lba} for comparable data from CMS. A summary of the collider bounds mentioned above is given in Table~\ref{tbl:mass.bounds}.

\begin{table}[htdp]
\begin{center}
\begin{tabular}{|c|c|}
\hline
Particle(s) & Lower Bound
\\
\hline
\hline
Left-handed sneutrinos & 45.6 GeV
\\
$\quad$Charginos, sleptons $\quad$& 100 GeV
\\
Squarks, except for stop or sbottom LSP's & 1000 GeV
\\
Stop LSP (admixture)							& 450 GeV
\\
Stop LSP (right-handed)							& 400 GeV
\\
Sbottom LSP							& 500 GeV
\\
Gluino & 1300 GeV
\\
$Z_R$ & 2500 GeV
\\
\hline
\end{tabular}
\end{center}
\caption{The different types of SUSY particles and the lower bounds implemented in this paper.}
\label{tbl:mass.bounds}
\end{table}%

\subsection{Constraints from Flavor and CP-Violation}
\label{sec:857}
\begin{figure}[h]
\centering
	\includegraphics[scale=0.6]{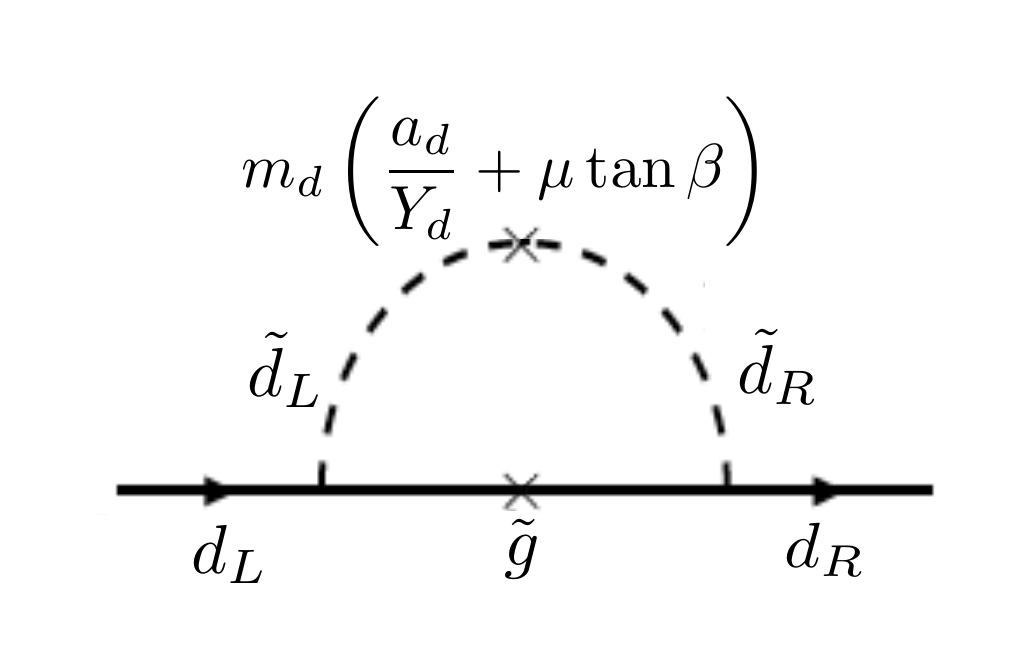}
	\caption{Radiative contribution of the gluino to the down quark mass. Similar contributions exist for the other fermions.}
	\label{fig:d.mass}
\end{figure}

A large number of low-energy experiments exist which place constraints on the SUSY parameter space. Some of the oldest and most well-known are the constraints placed on flavor changing neutral currents from the analyses in references~\cite{Ellis:1981ts, Barbieri:1981gn,Campbell:1983bw}--for example, those arising from $K-\bar K$ oscillation--and on CP violation~\cite{Ellis:1982tk, Buchmuller:1982ye, Polchinski:1983zd,del Aguila:1983kc, Nanopoulos:1983xd}--for example, from electric dipole moment measurements. Generically, the implication of these constraints are, approximately, as follows:
\begin{itemize}
	\item Soft sfermion mass matrices are diagonal.
	\item The first two generations of squarks are degenerate in mass.
	\item The trilinear $a$-terms are diagonal.
	\item The gaugino masses and trilinear $a$-terms are real.
\end{itemize}
In addition, it is typically assumed that the soft trilinear $a$-terms are proportional to the Yukawa couplings, that is, generically $a = Y A$ for each fermions species. Each $A$ is a dimensionful real number on the order of a TeV, while each $Y$ factor is a dimensionless matrix in flavor space. This condition effectively makes all non-third generation trilinear terms insignificant. Note that this assumption does not immediately follow from the above experimental constraints. However, significant radiative corrections to fermion masses, proportional to the $a$-term, can arise in SUSY, as first discussed for fermions in references~\cite{Hall:1993gn, Carena:1994bv, ArkaniHamed:1995fq}. For example, a down quark mass is modified by gluino exchange, through the diagram in Fig.~\ref{fig:d.mass}, as follows:
\beq
	\Delta M_d = M_d M_{\tilde g} \frac{2 \alpha_3}{3 \pi} \left(\frac{a_d}{Y_d} + \mu \tan \beta \right)
	I \left(m_{\tilde b_L}^2, m_{\tilde b_R}^2, M_{\tilde g}^2 \right),
\eeq
where
\beq
	I(x,y,z) = \frac{xy \ln \frac{x}{y} + yz \ln \frac{y}{z} + xz \ln \frac{z}{x}}{(x-y)(y-z)(x-z)},
\eeq
and $M_{\tilde g}$ is the gluino mass. If $a_d$ is on the order of a TeV, this radiative correction can be quite large, possibly larger than the down quark mass. If this were the case, the radiative correction would have to be fine-tuned against the tree-level contribution to reproduce the correct down quark mass. This motivates allowing only the third generation $a$-terms to be significant. Hence, our assumption that 
\begin{eqnarray}
	a(M_\i)=Y(M_\i)A(M_\i) \ .
\end{eqnarray}
This makes all $a$-parameters, except for those associated with $t, b$ and $\tau$,  insignificant. For simplicity, we will choose all other $a$-parameters to be zero.

Summarizing the above, we employ the following constraints motivated by low-energy physics:
\begin{align}
\begin{split}
	& m_{\tilde q}^2 = \text{diag}\left(m^{2}_{\tilde q_1},m^{2}_{\tilde q_1}, m^{2}_{\tilde q_3} \right)~~,~~
	\tilde q = \tilde Q, \, \tilde U, \,  \tilde D \ ,
	\\
	&  m_{\tilde \ell}^2 = \text{diag}\left(m^{2}_{\tilde \ell_1},m^{2}_{\tilde \ell_2}, m^{2}_{\tilde \ell_3} \right)~,~~
	 \tilde \ell = \tilde L, \, \tilde E \ , \tilde \nu^c \ ,
	\\
	& a_f(M_\i) = Y_f(M_\i)A_f(M_\i) ~~~~,~~ \ f = t,\, b, \, \tau \ .
\end{split}
\end{align}
Note that all of these constraints can be implemented at the high scale, since RG evolution to the SUSY scale will not spoil these relations. Furthermore, we do not assume here that the first and second generation slepton masses are degenerate--unlike the squark masses-- since this is not required by low energy experiments. The degeneracy or non-degeneracy of these states will not, however, greatly effect the results of this paper. 
%
\section{The Parameter Space and Scan}
\label{sec:1141}
%

The previous section reviewed the framework used in this paper for connecting the high scale to LHC accessible physics. It remains at this point to discuss the input values for the SUSY breaking parameters. In this section, we introducing a novel way to analyze the initial parameter space of a SUSY model. While there have been many studies of specific, fixed boundary conditions at the high scale, and some recent interesting discussions of random parameter scans at the TeV scale~\cite{Berger:2008cq,Conley:2010du}, a study that combines both has not--until now-- been undertaken. Specifically, our approach in this paper is to make a statistical scan of input parameters at the high scale--followed by a RG evolution of each set of those parameters to the TeV scale and an analysis of which of these high scale initial conditions lead to realistic physics. While the soft SUSY breaking sector contains over 100 dimensionful parameters, the constraints of low energy experiments discussed in the previous section only allow collider significant values for about a fifth of these-- 24 to be specific. These, along with $\tan \beta$ and a discussion of the sign of certain parameters, are presented in Table~\ref{tbl:scan}.

The high scale initial values of the 24 relevant SUSY breaking parameters are determined as follows. To conduct our scan, we make the assumption that there is only one overall scale associated with SUSY breaking. This assumption does not require that the soft mass parameters be equal to each other, or even have similar values. It does, however, require that these parameters be at least within an order of magnitude, or so, of each other. To quantify this, we demand that any dimension one soft SUSY breaking parameter be chosen at random within the range
\beq
	(\frac{M}{f}, Mf) \ ,
\eeq
where $M$ is the mass setting the scale of SUSY breaking and $f$ is a dimensionless number satisfying $1\leq f  \lesssim 10$. We will further insist that any such parameter be evenly scattered around $M$; that is, that $M$ be the average of the randomly generated values. Clearly, this will not be the case if parameters are chosen from a uniform probability distribution in the range $(\frac{M}{f}, Mf)$--referred to as a ``flat prior''. Instead, a ``log prior'' is adopted. This means that the natural logarithm of a given soft SUSY breaking parameter is chosen from a uniform distribution in the range
\beq
	(\ln \left({\frac{M}{1~{\rm GeV}}}{\frac{1}{f}}\right), \ln \left({\frac{M}{1~{\rm GeV}}}f\right)) \ .
\eeq
With a log prior distribution, $M$ is the geometric mean of the randomly generated parameters. In addition to the dimensionful soft masses, we also scan $\tan \beta$ as a flat prior in the range (1.2, 65), thus selected so that all Yukawa couplings remain perturbative through the entire range. Furthermore, we randomly generate the signs of $\mu$, the three tri-scalar couplings $a_{t,b,\tau}$, and the four gaugino masses $M_{3,2,R,BL}$.

\begin{figure}[!htbp]
	\centering
	\includegraphics[scale=1]{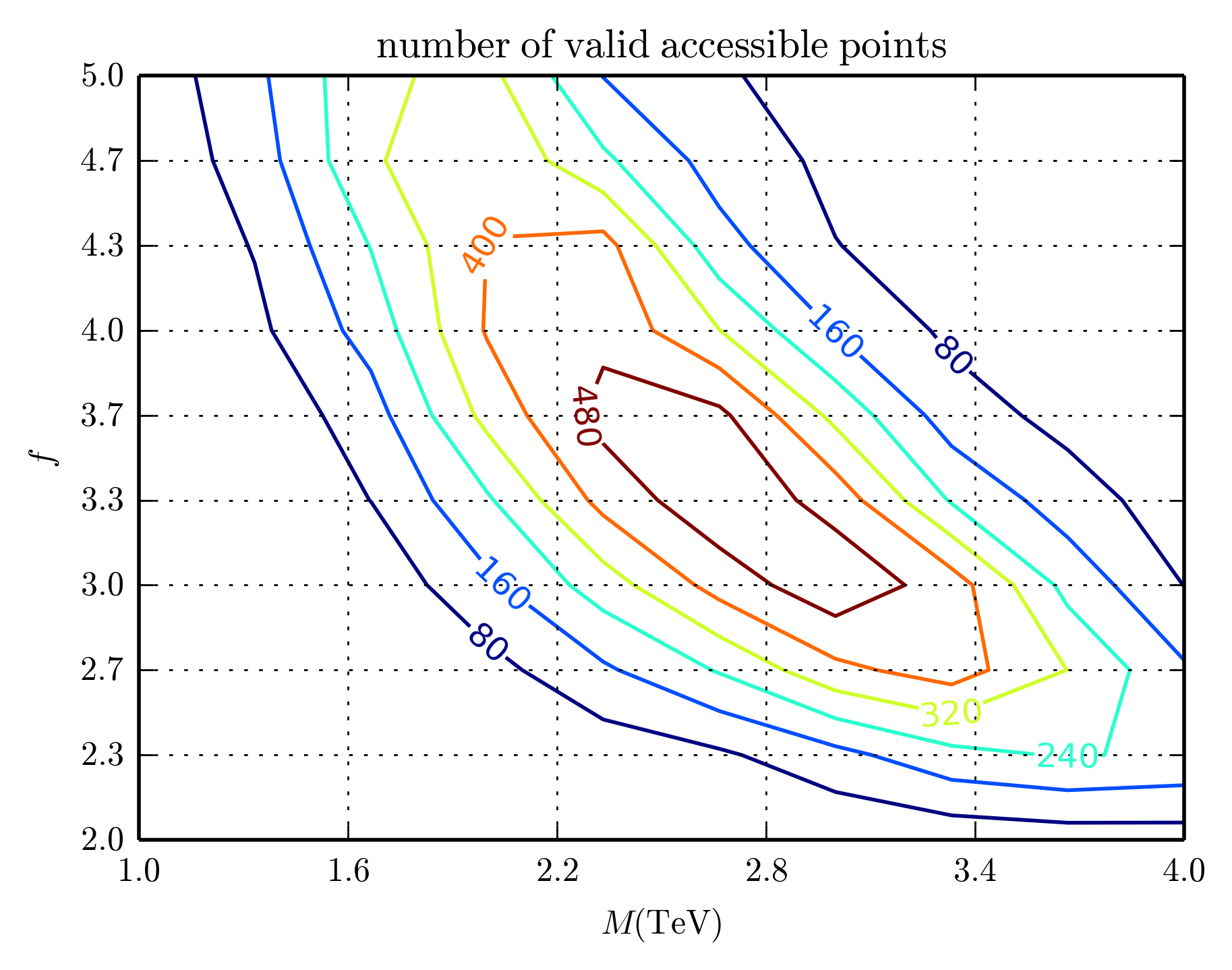}
	\caption{\small A contour plot in the $M$ - $f$ plane of the number of valid accessible points; that is, points that meet all experimental constraints given in the previous section and for which all sparticles are below 10 TeV. A broad peak or plateau is evident around $M=2700$ GeV and $f=3.3$.}
	\label{fig:948}
\end{figure}

In this paper, we are interested in the low energy spectra being accessible at the LHC or a next generation collider. Therefore, in addition to the experimental constraints mentioned in the previous section, we further demand that all sparticle masses be lighter than 10 TeV. We call any point that satisfies this, as well as all previous criteria, a ``valid accessible'' point. 
The parameters $M$ and $f$ are chosen in such a way so as to maximize the number of such points. To determine the values of $M$ and $f$ which yield the greatest number of valid accessible points, we begin by making a ten by ten grid in the $M-f$ plane. At each of these hundred points, we randomly generate one hundred thousand initial points in the 24-dimensional parameter space discussed above, RG scale them to low energy, and count the subset that satisfies the experimental checks discussed above. We then plot curves corresponding to a constant number of valid accessible points in Fig.~\ref{fig:948}. The plot shows a broad peak or plateau, the center of which maximizes the number of such points. This maximum occurs approximately for
\begin{eqnarray}
	M=2700 \text{ GeV} ,\ f=3.3 \ .
\label{cat1}
\end{eqnarray}
These values will be used to generate the results in the remainder of this paper. Note that for these values, the smallest soft parameter is maximally about an order of magnitude away from the largest soft parameter. Specifically, for $M$ and $f$ in Eq. (\ref{cat1}), the ranges for the random scan of each parameter are given in Table~\ref{tbl:scan}.

\begin{table}[htdp]
\begin{center}
\begin{tabular}{|c|c|c|}
	\hline
	Parameter & Range & \ Prior \
	\\
	\hline
	\hline
	\quad $m_{\tilde q_1} = m_{\tilde q_2}, \ m_{\tilde q_3}: \quad \tilde q = \tilde Q, \tilde u^c, \tilde d^c$ \quad & (820, 8900) GeV & log
	\\
	$m_{\tilde \ell_1}, m_{\tilde \ell_2}, \ m_{\tilde \ell_3}: \quad \tilde \ell = \tilde L, \tilde e^c, \tilde \nu^c$
		& \ (820, 8900) GeV \ & log
	\\
	$m_{H_u}, m_{H_d}$ & (820, 8900) GeV & log
	\\
	$\left|A_f\right|: \quad f = t,b, \tau$ & (820, 8900) GeV & log
	\\
	$\left|M_a\right|: \quad a = R, BL, 2, 3$ & (820, 8900) GeV & log
	\\
	$\tan \beta$ & (1.2, 65) & flat
	\\
	Sign of $\mu, a_f, M_a: \quad f=t,b,\tau \quad a=R, BL, 2, 3$ & [-,+] & flat
	\\
	\hline
\end{tabular}
\end{center}
\caption{The parameters and their ranges scanned in this study, as well as the type of prior. The ranges for the soft SUSY breaking parameters are optimized to produce the greatest number of valid points with all masses below 10 TeV.}
\label{tbl:scan}
\end{table}%

\begin{figure}[!htbp]
	\centering
	\includegraphics[scale=1]{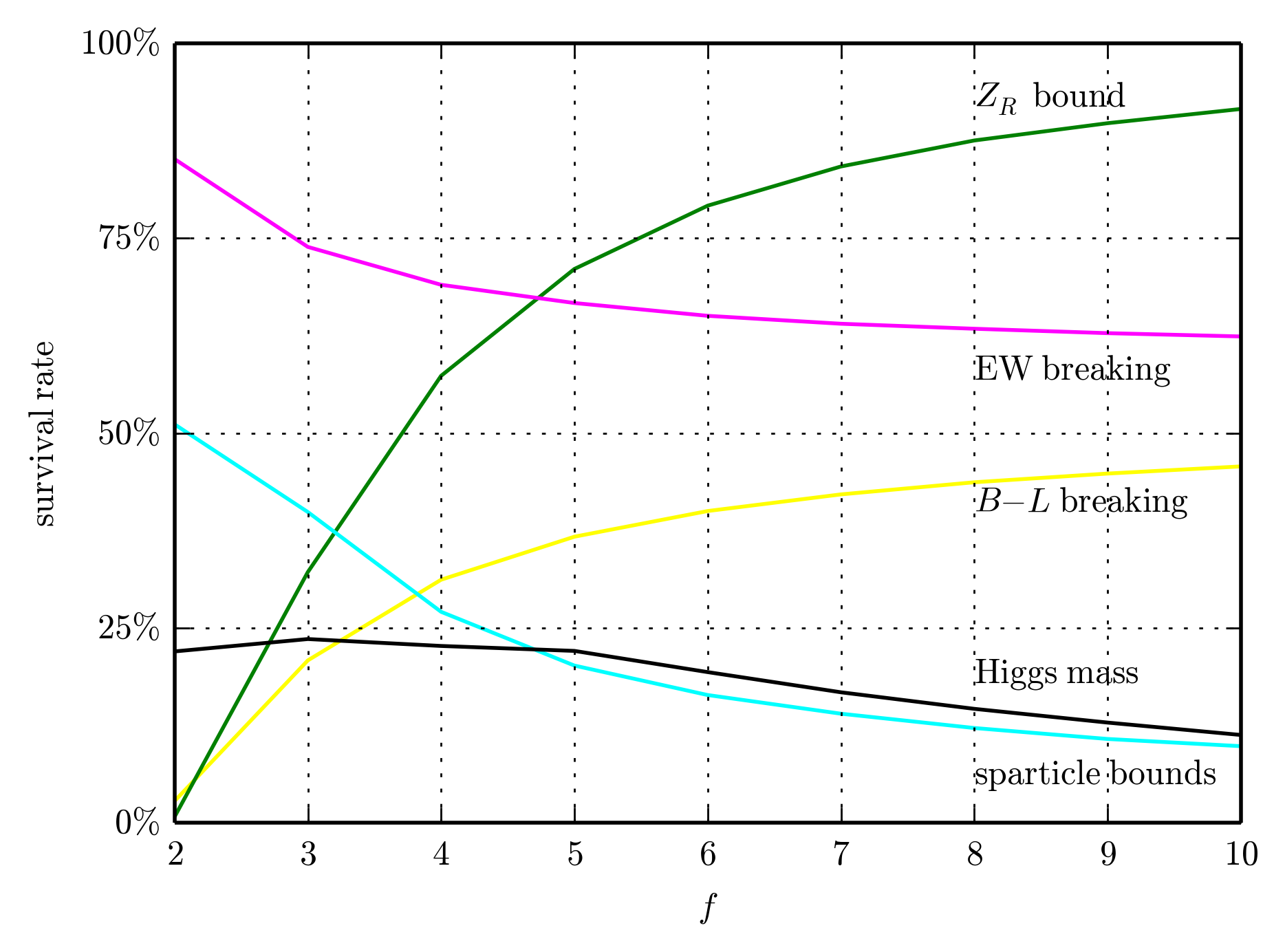}
	\caption{\small Survival rates of the various checks as a function of $f$ for $M=2700$ GeV. The $B-L$ breaking and $Z_R$ lower bound checks favor larger $f$ while the others favor small $f$. All of the checks taken together favor a moderate value of $f\sim 3.3$.}
	\label{fig:948B}
\end{figure}

The existence of a peak in Fig.~\ref{fig:948} around moderate values of $f$ is a consequence of combining the various experimental checks we apply to each of the randomly generated points. For a fixed value of $M$, some individual checks favor higher values of $f$, while others favor lower values. This is analyzed in terms of the ``survival rate''. The survival rate for a given check is defined as the number of points in the 24-dimensional initial parameter space surviving that check as a percentage of the number of points that survived all previous checks. This will be discussed in detail in the next section for the fixed values of $M$ and $f$ given in Eq. (\ref{cat1}). Here, for $M=2700$ GeV, we analyze the impact of the parameter $f$ on the various survival rates. The peak around moderate values of $f$ shown Fig.~\ref{fig:948} can be understood by observing how the survival rates for different checks depend on $f$. This is shown in Fig.~\ref{fig:948B}. The $B-L$ symmetry breaking check and the $Z_R$ lower bound check both prefer higher values of $f$. This is because such values favor larger $S$-terms and thereby promote $B-L$ symmetry breaking. The $EW$ symmetry breaking check favors lower values of $f$. Intuitively, this is not surprising since universal boundary conditions (which correspond to $f=1$) in the MSSM allow electroweak symmetry breaking. The sparticle lower bounds check favors low $f$. This is because larger $f$ leads to larger $S$-terms which, in turn, can drive some sparticles masses to be light through the RGE's. The Higgs mass check also favors low $f$ because larger $S$-terms may drive the stop masses away from the $\sim TeV$ value favored by the Higgs mass. With some checks favoring large $f$ and others small $f$, it is not surprising that all checks taken together favor a moderate values of $f$.

%
\section{Results}
\label{sec:1058}
%

All of the following results arise from a scan consisting of ten million randomly generated points with $M=2700$ GeV and $f=3.3$. We will refer to this as the ``main scan''. 
Recall that in the previous section, in addition to the experimental constraints, we imposed an extra condition that all masses be lighter than $10$ TeV. Technically, this was done to ensure the maximum number of valid accessible points--that is, valid points with masses accessible to the LHC or a next generation collider. Having done this, we, henceforth, remove this additional condition. That is, the only constraints in the main scan are the experimental ones given in Section~\ref{sec:1115}. All valid points must satisfy these constraints as well as some other checks, which are reviewed here briefly.

In order to be valid, a point must break $B-L$ symmetry, the $Z_R$ mass must be above the lower bound, electroweak symmetry must be broken and since the stops play a crucial role in electroweak breaking, we designate a check that the stops are not tachyonic. Since our numerical analysis uses an iterative process to solve for the SUSY and $B-L$ scales, it is possible that a point may pass a check on the first iteration but fail it on the final iteration, although this is very uncommon. Therefore, we include a ``spill'' check of the $Z_R$ bound, electroweak breaking check and non-tachyonic stop checks. Points that pass these spill checks did so on the final iteration of solving for the $B-L$ and SUSY scales. Furthermore, a valid point must have $B-L$ and SUSY scales that converge to a value in the iterative solution process. We also 
check that--in addition to the stops--all other SUSY sparticles are not tachyonic and satisfy the imposed mass bounds. Finally, we check that the Higgs mass matches its experimental value.

All of these conditions are listed in the first column of Table~\ref{tab:320}. The second column lists the number of points in the main scan that passed that check, out of ten million. The third column is the same information listed as a percent of the number of points in the main scan. The fourth column is the same information listed as a percent of the number of points that passed the previous checks. We refer to this quantity as the rate of survival for each check. This is an interesting quantity because it quantifies how easy or hard it is for a randomly generated point to pass that specific check.

\begin{table}[!htbp]
\begin{center}
\begin{tabular}{|c|c|c|c|}
\hline
check&number surviving&percent surviving&rate of survival\\
\hline
$B-L$ breaking	&2,225,704	&22.3	\%	&22.3	\%\\
$Z_R$ bound	&919,117	&9.19	\%	&41.3	\%\\
EW breaking	&722,750	&7.23	\%	&78.6	\%\\
non-tachyonic stops	&619,668	&6.2	\%	&85.7	\%\\
$Z_R$ bound spill	&597,988	&5.98	\%	&96.5	\%\\
EW breaking spill	&565,272	&5.65	\%	&94.5	\%\\
non-tachyonic stops spill	&553,592	&5.54	\%	&97.9	\%\\
convergence	&553,150	&5.53	\%	&99.9	\%\\
sparticle bounds	&276,676	&2.77	\%	&50	\%\\
Higgs mass	&58,096	&0.581	\%	&21	\%\\
\hline
\end{tabular}
\end{center}
\caption{\small This table shows all of the checks applied to the randomly generated points. It specifies the number of such points passing each check, as well as their percent of survival. The fourth column is the most informative because it provides insight into how likely it is that an individual check is satisfied by a randomly generated point. Because the SUSY and $B-L$ scales are solved for iteratively, it is possible to pass a check in the first iteration and fail it later. A passed ``spill'' check indicates that that check was passed in the final iteration.}
\label{tab:320}
\end{table}

A striking feature of Table~\ref{tab:320} is that $B-L$ breaking happens robustly. This was one of the central questions that this paper sought to answer. Our analysis demonstrates that, for $M=2700$ GeV and $f=3.3$, no special tuning or choice of parameters is required at the $M_{\i}$ scale to achieve $B-L$ symmetry breakdown. 
Further analysis--for other values of $M$ and $f$--shows that the percentage of points that break $B-L$ is, in general, independent of $M$. This is because $B-L$ breaking is not a question of generating a specific scale. Rather, it involves having soft masses aligned in such a way as to allow the $S$-terms to drive the third family right-handed sneutrino tachyonic, see Eq.~(\ref{eq:RGE.snu}). On the other hand, $B-L$ breaking is dependent on the choice of $f$. In the limit $f \to 1$, $S_{BL,R} \to 0$ and $B-L$ breaking becomes impossible--whereas increasing $f$ will allow $B-L$ breaking to occur. A second feature of Table~\ref{tab:320} is that, for $M=2700$ GeV and $f=3.3$, $B-L$ breakdown where the $Z_{R}$ mass exceeds the experimental lower bound, although less prevalent, is still rather robust. In contrast to arbitrary $B-L$ breaking, however, since $M_{Z_R} \simeq \sqrt 2|m_{\tilde \nu^c}|$ it follows that passing the $Z_R$ mass bound check is sensitive to the choice of $M$--with the survival rate increasing with $M$. 

A third important conclusion drawn from Table~\ref{tab:320} is that, for the main scan, a large percentage of the initial points that have $B-L$ breaking consistent with the $Z_{R}$ mass lower bound, also lead to the radiative breakdown of electroweak symmetry. Note that the $Z$ mass can always be adjusted--albeit by low energy tuning--to its experimental value of $91.2$ GeV. Again, further analysis--for other values of $M$ and $f$--shows that electroweak breaking, like $B-L$ breaking, is also roughly independent of $M$. However, unlike $B-L$, small $f$ favors electroweak symmetry breaking. As is well known from the literature~\cite{Ibanez:1982fr, Ellis:1982wr, AlvarezGaume:1983gj}, electroweak breaking occurs for universal boundary conditions--that is, for $f = 1$. On the other hand, as $f$ increases, the randomly generated parameter $m^{2}_{H_u}(M_\i)$ can be considerably larger than the square of the initial stop masses. In this case, the RGE evolution to the SUSY scale may be insufficient to render $m^{2}_{H_u}(M_\susy)$ tachyonic. However, since the initial soft masses are randomly generated, the electroweak breaking survival rate will decrease with increasing $f$, but will not go to zero.

Whether or not stop masses remain non-tachyonic at low scale depends on the randomly chosen values of several of the initial parameters. As can be seen from Table~\ref{tab:320}, for the values of $M$ and $f$ chosen for the main scan, non-tachyonic stops are very common. To remind the reader, the checks labeled spill are repeats of earlier checks that are conducted after the final iteration of solving for the SUSY scale. Since this iterative process usually only affects the relevant checks logarithmically, the spill bins are expected to have high survival rates. The survival rate for convergence of the iterative process of finding values of $M_{B-L}$ and $M_\susy$ is almost 100\%--since the soft masses have a logarithmic dependence on the scale. The survival rate for the SUSY particle mass bounds check is, for $M=2700$ GeV and $f=3.3$, comparable to that of the $Z_R$ mass bound. Further analysis shows that this rate is also controlled by the choice of $M$--a higher value for $M$ resulting in a higher survival rate for this check. The Higgs mass survival rate for the main scan is, perhaps, surprisingly high--given that we were checking that the Higgs mass for a randomly generated point matches an experimentally measured value within an error of less than one percent. The reason this rate is so high is that the measured value of the Higgs mass is within the range expected for TeV scale supersymmetry breaking.

\begin{figure}[!htbp]
	\centering
	\includegraphics[scale=1]{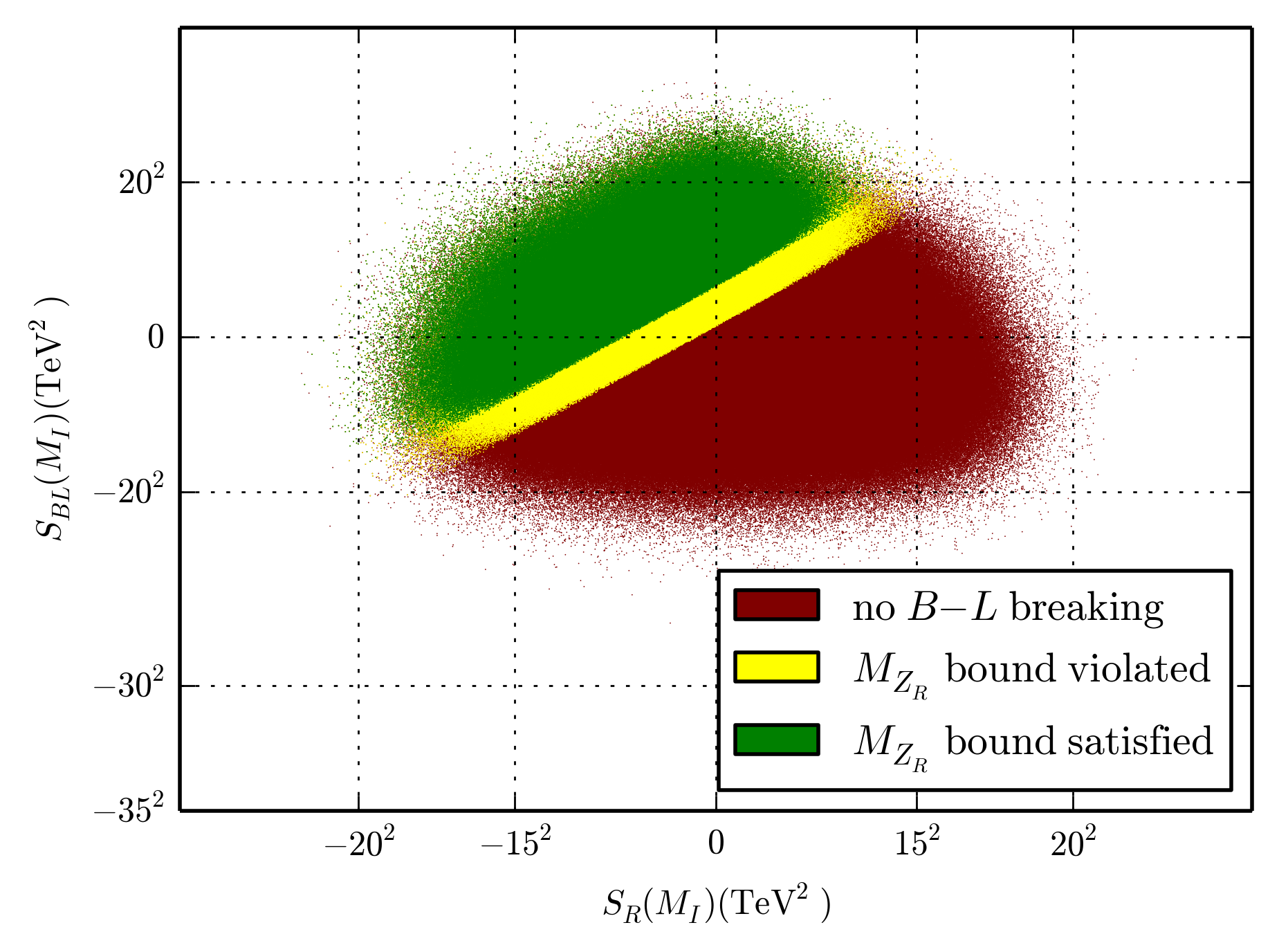}
	\caption{\small Points from the main scan in the $S_{BL}(M_\i)$ - $S_R(M_\i)$ plane. Red indicates no $B-L$ breaking, in the yellow  region $B-L$ is broken but the $Z_R$ mass is not above its 2.5 TeV  lower bound, while green points have $M_{Z_R}$ above this bound. The figure expresses the fact that, despite there being 24 parameters at the UV scale scanned in our work, $B-L$ physics is essentially dependent on only two combinations of them--the two $S$-terms. 
Note that the green points obscure a small density of the yellow and red points behind them. Similarly the yellow points obscure some red points.}
	\label{fig:1204}
\end{figure}

Since the initial soft SUSY breaking parameter space is 24-dimensional, graphically displaying the subspaces associated with each survival check in Table~\ref{tab:320} is very difficult. However, as can be seen from the RGEs and has been discussed in the text, much of the scaling behavior of the parameters is controlled by the two $S$-terms, $S_R$ and $S_{BL}$, defined in Eqs.~(\ref{eq:S.R}) and (\ref{eq:S.BL}). It follows that the results in Table~\ref{tab:320} can be reasonably displayed in the two-dimensional  $S_{BL}(M_\i)$ - $S_R(M_\i)$ plane. We begin by presenting in Fig.~\ref{fig:1204} the initial points in the $S_{BL}(M_\i)$ - $S_R(M_\i)$ plane that satisfy, sequentially, the first two fundamental checks in Table~\ref{tab:320}; that is, $B-L$ breaking and the experimental $Z_{R}$ mass lower bound. Points that do not break $B-L$ are shown
in red, points that satisfy $B-L$ breaking but not the $Z_R$ mass bound are in yellow, and points that break $B-L$ symmetry and satisfy the $Z_R$ mass bound are shown in green.
%
\begin{figure}[!htbp]
        \centering
        \includegraphics[scale=1]{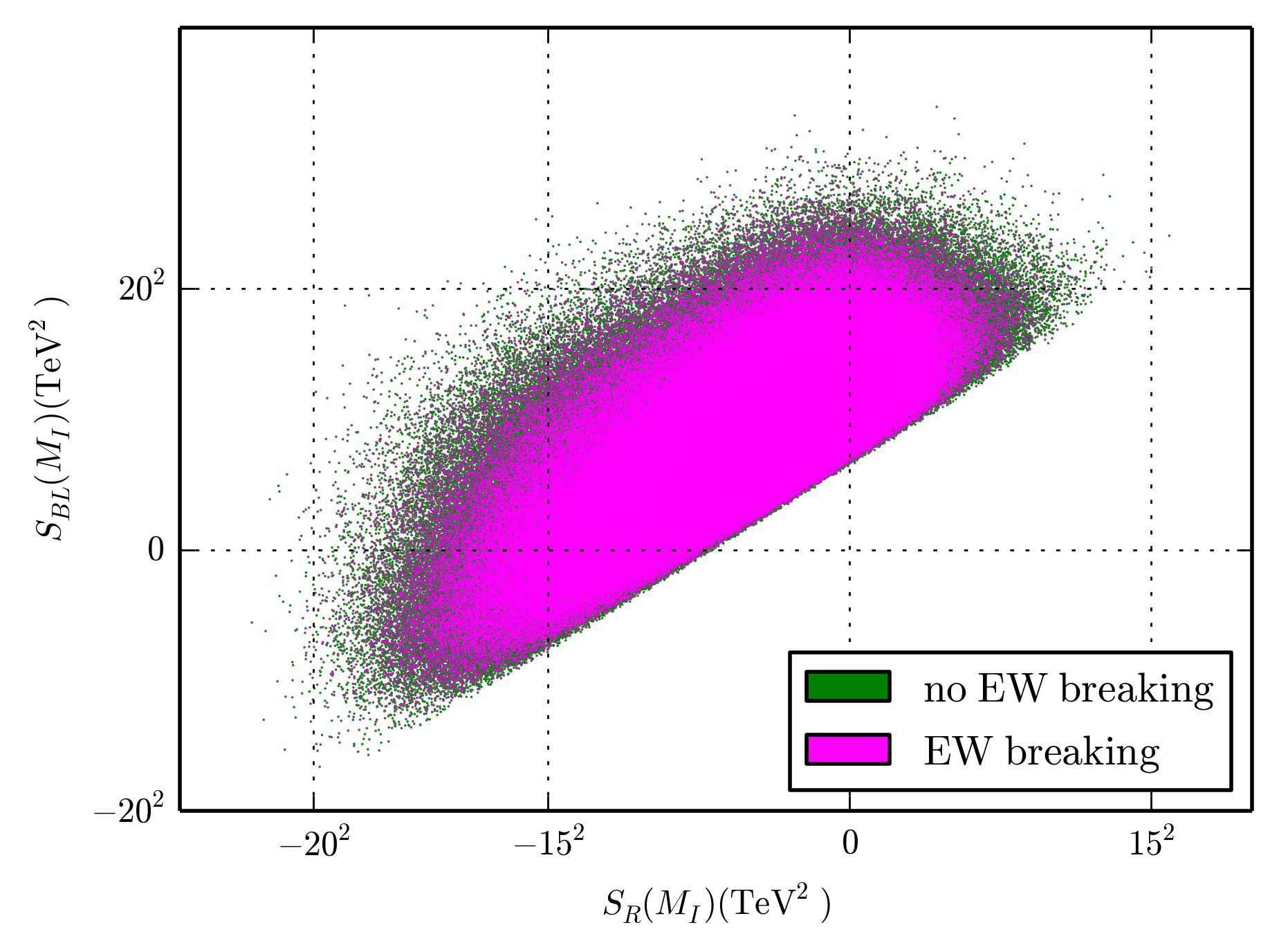}
	\caption{\small A plot encompassing the green region in Fig~\ref{fig:1204}. The green points in this plot correspond to those which appropriately break $B-L$ symmetry, but which do not break electroweak symmetry. However, the purple points, in addition to breaking $B-L$ symmetry with an appropriate $Z_{R}$ mass, also break EW symmetry. Note that a small density of green points that do not break EW symmetry are obscured by the purple points.}
        \label{fig:1205}
\end{figure}
\begin{figure}[!htbp]
	\centering
	\includegraphics[scale=1]{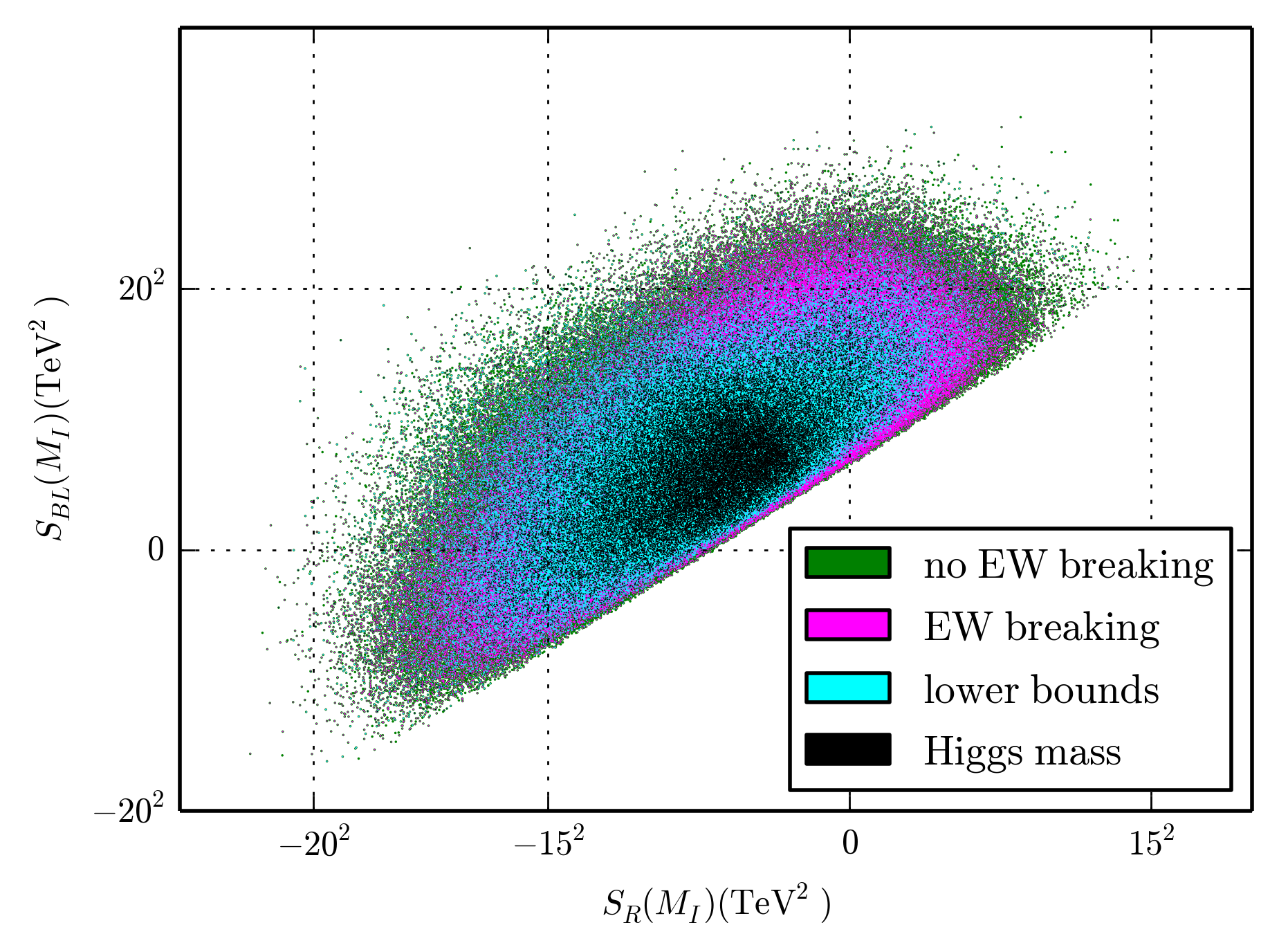}
	\caption{\small A plot of the valid points in our main scan. The green and purple points correspond to the green and purple points in Fig~\ref{fig:1205}. The cyan points additionally have non-tachyonic stops, pass all spill checks and convergence, and satisfy all sparticle mass lower bounds. The black points are fully valid. That means that, in addition to satisfying all previous checks, they reproduce the correct Higgs mass within the stated tolerance. The distribution of points indicates that while $B-L$ breaking prefers large $S$-terms, sfermion mass constraints prefer them to be not too large. Again, the cyan and black points may obscure a low density of other points not satisfying their constraint.}
	\label{fig:1148}
\end{figure}
This plot reaffirms the conclusion drawn from Table~\ref{tab:320} that $B-L$ breaking consistent with present experiments is a robust phenomena. Furthermore, it shows the strong dependence of $B-L$ breaking and the $Z_R$ mass on the values of the $S$-terms. There is a line in the $S_{BL}$ - $S_R$ plane--between the yellow and red regions--below which $B-L$ breaking is not possible. Note that this includes the origin, which corresponds to vanishing $S$-terms and, hence, universal soft masses. This shows that at least a small splitting from sfermion universality is required for  $B-L$ breaking. Another line exists--between the green and yellow regions--below which $Z_R$ is always lighter than its lower bound.

Proceeding sequentially, we present in Fig.~\ref{fig:1205} the initial points in the $S_{BL}(M_\i)$ - $S_R(M_\i)$ plane that, in addition to breaking $B-L$ with a $Z_{R}$ mass above the experimental bound, also break EW symmetry. The entire colored region encompasses the green points shown in Fig.~\ref{fig:1204}. Those points that also break $EW$ symmetry are displayed in purple. This plot reaffirms the conclusion drawn from Table~\ref{tab:320} that most of the points that break $B-L$ with a $Z_{R}$ mass above the experimental bound, also break EW symmetry. Note that a small density of green points that do not break EW symmetry are obscured by the purple points.

In Fig.~\ref{fig:1148}, we reproduce Fig.~\ref{fig:1205} but now, in addition, sequentially indicate the points that are consistent with the remaining checks--that is, non-tachyonic stops/spill checks/convergence/all lower bounds on sparticles masses satisfied and, finally, that they reproduce the Higgs mass within the experimental uncertainty. Points that appropriately break $B-L$ symmetry but do not satisfy electroweak symmetry breaking are still shown in green. Points that, additionally, do break electroweak symmetry are again shown in purple. Such points that also  have non-tachyonic stops, pass all spill checks and convergence, and which satisfy all lower bounds on sparticles masses, but do not match the known Higgs mass, are now indicated in cyan. Finally, points that satisfy all checks, including the correct Higgs mass, are shown in black. These are the valid points. The density of black points indicates, as observed above, that there is a surprisingly high number of initial parameters that satisfy all present low energy experimental constraints. The distribution of black points can be explained from the fact that, while $B-L$ breaking favors non-zero $S$-terms, very large $S$-terms can effect the RGE evolution of sfermion masses adversely. Since the effect of the $S$-terms depends on the charge of the sfermion in question, some sfermions will become quite heavy while others light or tachyonic. Therefore, in general, the valid points in our scan are a compromise between large $S$-terms, needed for a $Z_R$ mass above its lower bound, and small $S$-terms needed to keep the sfermion RGEs under control.

\begin{figure}[!htbp]
	\centering
	\includegraphics[scale=1.0]{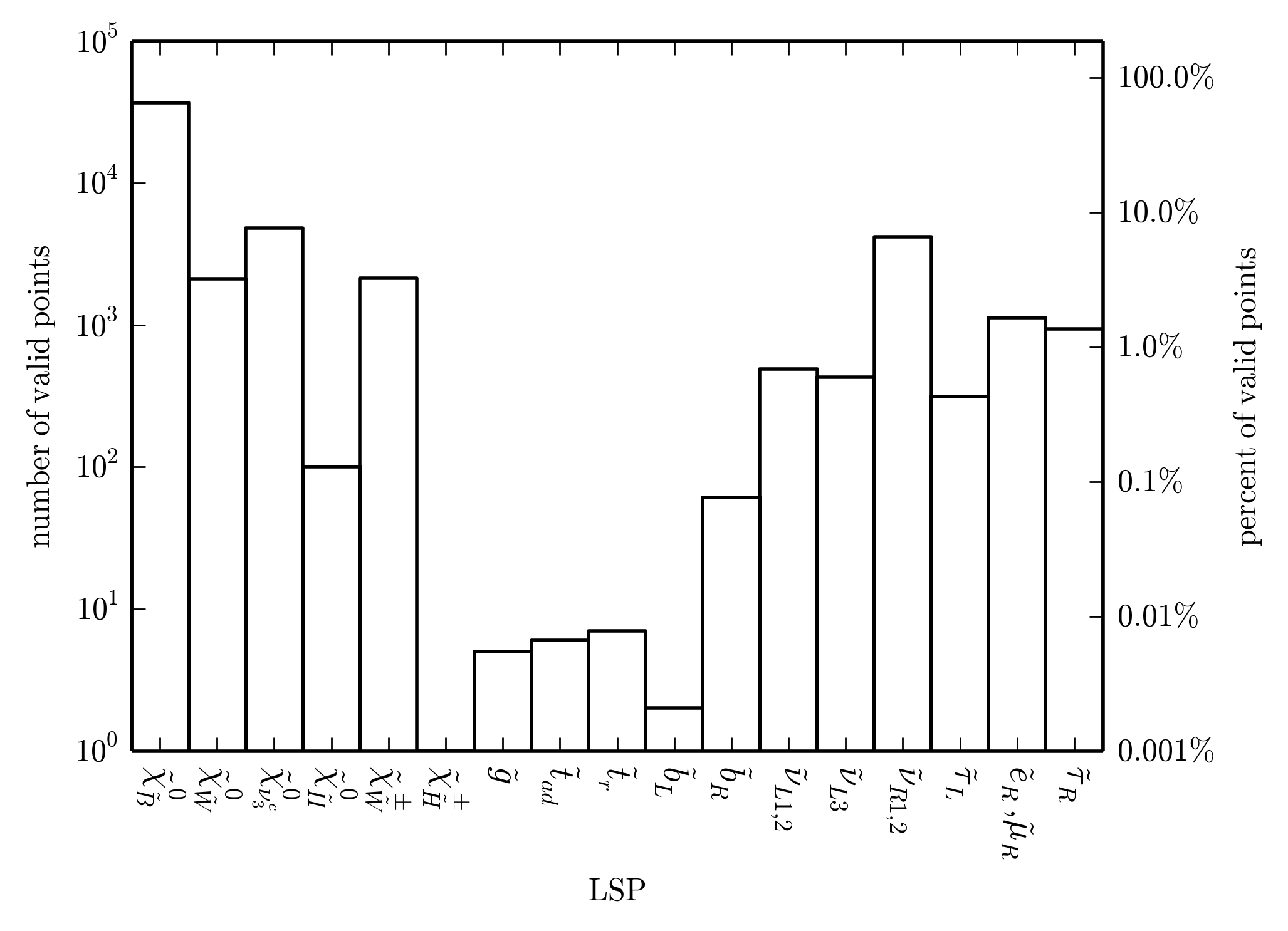}
	\caption{\small A histogram of the LSP's in the main scan showing the percentage of valid points with a given LSP. Sparticles which did not appear as LSP's are omitted. The y-axis has a log scale. The dominant contribution comes from the lightest neutralino, as one might expect. The notation for the various states, as well as their most likely decay products, are given in Table~\ref{tbl:LSP}. Note that we have combined left-handed first and second generation sneutrinos into one bin, and that each generation makes up about 50\% of the LSP's. The same is true for the first and second generation right-handed sleptons and sneutrinos.
}
	\label{fig:1039}
\end{figure}

The most important property of the inital SUSY parameter space in determining low-energy phenomenology is the identity of the LSP. Recall that when $R$-parity is violated, no restrictions exist on the identity of the LSP; for example, it can carry color or electric charge. Our main scan provides an excellent opportunity to examine the possible LSP's and the probability of their occurence . To this end, a histogram of possible LSP's is presented in Fig.~\ref{fig:1039}--with the possible LSP's indicated along the horizontal axis, and $\log_{10}$ of the number of valid points with a given LSP on the vertical axis. The notation here is a bit condensed, but is specified in more detail in Table~\ref{tbl:LSP}. The notation is devised to highlight the phenomenology of the different LSP's, specifically their decays, which are also presented in Table~\ref{tbl:LSP}.

\begin{table}[htdp]
\begin{center}
\begin{tabular}{|c|c|c|}
\hline
\ Symbol \ & Description & Decay
\\
\hline
\hline
$\tilde \chi^0_{\tilde B}$ & A bino-like neutralino, mostly rino ($\tilde W_R$) or mostly blino ($\tilde B'$).
& \multirow{4}{*}{$\ell^\pm W^\mp$, $\nu Z$, $\nu h$}
\\
$\tilde \chi^0_{\tilde W}$ & Mostly wino neutralino. &
\\
$\tilde \chi_{\nu^c}$ & Mostly third generation right-handed neutrino. &
\\
$\tilde \chi^0_{\tilde H}$ & Mostly Higgsino neutralino. &
\\
\hline
$\tilde \chi^{\pm}_{\tilde W}$ & Mostly wino charginos. & \multirow{2}{*}{$\nu W^\pm$, $\ell^{\pm} Z$, $\ell^{\pm} h$}
\\
$\tilde \chi^{\pm}_{\tilde H}$ & Mostly Higgsino charginos. & 
\\
\hline
$\tilde g$ & Gluino. & $t \bar t \nu$, $t \bar b \ell^-$
\\
\hline
 $\tilde t_{ad}$ & Left- and right-handed stop admixture. & $\ell^+  b$
\\
\hline
$\tilde t_{r}$ & Mostly right-handed stop (over 99\%). & $t \nu$, $\tau^+ b$
\\
\hline
$\tilde q_R$ & Right-handed first and second generation squarks. & $\ell^+j$, $\nu j$
\\
\hline
$\tilde b_L$ & Mostly left-handed sbottom. & $b \nu$
\\
\hline
$\tilde b_R$ & Mostly right-handed sbottom. & $b \nu$, $\ell^- t$
\\
\hline
\multirow{2}{*}{$\tilde \nu_{L_{1,2}}$} & First and second generation left-handed sneutrinos.
				& \multirow{3}{*}{\parbox[t]{3cm}{$b \bar b$, $W^+ W^-$, $ZZ$, \\ $t \bar t$, $\ell'^+ \ell^-$, $hh$, $\nu \nu$}}
\\
								&	LSP's are split evenly among these two generations. &
\\
$\tilde \nu_{L_3}$ & Third generation left-handed sneutrino. & 
\\
\hline
$\tilde \nu_{R_{1,2}}$ & First and second generation right-handed sneutrinos. & $\nu \nu$
\\
\hline
\multirow{2}{*}{$\tilde \tau_L$} &  \multirow{2}{*}{Third generation left-handed stau.} & $t \bar b$, $W^- h$,
\\
			&		& $e \nu$, $\mu \nu$, $\tau \nu$
\\
\hline
\multirow{2}{*}{$\tilde e_R, \mu_R$} &  First and second generation right-handed sleptons. & \multirow{2}{*}{$e \nu$, $\mu \nu$}
\\
	& LSP's are split evenly between these two generations.	&
\\
\hline
$\tilde \tau_R$ &  Third generation right-handed stau. & $t \bar b$, $e \nu$, $\mu \nu$, $\tau \nu$
\\
\hline
\end{tabular}
\end{center}
\caption{The notation used for the states in Fig.~\ref{fig:1039} and their probable decays. More decays are possible in certain situations depending on what is kinematically possible and the parameter space. Gluino decays are especially dependent on the NLSP, here assumed to be a neutralino. Here, the word ``mostly'' means it is the greatest contribution to the state. The symbol $\ell$ represents any generation of charged leptons. The left-handed sneutrino decay into $\ell'^+ \ell^-$ indicates a lepton flavor violating decay--that is, $\ell'^+$ and $\ell^-$ do not have the same flavor.  Note that $j$ is a jet--indicating a light quark.}
\label{tbl:LSP}
\end{table}
The most common LSP in our main scan is the lightest neutralino, $\tilde \chi_1^0$. However, not all $\tilde \chi_1^0$ states are created equal. LHC production modes for the lightest neutralino depend significantly on the composition of the neutralino--a bino LSP cannot be directly produced at the LHC, but the other neutralino LSP's can. This is the basis we use for the division of these states. The state $\tilde \chi_{\tilde B}^0$ designates a mostly rino or mostly blino neutralino, $\tilde \chi_{\tilde W}^0$ a mostly wino neutralino and $\tilde \chi_{\tilde H}^0$ a mostly Higgsino neutralino. Here, the word mostly indicates the greatest contribution to that state. As an unrealistic example, if $\tilde \chi_1^0$ is 34\% wino, 33\% bino and 33\% Higgsino, it is still labeled $\tilde \chi_{\tilde W}^0$. The chargino LSP's are similarly separated into wino-like and higgsino like charginos, and the stops and sbottom divisions are as in our earlier papers, references~\cite{Marshall:2014kea,Marshall:2014cwa}. Note that this notation for the stops, $\tilde t_{ad}$ and $\tilde t_r$, are only used to describe stop LSP's. For non-LSP stops, we use the conventional notation $\tilde t_1$ and $\tilde t_2$.

To make Fig.~\ref{fig:1039} more readable, we have made an effort to combine bins that have similar characteristics. The first and second generation left-handed sneutrinos are combined into one bin, where about 50\% of the LSP's are first generation sneutrinos. The same holds true for the first and second generation right-handed sleptons, while the first generation right-handed sneutrino is always chosen to be lighter than the second generation right-handed sneutrino. This similarity between the first and second generation sleptons is expected, since their corresponding Yukawa couplings are not large enough to distinguish them through the RG evolution. For both sleptons and squarks, more LSP's exist for the third generation--as expected from the effects of the third generation Yukawa couplings, which tend to decrease sfermion masses in RGE evolution.

The myriad of possible LSP's leads to a rich collider phenomenology. This phenomenology is not the main focus of this paper, but it is worthwhile to briefly review it here. In models where $R$-parity is parameterized by bilinear $R$-parity, such as the $B-L$ MSSM, SUSY particles are still pair produced and cascade decay to the LSP. At this point, the bilinear $R$-parity violating terms allow the LSP to decay. While only a few studies have been done on the phenomenology of the minimal $B-L$ MSSM~\cite{FileviezPerez:2012mj, Perez:2013kla,Marshall:2014kea,Marshall:2014cwa}, there have been several works on the phenomenology of explicit bilinear $R$-parity violation, which has some similarities to this model. References to such papers are mentioned below although see~\cite{Porod:2000hv, Hirsch:2003fe, Graham:2012th, Graham:2014vya} for general discussions. Table~\ref{tbl:LSP} provides some basic information on the most probable decay modes of each of the possible LSP's. Note that $\ell$ signifies a charged lepton of any generation and $j$ a jet--implying a light quark. Interesting aspects of Table~\ref{tbl:LSP} are the following.

\underline{LSP Phenomenology}
\begin{itemize}
	\item \textbf{Neutralinos}: Only neutralinos with non-significant blino or rino components can be significantly produced at the LHC. Note that in addition to the usual possibilities, a mostly right-handed third generation neutrino is also a possible lightest neutralino component here, because of $R$-parity violation. This can be pair produced through the $Z_R$ resonance. Due to the Majorana nature of the neutralinos, they can lead to same-sign dilepton signals--a clear sign of lepton number violation. This is true whether they are directly produced or occur at the end of a cascade decay. The generation of $\ell$ depends on the neutrino mass hierarchy, as discussed in~\cite{Marshall:2014kea,Marshall:2014cwa}. In the normal hierarchy, muons and taus are most likely, while in the inverted hierarchy all charged leptons are possible.
	\item \textbf{Gluino}: Most of the LSP decay products mentioned in Table~\ref{tbl:LSP} mimic well-known hypothetical states--for example, neutralinos decay like TeV scale right-handed neutrinos and squarks decay like leptoquarks. The same can not be said of the gluino, making it an interesting candidate for further study. However, its decays depend strongly on the identity of the next to lightest supersymmetric particle (NLSP). Also, bounds on the gluino are the strongest because of its large production cross section. Therefore, when the gluino is the LSP, it is likely that it is the only LHC-accessible SUSY particle. As with the neutralinos, the gluino's Majorana nature allows same-sign dilepton final states--indicating lepton number violation.
	\item \textbf{Squarks}: All squark LSP's act like leptoquarks in this model, meaning they are pair-produced and decay into a lepton and a quark. Stop and sbottom LSP's in this model were studied in \cite{Marshall:2014kea,Marshall:2014cwa}. For both a down- and an up-type non-third generation squark LSP, there will be two highly degenerate LSP states--either a degenerate down and strange squark pair or a degenerate up and charm squark pair--as required by phenomenology. In the inverted hierarchy, these can decay into an electron and jet or a neutrino and a jet, making them tempting explanations for the recent CMS excess, see~\cite{CMS:2014qpa}, in the $eejj$ and $e\nu jj$ channels~\cite{ Chun:2014jha}. However, the branching ratios seem to be inconsistent with the cross section~\cite{Queiroz:2014pra,Allanach:2015ria}. See reference~\cite{Evans:2012bf} for a study of stop in trilinear R-parity violation.
	\item \textbf{Left-handed sneutrinos}: Left-handed sneutrinos decay like heavier neutral Higgs bosons, that is, $H^{0}$ and $A^{0}$, due to their $R$-parity violating mixing with the Higgs sector. In general, decays into heavier Higgses are also possible but, of course, this depends on kinematics. The final state $\ell'^+ \ell^-$ represent a lepton flavor violating final state, such as $\mu^+ e^-$. Sneutrinos LSP decays were studied in the case of explicit bilinear $R$-parity violation, which has some similarities to the $B-L$ MSSM, in reference~\cite{Aristizabal Sierra:2004cy, AristizabalSierra:2012qa}. 
	\item \textbf{Right-handed sneutrinos}: These states decay into missing energy and, therefore, cannot be easily distinguished from the $R$-parity conserving MSSM. However, since the sneutrino is spin 0, as opposed to spin half neutralinos, a detailed collider study might reveal some differences. It is also interesting to note that it may be possible to pair produce right-handed sneutrinos through a $Z_R$ resonance, although the cross section would probably be small.
	\item \textbf{Sleptons}: Both left-handed and right-handed charged sleptons decay like charged Higgs bosons, with which the sleptons mix due to $R$-parity violation. The left-handed sleptons have more channels open to them because of their isospin charge. Each left-handed slepton comes in an $SU(2)$ doublet with the associated left-handed sneutrino. Splitting of this doublet is mainly due to electroweak $D$-term contributions to the mass, which push the associated left-handed sneutrino to lighter mass values, making it the LSP. In the case of the left-handed stau, however, mixing effects through the Yukawa and tri-scalar couplings (see Appendix~\ref{sec:546}) have the potential to make its mass lighter than the third-family left-handed sneutrino. Therefore, the left-handed stau is the only left-handed charged slepton capable of being the LSP. Slepton LSPs with explicit $R$-parity violation were discussed in reference~\cite{Bartl:2003uq}.
\end{itemize}

To get a sense of the non-LSP spectrum, we produce histograms of the masses of the sparticles from the main scan. In the following histograms, there will be quite a few pairs of fields that will be highly degenerate; these will be represented by only one curve. This includes $SU(2)_L$ sfermion partners, which are only split by small electroweak terms. First generation squarks are also degenerate with second generation squarks with the same isospin, due to phenomenological constraints. A consequence of this is that all first and second generation left-handed squarks are highly degenerate.
In viewing these histograms, it is helpful to remember that aside from the usual RGE effects of the MSSM, there are two additional effects involved. The first of these is the boundary conditions at the $B-L$ scale, corresponding to the $B-L$ and $I_{3R}$ $D$-terms which are given in Eq.~(\ref{eq:BC.BL}). The second is the new RGE effects of the $S_R$ and $S_{BL}$ terms. Although the signs of these terms are not fixed, Fig.~\ref{fig:1204} shows that $S_{BL}$ is typically positive while $S_R$ is typically negative. This indicates that $S_R$ will tend to increase (decrease) sfermion masses for sfermions with a positive (negative) $I_{3R}$ charge, while $S_{BL}$ tends to increase (decrease) sfermion masses for sfermions with negative (positive) $B-L$. 

\begin{figure}[!htbp]
\centering
	\includegraphics[scale=0.6]{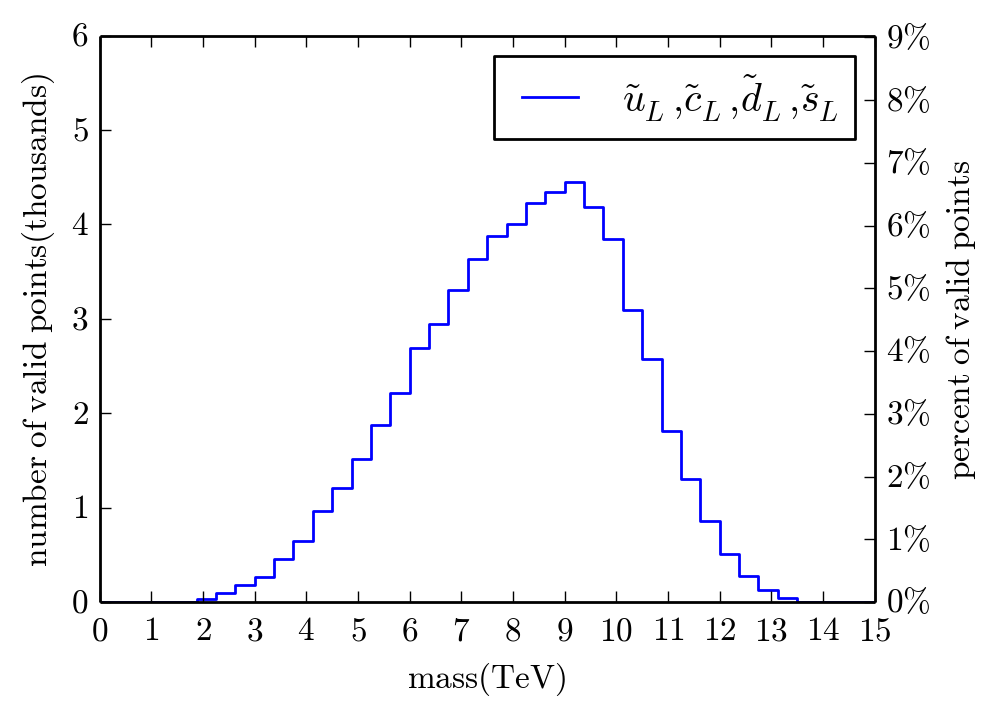}
	\includegraphics[scale=0.6]{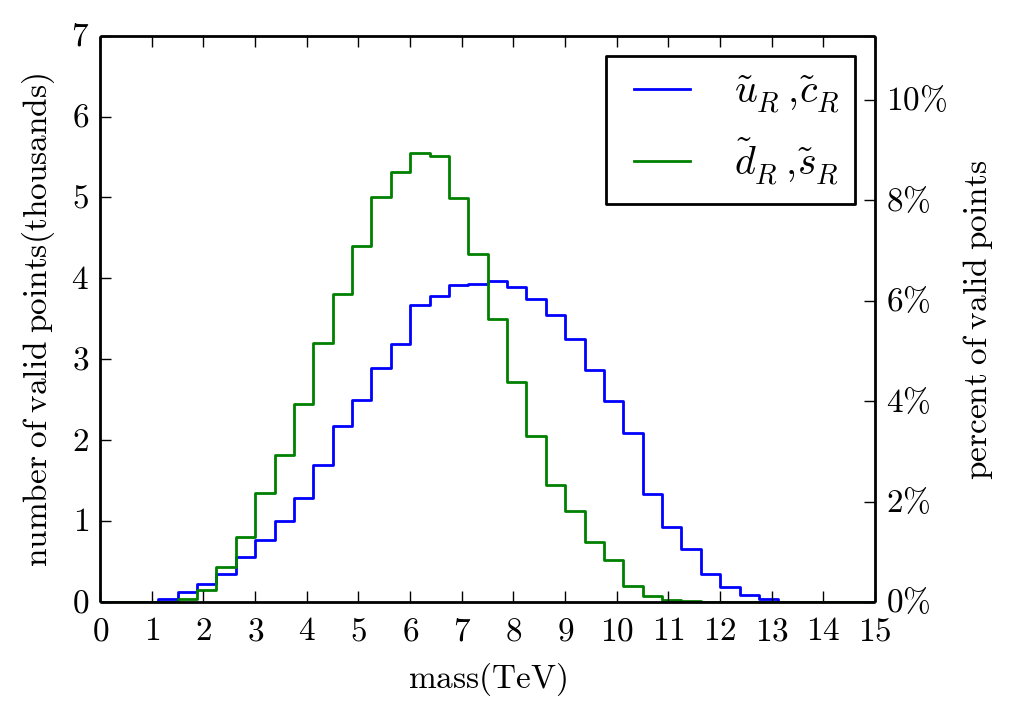}
	\includegraphics[scale=0.6]{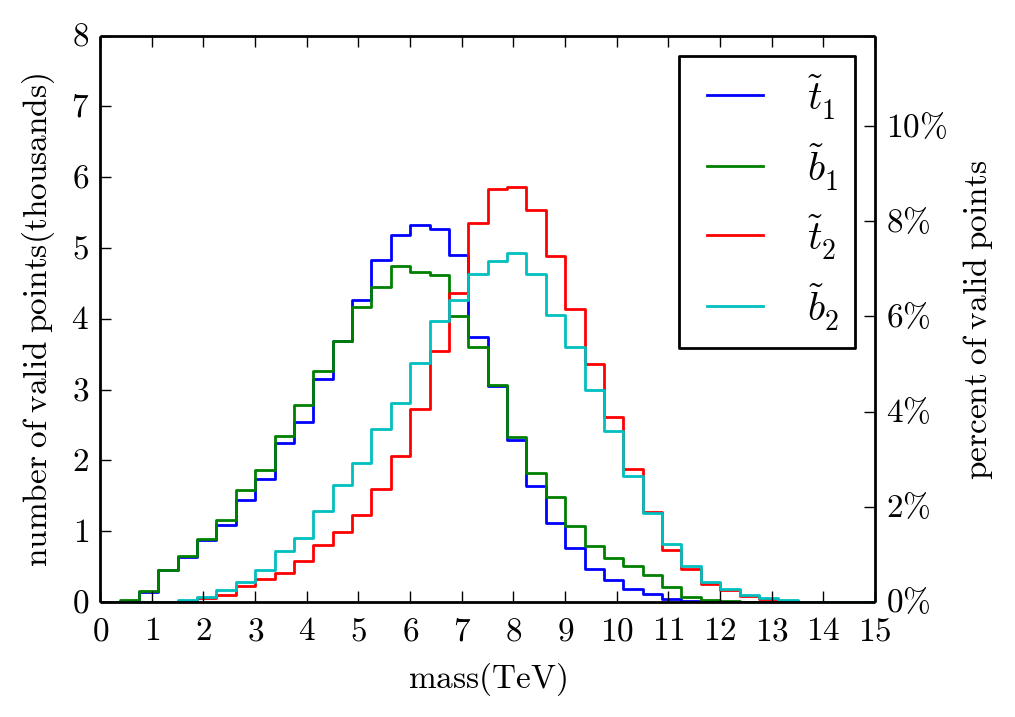}
\caption{\small Histograms of the squark masses from the main scan. The first- and second-family left-handed squarks are shown in the top-left panel. Because they come in $SU(2)$ doublets, and the first- and second-family squarks must be degenerate, all four of these squarks have nearly identical mass and the histograms coincide. The first- and second-family right-handed squarks are shown in the top-right panel. The right-handed down squarks are generally lighter than their up counterparts because of the effect of the $U(1)_{3R}$ charge in the RGEs. The third family squarks are shown in the bottom panel. }
\label{fig:1052}
\end{figure}

Figure~\ref{fig:1052} shows histograms of the squark masses. Because they come in $SU(2)$ doublets and the first- and second-family squarks must be degenerate, all four of the first- and second-family left-handed squarks have nearly identical mass and the histograms coincide. The degeneracy of first- and second-family squarks is also evident in the right-handed squark masses. The first and second family right-handed down squarks are generally lighter than their up counterparts because of the effect of the $U(1)_{3R}$ charge in the RGEs.

\begin{figure}[!htbp]
\centering
	\includegraphics[scale=0.6]{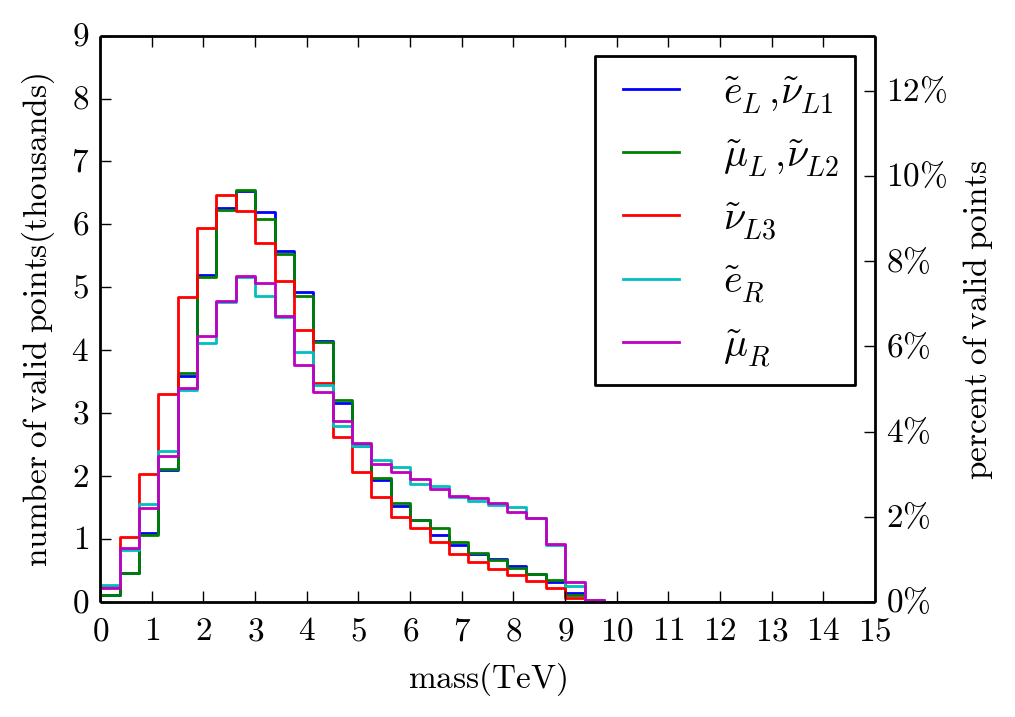}
	\includegraphics[scale=0.6]{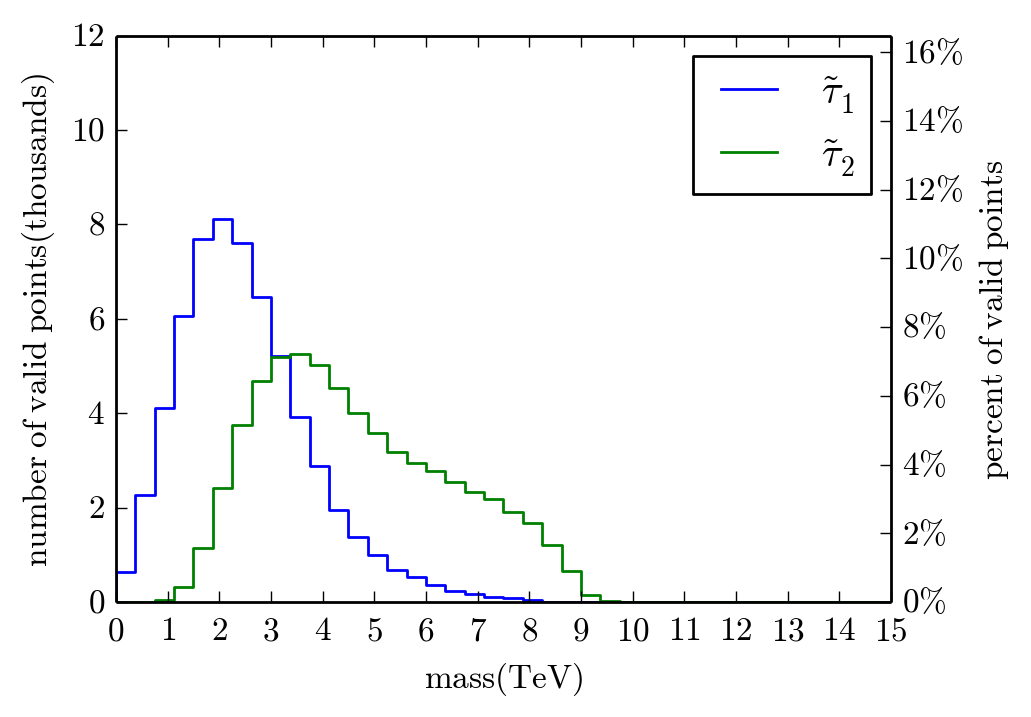}
	\includegraphics[scale=0.6]{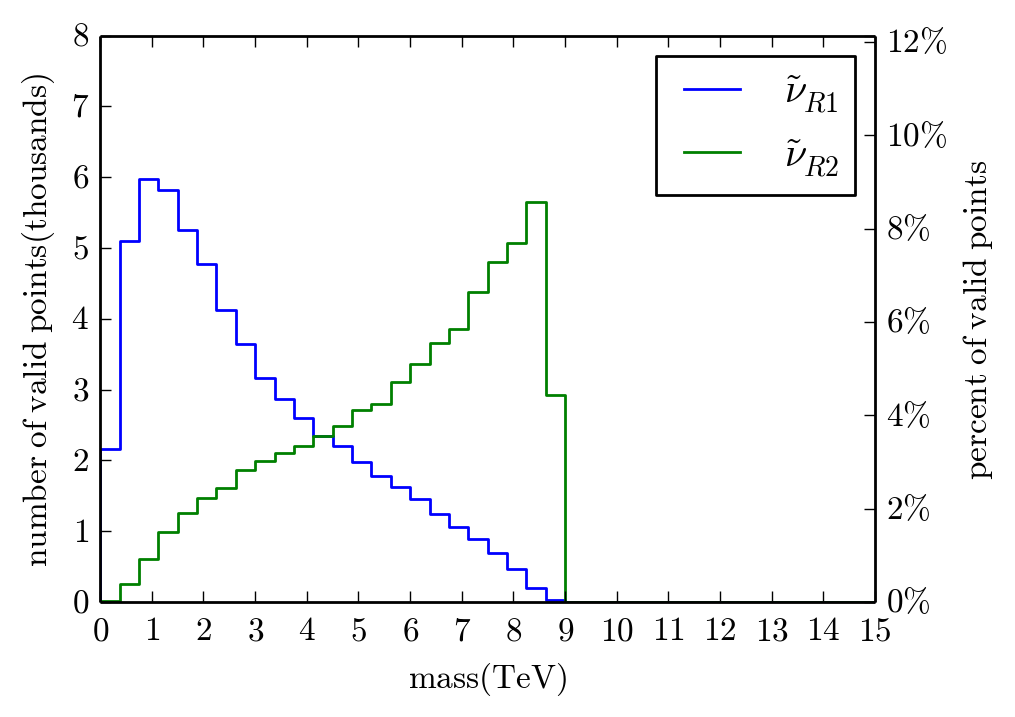}
\caption{\small Histograms of the sneutrino and slepton masses in the the main scan. First- and second-family entries are in the top-left panel, along with the third family left-handed sneutrino. Staus are in the top-right panel with mass-ordered labeling. In the bottom panel, the first- and second-family right-handed sneutrinos are labeled such that $\tilde \nu_{R1}$ is always lighter than $\tilde \nu_{R2}$.}
	\label{fig:hist.sleptons}
\end{figure}

Figure~\ref{fig:hist.sleptons} shows histograms of the masses of the sneutrinos and sleptons. The third-family sleptons and left-handed sneutrinos tend to be the lighter because of the influence of the tau Yukawa couplings. The right-handed sneutrinos are labeled such that $\tilde \nu_{R_1}$ is always lighter than $\tilde \nu_{R_2}$.

\begin{figure}[!htbp]
\centering
	\includegraphics[scale=0.6]{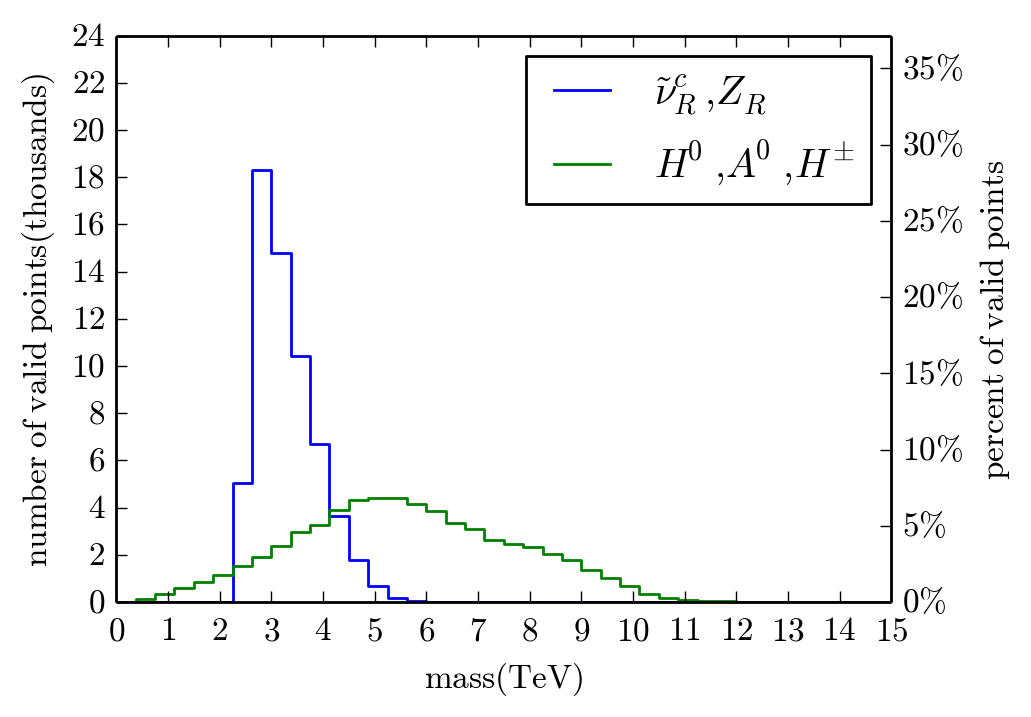}
	\includegraphics[scale=0.6]{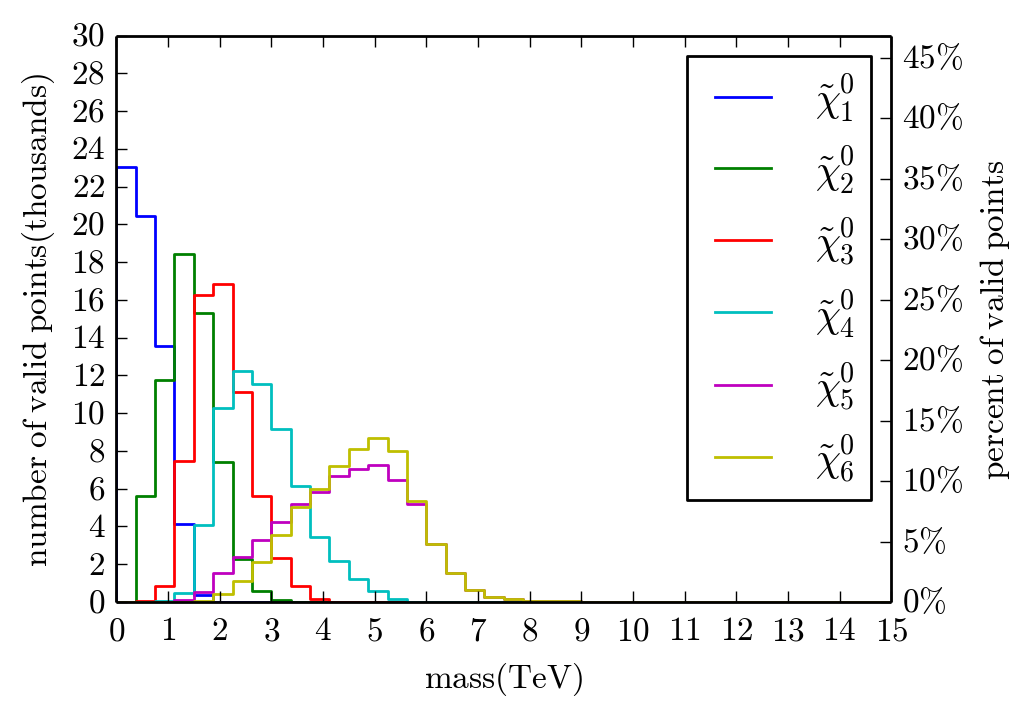}
	\includegraphics[scale=0.6]{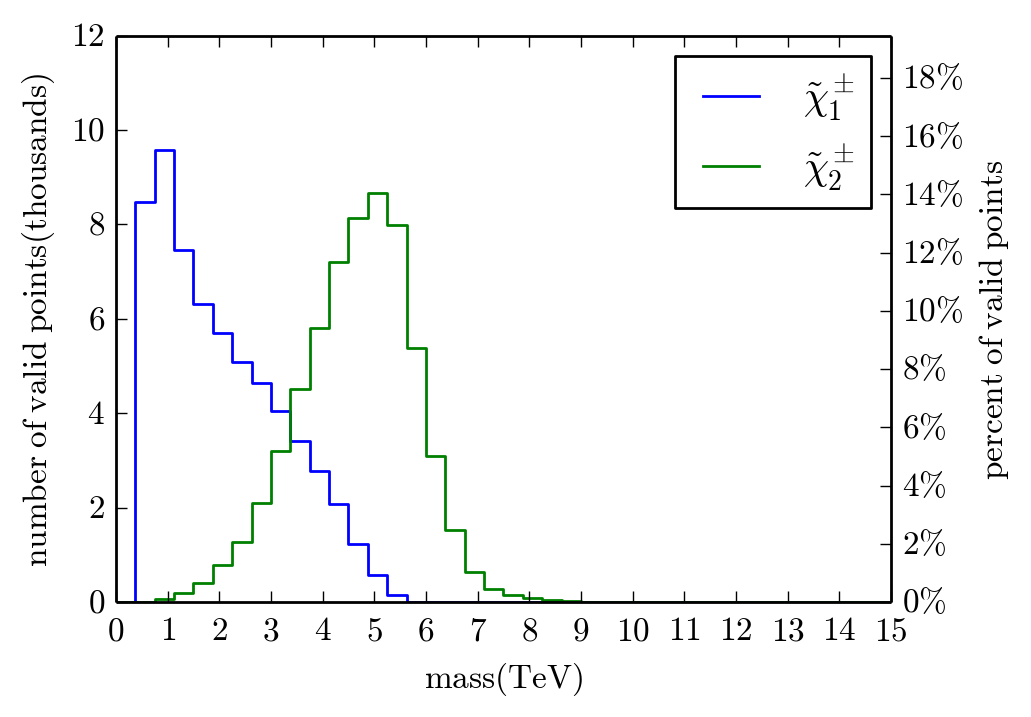}
	\includegraphics[scale=0.6]{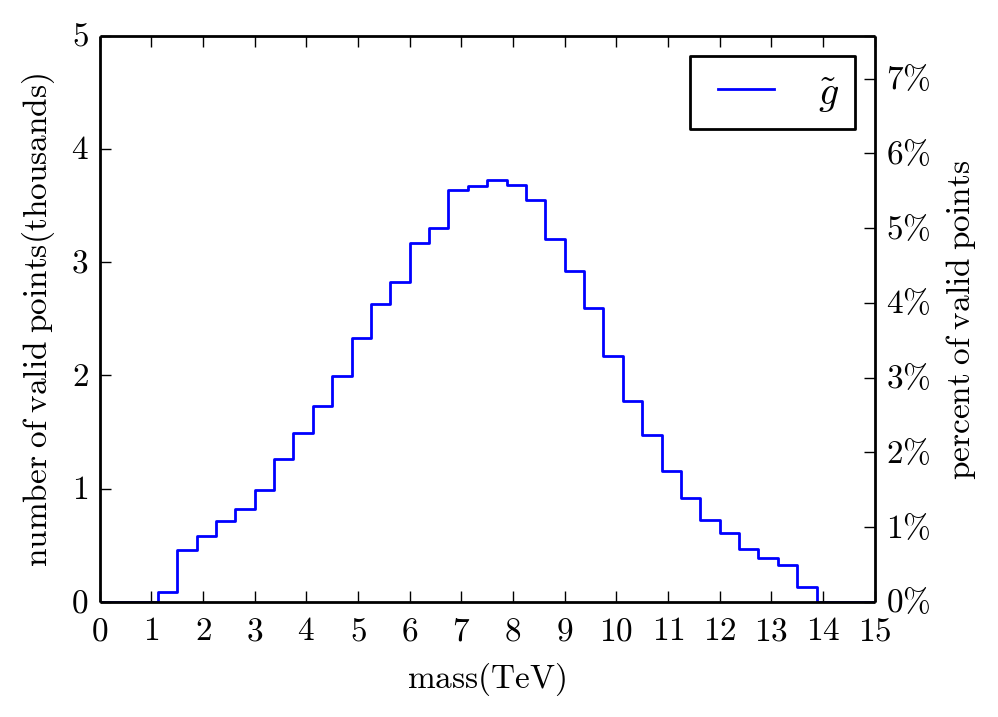}
\caption{\small The CP-even component of the third-family right-handed sneutrino, heavy Higgses, neutralinos, charginos  and the gluino in the valid points from our main scan. The CP-even component of the third generation right-handed sneutrino is degenerate with $Z_R$. The $\tilde \chi^0_5$ and $\tilde \chi^0_5$ are typically Higgsinos.}
\label{fig:1054}
\end{figure}

Figure~\ref{fig:1054} presents histograms of the CP-even component of the third generation right-handed sneutrino, the heavy Higgses, the neutralinos, the charginos, and the gluino. The CP-even component of the third generation right-handed sneutrino is degenerate with $Z_R$. It is always heavier than 2.5 TeV because we have imposed the collider bound on $Z_R$. The neutralinos and charginos are labeled from lightest to heaviest as is canonical in SUSY models. The $\tilde \chi^0_5$ and $\tilde \chi^0_6$ are typically Higgsinos.

We emphasize that all of the above histograms are calculated using our main scan; that is, for the choice of $M=2700$ GeV and $f=3.3$. We remind the reader that these values were chosen so as to maximize the number of valid accessible points.  However, the mass scale of these histograms is heavily dependent on the choice of $M$. Smaller values for $M$ will move the above distributions distinctly toward lighter sparticle masses.

\begin{figure}[!htbp]
\centering
	\includegraphics[scale=0.6]{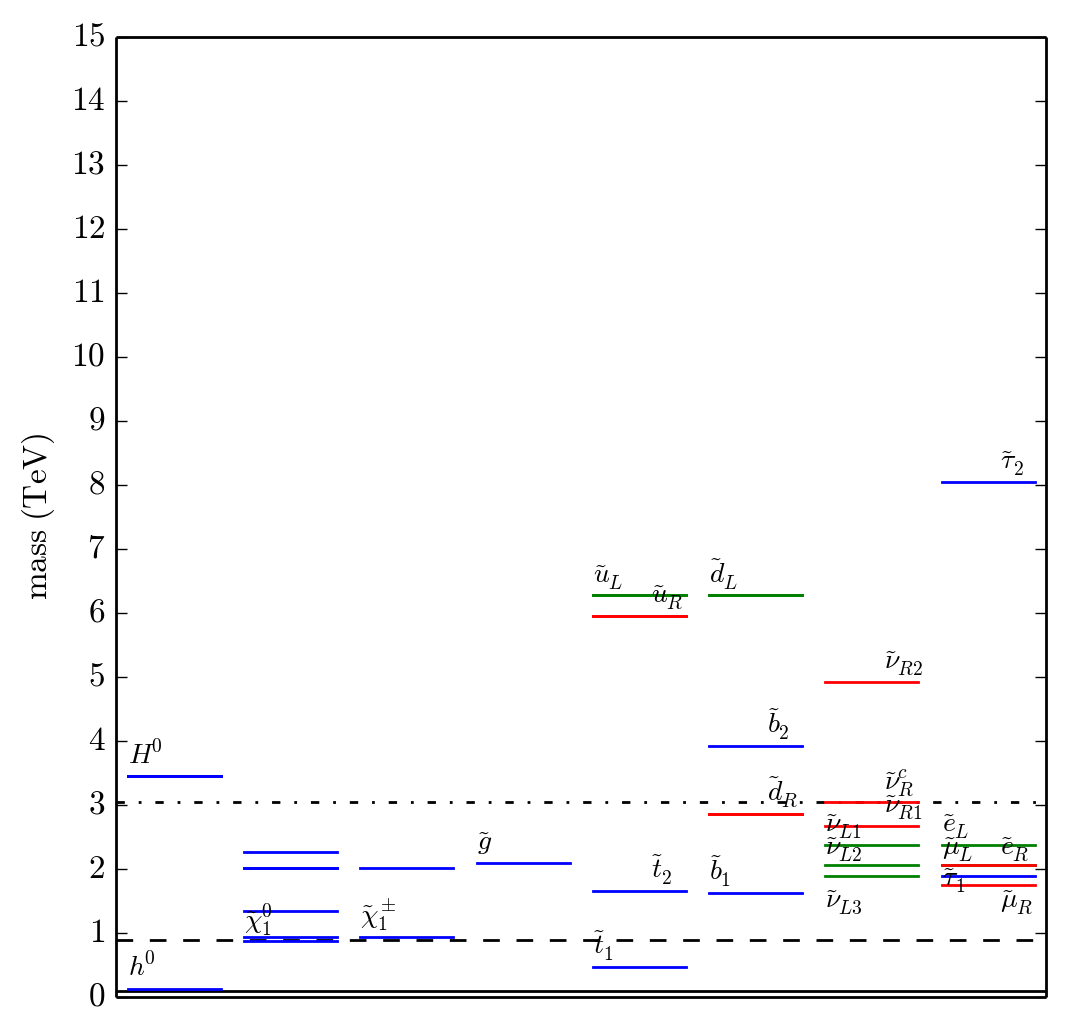}
	\includegraphics[scale=0.6]{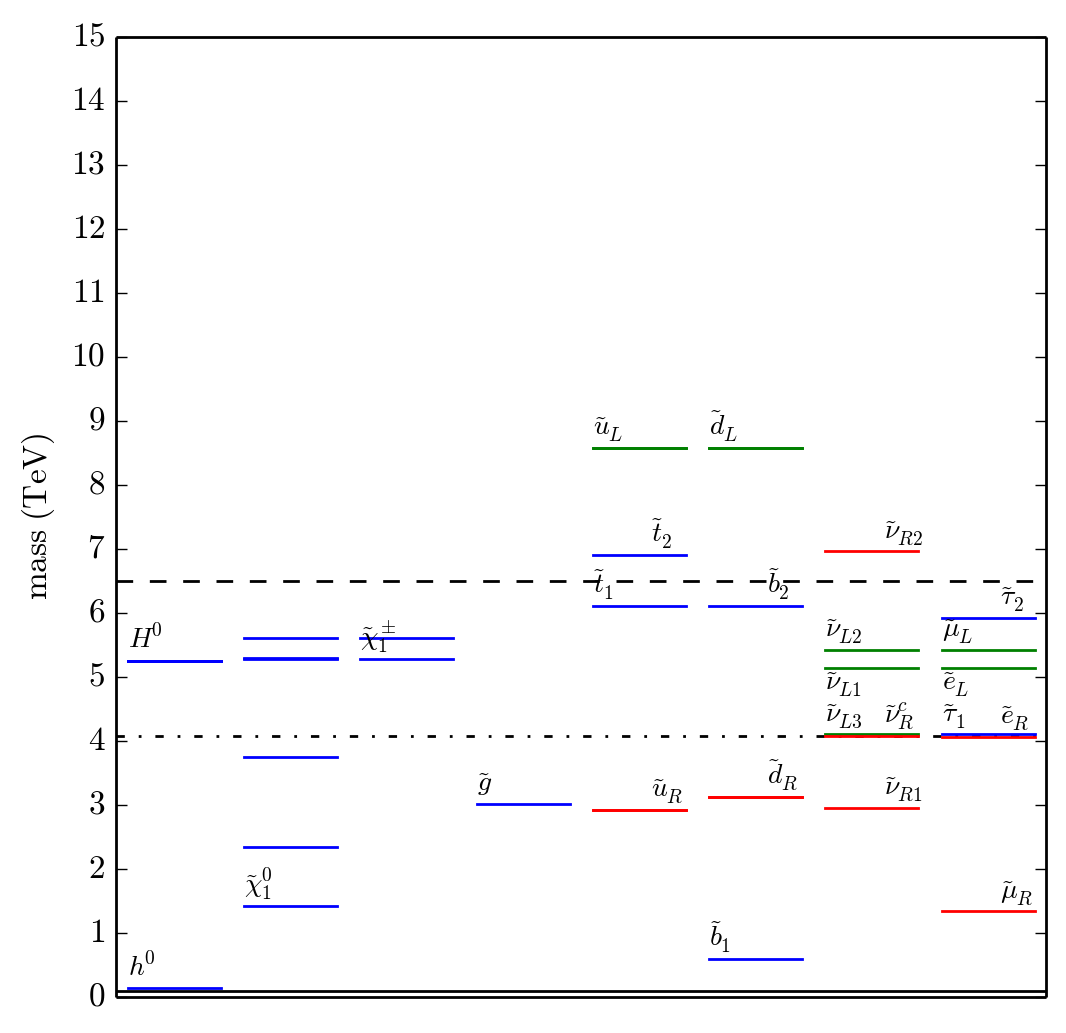}
	\caption{\small Two sample physical spectra with an admixture stop LSP and right-handed sbottom LSP. The $B-L$ scale is represented by a black dot-dash-dot line. The SUSY scale is represented by a black dashed line. The electroweak scale is represented by a solid black line. The label $\tilde u_L$ is actually labeling the nearly degenerate $\tilde u_L$ and $\tilde c_L$ masses. The labels $\tilde u_R$, $\tilde d_L$ and $\tilde d_R$ are similarly labeling the nearly degenerate first- and second- family masses.}
	\label{fig:154}
\end{figure}

\begin{figure}[!htbp]
\centering
	\includegraphics[scale=0.6]{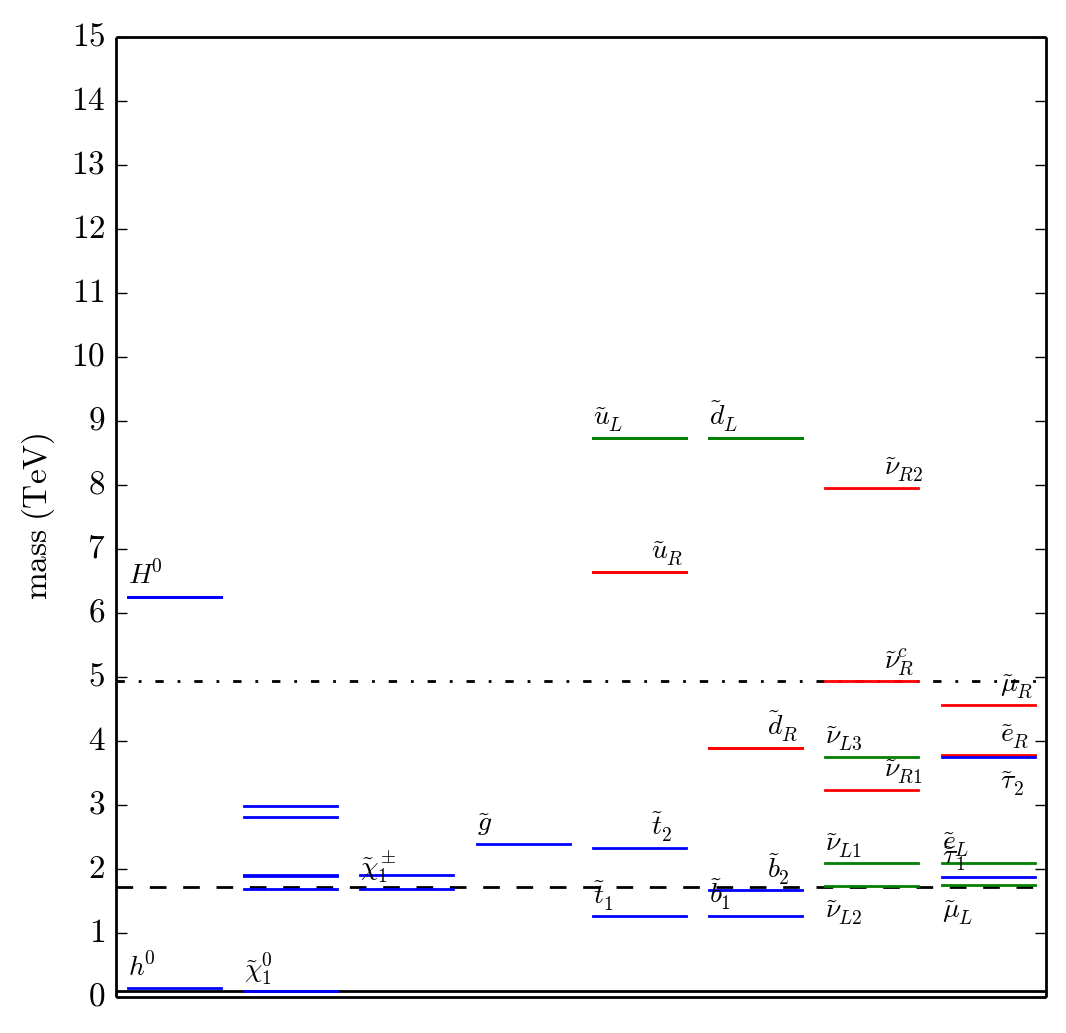}
	\includegraphics[scale=0.6]{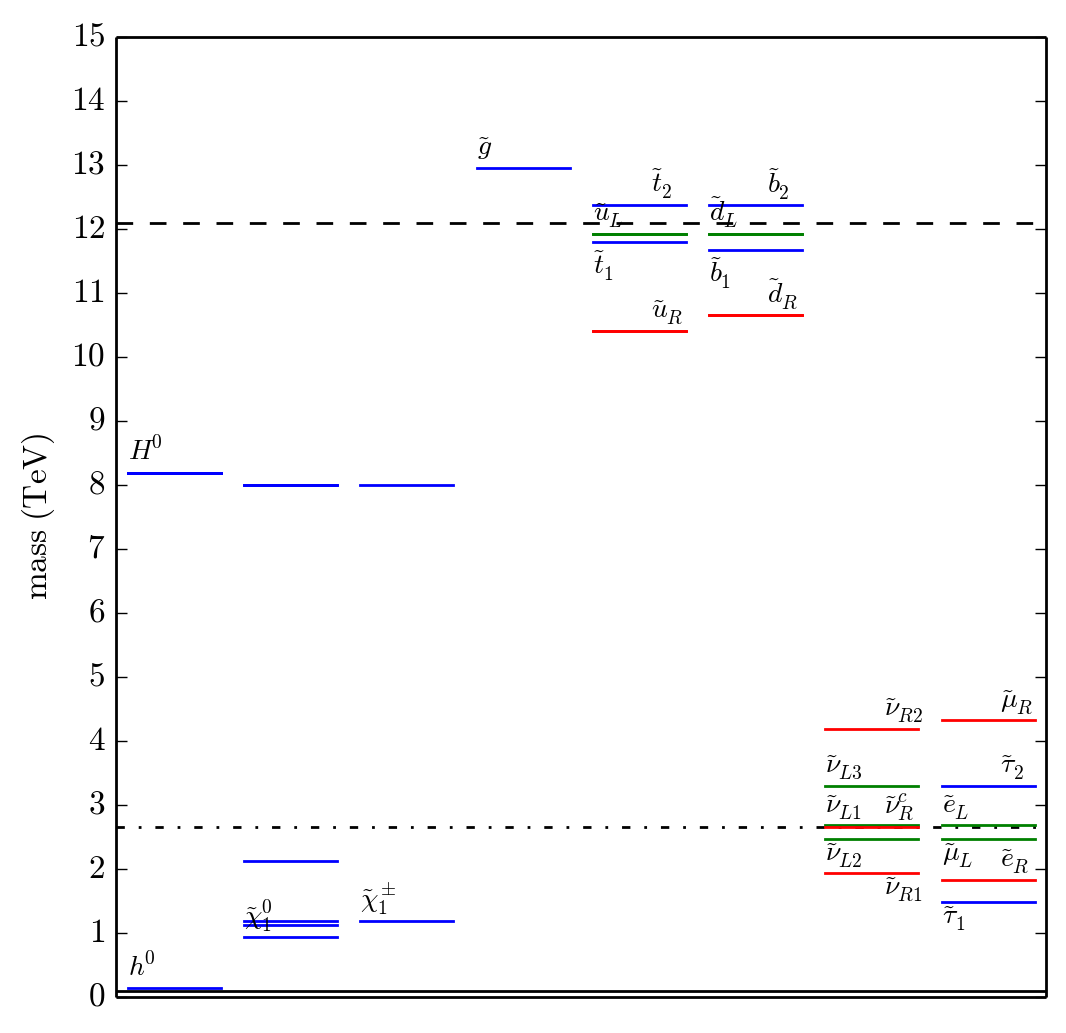}
	\caption{\small Two sample physical spectra with a right-side-up hierarchy and upside-down hierarchy. The $B-L$ scale is represented by a black dot-dash-dot line. The SUSY scale is represented by a black dashed line. The electroweak scale is represented by a solid black line. The label $\tilde u_L$ is actually labeling the nearly degenerate $\tilde u_L$ and $\tilde c_L$ masses. The labels $\tilde u_R$, $\tilde d_L$ and $\tilde d_R$ are similarly labeling the nearly degenerate first- and second- family masses.}
	\label{fig:156}
\end{figure}

Plots of the physical particle spectra for four valid points are presented in Figs.~\ref{fig:154} and \ref{fig:156}. 
These four points are automatically selected from the pool of valid points from the main scan based on simple criteria. The first is the spectrum with an admixture stop LSP with the largest gap between stop LSP and the next lightest sparticle. The second is similar; now, however, with a right-handed sbottom LSP. The third and fourth are the valid points with the largest right-side-up and upside-down hierarchy respectively; that is, the largest splittings between the $B-L$ and SUSY scales in the two possible hierarchies.

\comment{
\begin{figure}[!htbp]
\centering
	\includegraphics[scale=1]{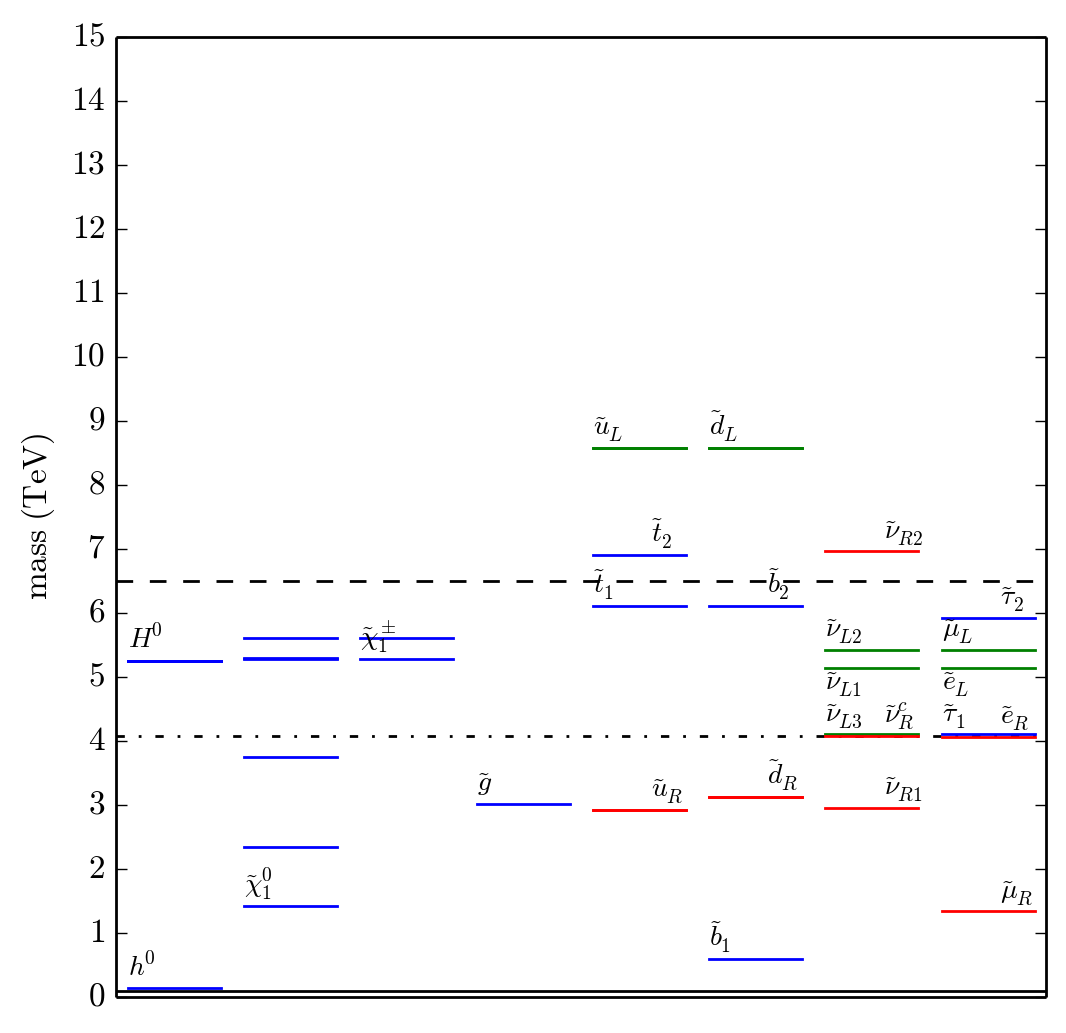}
	\caption{\small An example physical spectrum with a right-handed sbottom LSP. The $B-L$ scale is represented by a black dot-dash-dot line. The SUSY scale is represented by a black dashed line. The electroweak scale is represented by a solid black line. The label $\tilde u_L$ is actually labeling the nearly degenerate $\tilde u_L$ and $\tilde c_L$ masses. The labels $\tilde u_R$, $\tilde d_L$ and $\tilde d_R$ are similarly labeling the nearly degenerate first- and second- family masses.}
	\label{fig:155}
\end{figure}
}

\comment{
\begin{figure}[!htbp]
\centering
	\includegraphics[scale=1]{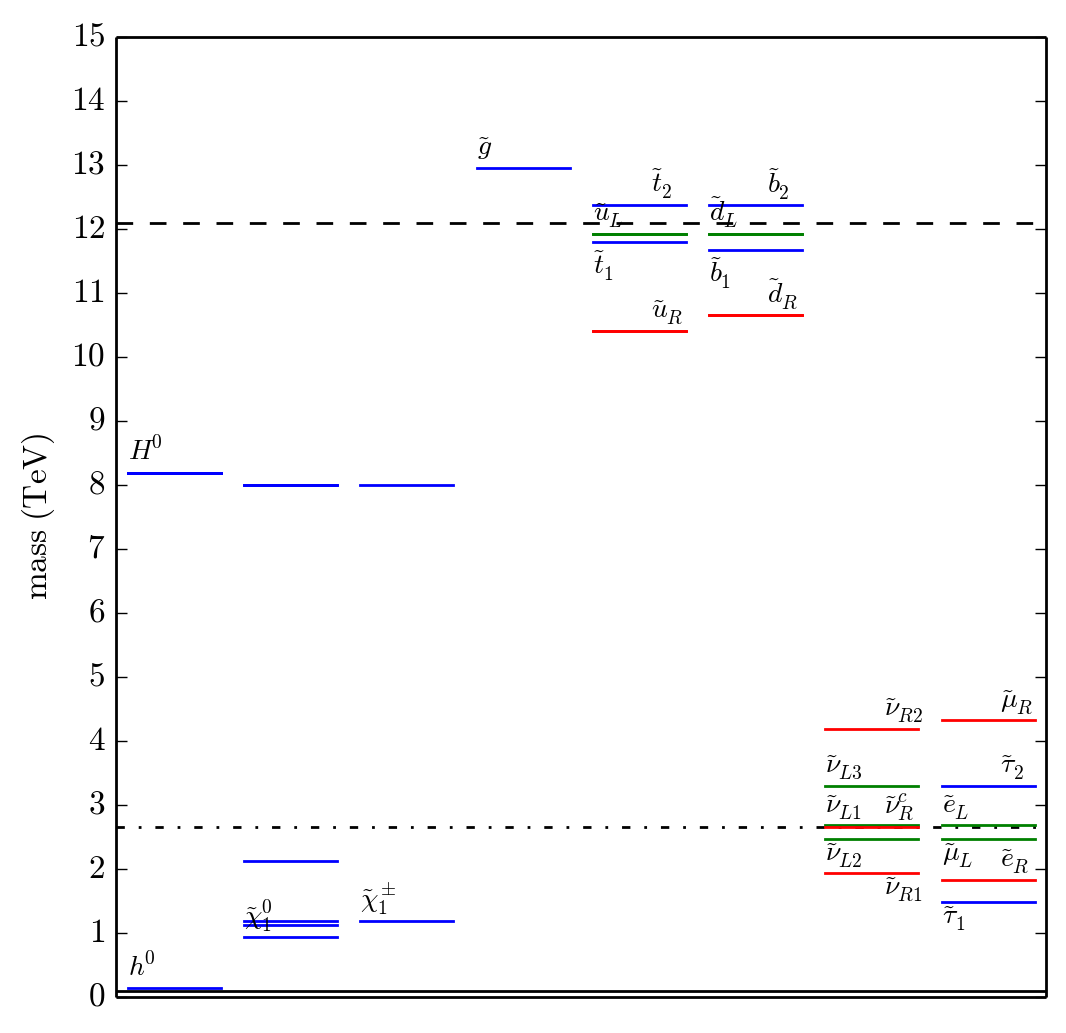}
	\caption{\small An example physical spectrum with an upside-down hierarchy. The $B-L$ scale is represented by a black dot-dash-dot line. The SUSY scale is represented by a black dashed line. The electroweak scale is represented by a solid black line. The label $\tilde u_L$ is actually labeling the nearly degenerate $\tilde u_L$ and $\tilde c_L$ masses. The labels $\tilde u_R$, $\tilde d_L$ and $\tilde d_R$ are similarly labeling the nearly degenerate first- and second- family masses.}
	\label{fig:157}
\end{figure}
}
\comment{\small Plots of physical sparticle spectrum from four example valid points from our main scan. Plot (a) shows an example of an admixture stop LSP, while (b) shows a right-handed sbottom LSP. Plots (c) and (d) show the largest right-side-up and upside-down hierarchy respectively. The $B-L$ scale is represented by a black dot-dash-dot line. The SUSY scale is represented by a black dashed line. The electroweak scale is represented by a solid black line. In all four plots the vertical axis runs from zero to 15 TeV.}

\begin{figure}[!htbp]
\centering
	\includegraphics[scale=0.6]{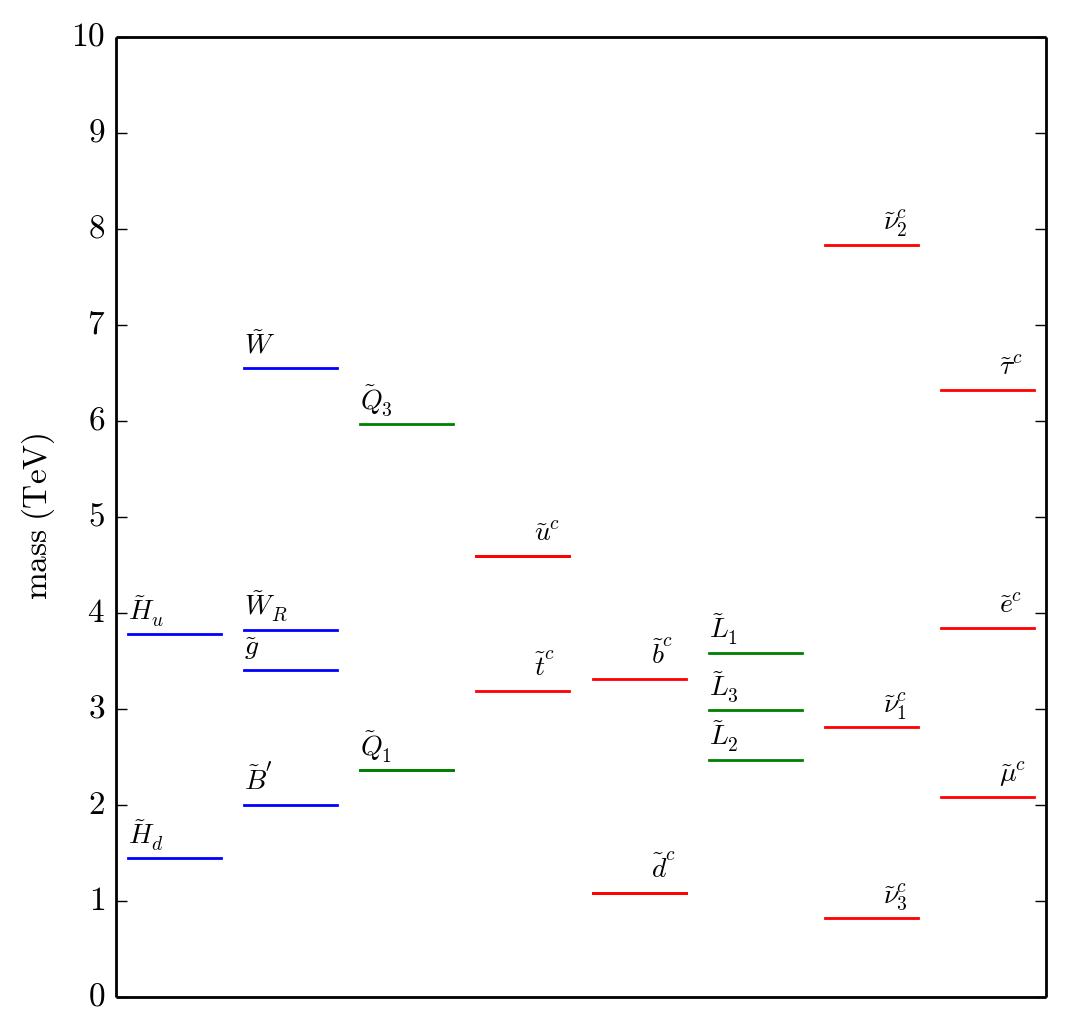}
	\includegraphics[scale=0.6]{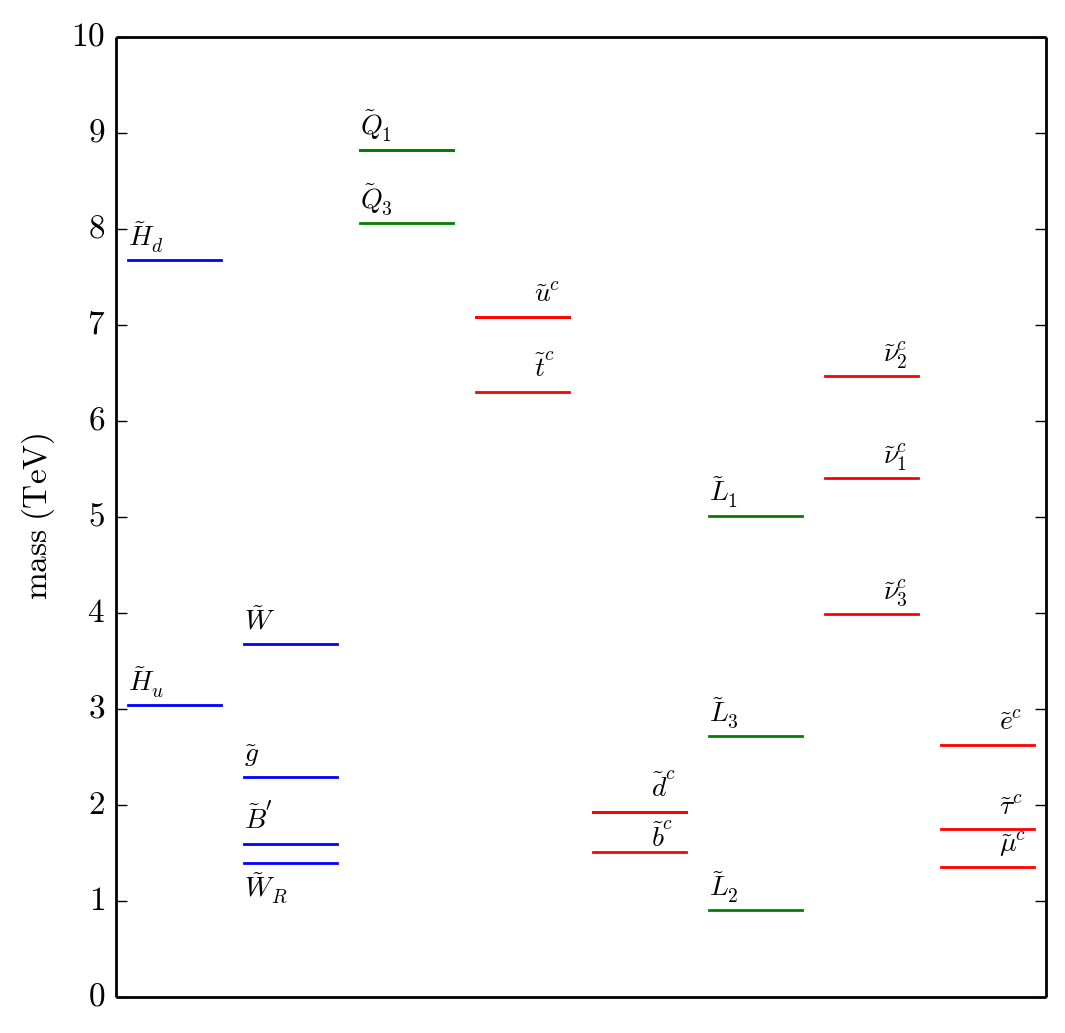}
	\caption{\small Example high-scale boundary conditions for the two valid points with the lightest and heaviest initial value of the third-family right-handed sneutrino soft mass. The label $\tilde Q_1$ is actually labeling the nearly degenerate $\tilde Q_1$ and $\tilde Q_2$ soft masses. The labels $\tilde u^c$ and $\tilde d^c$ are similarly labeling the nearly degenerate first and second family masses.}
	\label{fig:158}
\end{figure}

\begin{figure}[!htbp]
\centering
	\includegraphics[scale=0.6]{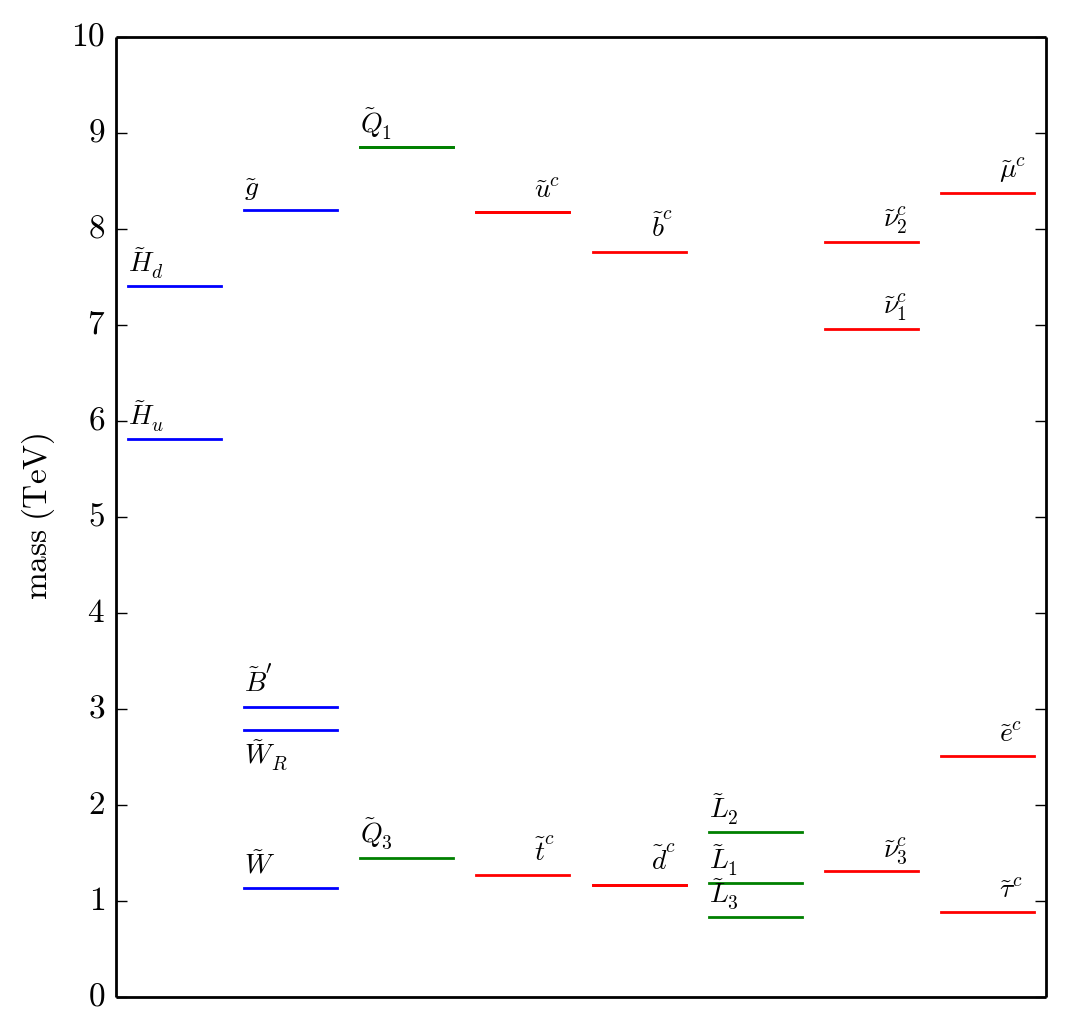}
	\includegraphics[scale=0.6]{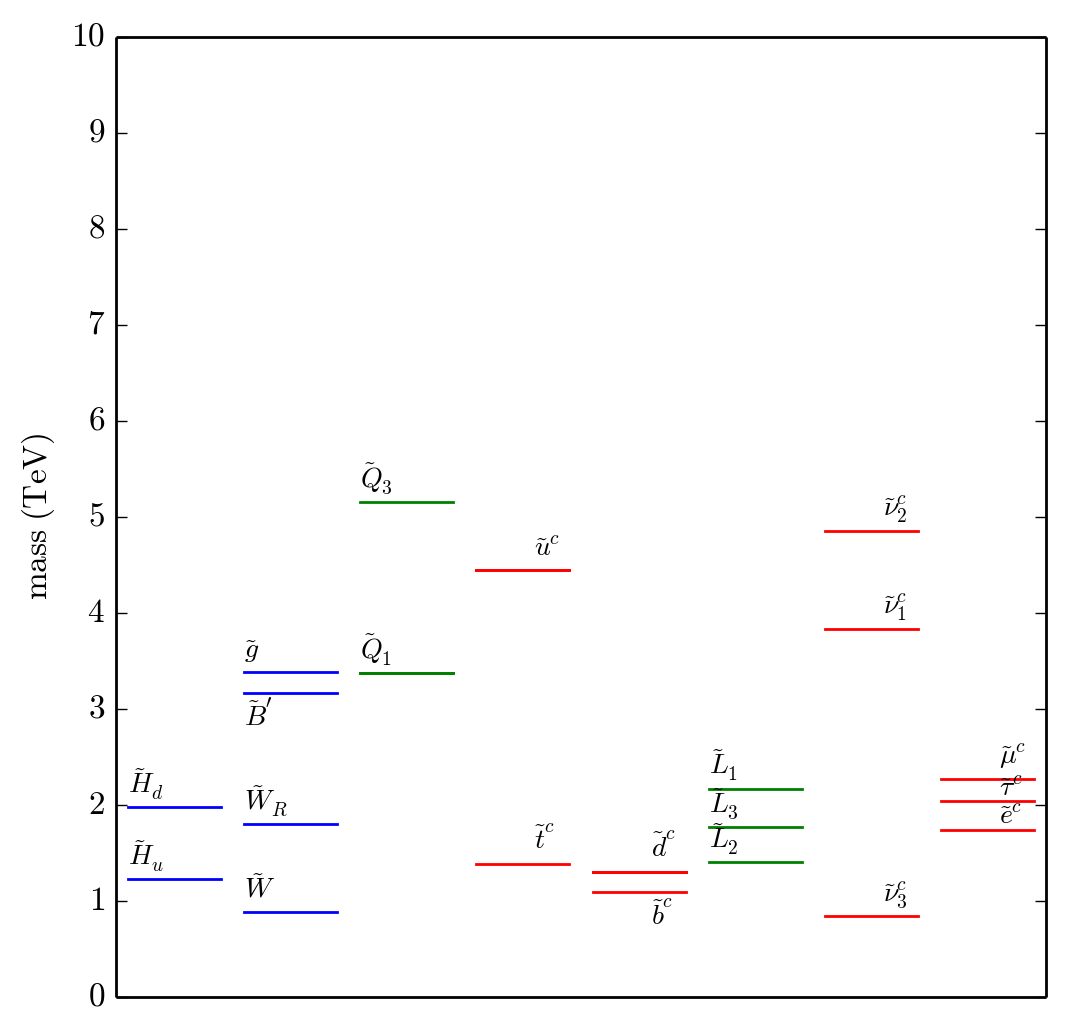}
	\caption{\small Example high-scale boundary conditions for the two valid points with the largest and smallest amount of splitting. The label $\tilde Q_1$ is actually labeling the nearly degenerate $\tilde Q_1$ and $\tilde Q_2$ soft masses. The labels $\tilde u^c$ and $\tilde d^c$ are similarly labeling the nearly degenerate first and second family masses.}
	\label{fig:160}
\end{figure}

Plots of the high-scale boundary values for four sample valid points from our main scan are presented in Figs.~\ref{fig:158} and \ref{fig:160}. While these look like Figs.~\ref{fig:154} and \ref{fig:156}, they do not correspond to physical masses but, rather, mass parameters at $M_\i$. These four valid points are automatically selected from the pool of valid points from the main scan based on simple criteria. The first two are those with the lightest and heaviest initial value of the third-family right-handed sneutrino mass. These show that it is not necessary to artificially choose a very light initial mass for the third-family right-handed sneutrino to effect the destabilizing of its potential and $B-L$ symmetry breaking. Note when reading these plots that the lightest right-handed sneutrino is always, without loss of generality, defined to be the third-family. The next two plots show the valid points with the largest and smallest amount of splitting in the initial values of the scalar soft mass parameters. The amount of splitting is defined as the standard deviation of the initial values of the 20 scalar soft mass parameters.

\comment{
\begin{figure}[!htbp]
\centering
	\includegraphics[scale=1]{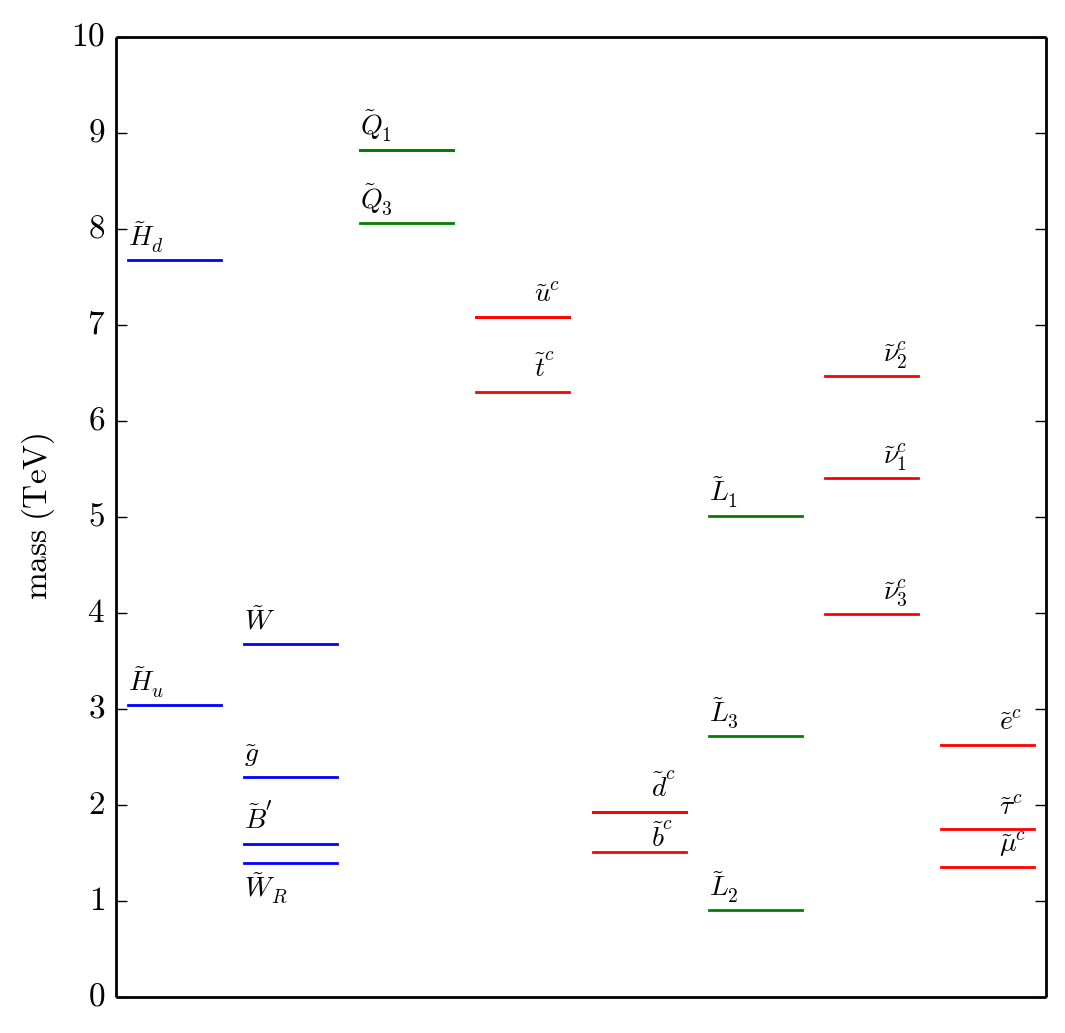}
	\caption{\small Example high-scale boundary conditions for the valid point with the heaviest initial value of the third-family right-handed sneutrino soft mass. The label $\tilde Q_1$ is actually labeling the nearly degenerate $\tilde Q_1$ and $\tilde Q_2$ soft masses. The labels $\tilde u^c$ and $\tilde d^c$ are similarly labeling the nearly degenerate first and second family masses.}
	\label{fig:159}
\end{figure}
}

\comment{
\begin{figure}[!htbp]
\centering
	\includegraphics[scale=1]{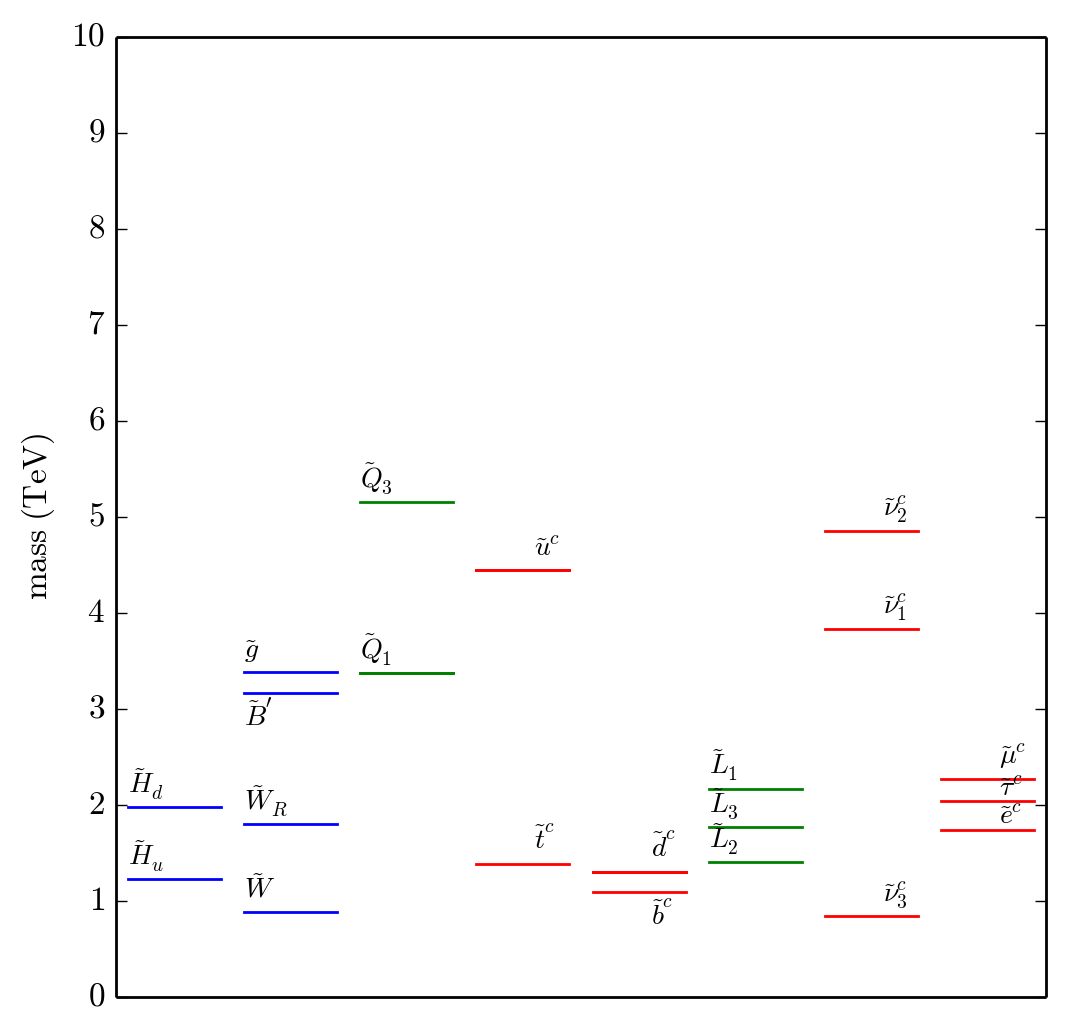}
	\caption{\small Example high-scale boundary conditions for the valid point with the smallest amount of splitting. The label $\tilde Q_1$ is actually labeling the nearly degenerate $\tilde Q_1$ and $\tilde Q_2$ soft masses. The labels $\tilde u^c$ and $\tilde d^c$ are similarly labeling the nearly degenerate first and second family masses.}
	\label{fig:161}
\end{figure}
}

\comment{\small Plots of the high-scale boundary values for four example valid points from our main scan. Plots (a) and (b) show the valid points with th e lightest and heaviest initial value of the third-family right-handed sneutrino, respectively. Plots (c) and (d) show the valid points with the largest and smallest amount of splitting in the scalar soft mass parameters. In all four plots the vertical axis runs from zero to 10 TeV.}

%
\section{Fine-Tuning}
\label{sec:FT}
%
Fine-tuning in supersymmetric models arises from Eq. (\ref{eq:EW.mu}),
\begin{eqnarray}
\frac12 M_Z^2&=&\frac{m_{H_u}^2\tan^2\beta-m_{H_d}^2}{1-\tan^2\beta}-\mu^2.
\label{eq:1229}
\end{eqnarray}
In both the MSSM and the minimal $B-L$ extension of the MSSM, the soft masses $m_{H_u}^2$ and $m_{H_d}^2$ receive contributions from other soft masses. Most important are the contributions from stop and gluino soft masses that appear in the RGEs for $m_{H_u}^2$--see Eq. (\ref{eq:$B-L$ MSSM.Hu}). They must be TeV-scale to satisfy sparticle mass lower bounds and the measured Higgs mass. These large TeV-scale contributions must be almost exactly cancelled to yield the relatively small value of $M_Z^2$ on the left side of Eq. (\ref{eq:1229}). This cancellation can come either from other soft masses or the $\mu^2$ term. The delicate cancellation between the parameters on the right side to yield the smaller term on the left side is ``fine-tuning''. The necessity of such fine-tuning in supersymmetric models has been referred to as the ``little hierarchy problem''. 

Here we explain the little hierarchy problem within the context of the $B-L$ MSSM, using a rough analytic argument along the lines of that presented in \cite{Luty:2005sn}. Although we discuss it using the language and notation of the minimal $B-L$ extension of the MSSM, the same argument holds in the MSSM. The largest contributions to $m_{H_u}^2$ come through its RGE in the $B-L$ MSSM scaling regime, Eq. (\ref{eq:$B-L$ MSSM.Hu}). Focusing on just the stop and gluino soft mass contributions, we can write a solution to this equation to first-order in $\ln (M_\i/M_\susy)$. Such a solution is quantitatively innacurate because it neglects higher powers of the large logarithm $\ln (M_\i/M_\susy)$. Be that as it may, it can still provide insight into how various scales enter the problem. The solution is
\begin{eqnarray}
m_{H_u}^2=-\frac{6}{16\pi^2}Y_t^2(m_{Q_3}^2+m_{t^c}^2)\ln \left(\frac{M_\i}{M_\susy} \right)+\cdots,
\end{eqnarray}
where the ellipsis represents neglected higher order terms and terms due to other contributions in Eq. (\ref{eq:$B-L$ MSSM.Hu}). Additionally, there are corrections due to the boundary condition Eq. (\ref{eq:BC.BL}). The $m_{Q_3}^2$ and $m_{t^c}^2$ themselves receive large contributions through their RGEs, Eqs. (\ref{eq:$B-L$ MSSM.Q3}) and (\ref{eq:$B-L$ MSSM.tc}). Focusing on the contributions from the gluino mass yields 
\begin{eqnarray}
m_{H_u}^2=-\frac{6}{16\pi^2}Y_t^2\left(m_{Q_3}^2+m_{t^c}^2 + \frac{4}{3\pi^2}g_3^2M_3^2\ln \left( \frac{M_\i}{M_\susy} \right) \right)\ln \left( \frac{M_\i}{M_\susy} \right)+\cdots.
\label{eq:521}
\end{eqnarray}
As discussed in Section~\ref{sec:856}, the stops and gluino have relatively high mass bounds from LHC searches. Additionally, as discussed in Appendix~\ref{sec:915}, satisfying the observed value of the Higgs mass tends to require heavy stops. This means that the stop and gluino soft mass contributions in Eq. (\ref{eq:521}) must be relatively large and give large contributions to the right-hand side of Eq. (\ref{eq:1229}). For example, if $Y_t=0.9$, $g_3^2=1$, $M_3=m_{\tilde q_3}=m_{\tilde t^c}=1\mbox{ TeV}$, $M_\i=10^{15}\mbox{ GeV}$, and $M_\susy=1\mbox{ TeV}$, these contributions are approximately equal to $-(2\mbox{ TeV})^2$. This must be almost exactly cancelled to yield the relatively small value of $M_Z^2=(91.2\mbox{ GeV})^2$ on the left-hand side of Eq. (\ref{eq:1229}). The cancellation usually comes from the $\mu^2$, but can also come from the terms in the ellipsis or $m_{H_d}^2$.

As stated above, in the minimal $B-L$ extension of the MSSM there are additional contributions coming from the boundary condition on the Higgs soft masses at the $B-L$ scale, Eq. (\ref{eq:BC.BL}). Since $\tan\beta>1$, the most important are the contributions to the $H_u$ soft mass. These are proportional to $M_{Z_R}$. Since there is a lower bound of 2.5 TeV on $M_{Z_R}$, it is reasonable to suspect that these contributions to the $H_u$ soft mass necessitate more delicate cancellation--thus worsening the little hierarchy problem. Rewritten in terms of $M_{Z_R}$, the associated boundary condition is
\begin{eqnarray}
m_{H_u}^2(M_{B-L}^-) = m_{H_u}^2(M_{B-L}^+)-\frac{1}{2}\frac{g_R^2}{g_R^2+g_{BL}^2}M_{Z_R}^2 \ .
\label{blackboard1}
\end{eqnarray}
The gauge couplings here, and in the remainder of this Section, are evaluated at $M_{B-L}$ unless otherwise specified. Before concluding that this exacerbates the  fine-tuning problem, we should replace the physical mass, $M_{Z_R}$, with more fundamental parameters of the theory, such as the soft masses evaluated at the intermediate scale. All of the scalar soft masses share in the generation of $M_{Z_R}$ through the $S$-terms. Substituting Eq. (\ref{eq:315}) into Eq. (\ref{eq:MZR.mnuc}) allows us to write the $S$-term contribution to $M_{Z_R}$. It is given by
\begin{eqnarray}
M_{Z_R}^2&=&\frac{1}{7}\frac{g_R^2-g_R^2(M_\i)}{g_R^2(M_\i)}S_R(M_\i)\nonumber\\
&&-\frac{1}{8}\frac{g_{BL}^2-g_{BL}^2(M_\i)}{g_{BL}^2(M_\i)}S_{B-L}(M_\i)+\cdots.
\end{eqnarray}
Substituting this into equation Eq. (\ref{blackboard1}) yields $S$-term contributions to the $H_{u}$ soft mass. In addition, the $S$-terms also influence the running of the $H_u$ soft mass through the RGEs. Including both of these contributions, the value of $m_{H_u}^2$ at the SUSY scale is
\begin{eqnarray}
m_{H_u}^2(M_\susy)&=&\frac{g_R^2}{g_R^2+g_{BL}^2}\left(\frac{1}{14}\frac{g_R^2-g_R^2(M_\i)}{g_R^2(M_\i)}S_R(M_\i)-\frac{1}{16}\frac{g_{BL}^2-g_{BL}^2(M_\i)}{g_{BL}^2(M_\i)}S_{B-L}(M_\i)\right)\nonumber\\
&&-\frac{1}{14}\frac{g_R^2-g_R^2(M_\i)}{g_R^2(M_\i)}S_R(M_\i)+\cdots.
\end{eqnarray}
Consider a sample valid point with $S_R(M_\i)<0$ and $S_{B-L}(M_\i)=-2S_R(M_\i)$. This case fits within the valid black points in Fig.~\ref{fig:1148}. This arises physically if all scalar soft masses are universal with the exception that the the first- and second-family right-handed sneutrino soft masses--which are heavier. In this case, the $S$-term contributions to $m_{H_u}^{2}$ can be written as
\begin{eqnarray}
m_{H_u}^2(M_\susy)&=&\left[\frac{g_R^2}{g_R^2+\frac{3}{2}g_{BL}^2}\left(\frac{1}{14}\frac{g_R^2-g_R^2(M_\i)}{g_R^2(M_\i)}+\frac{1}{8}\frac{g_{BL}^2-g_{BL}^2(M_\i)}{g_{BL}^2(M_\i)}\right)\right.\nonumber\\
&&-\left.\frac{1}{14}\frac{g_R^2-g_R^2(M_\i)}{g_R^2(M_\i)}\right]S_R(M_\i)+\cdots.
\end{eqnarray}
Let us choose, for example, $M_\susy=1$ TeV and $M_{B-L}=2.5$ TeV. Then the dimensionless coefficient of $S_{R}({M\i})$ turns out to be $-0.022$.  Since this value is considerably smaller than unity, it follows that the $H_u$ soft mass is not--in fact--very sensitive to the fundamental parameters that set the $Z_R$ mass. This remains true for all values of $M_\susy$ and  $M_{B-L}$ associated with valid points. Therefore, there is not a significant amount of new fine-tuning introduced in this way.

Fine-tuning in supersymmetric models has historically~\cite{Antoniadis:2014eta,Ciafaloni:1996zh,de Carlos:1993yy,Casas:2003jx} been quantified using the Barbieri-Giudice (BG) sensitivity, introduced in \cite{Ellis:1986yg} and \cite{Barbieri:1987fn}. This quantifies the sensitivity of  some observable quantity to changes in any of the fundamental parameters of a theory. The delicate cancellation between TeV-scale supersymmetry parameters in Eq. (\ref{eq:1229}) results in the electroweak scale, $M_Z$, having a large BG sensitivity. The BG sensitivity of the electroweak scale is defined as
\begin{eqnarray}
F_{a_i}=\left |\frac{a_i}{M_Z^2}\frac{\partial M_Z^2}{\partial a_i}\right|,
\end{eqnarray}
where $a_i$ is any of the fundamental parameters of the theory. This says that a fractional change in $a_i$ would produce a fractional change in $M_Z^2$ that is $F_{a_i}$ times larger. The overall degree of fine-tuning is usually taken to be the largest of all the $F_{a_i}$'s ; that is,
\begin{eqnarray}
F=\mbox{max}(F_{a_i}) \ .
\end{eqnarray}
The BG sensitivity $F$ will be  used to quantify the fine-tuning required in the $B-L$ MSSM.

It is worth mentioning that some authors have pointed out drawbacks to the BG sensitivy and suggested other quantifications of fine-tuning. For example, as discussed in \cite{de Carlos:1993yy}, the BG sensitivity and the overall degree of fine-tuning depend on how the fundamental parameters of the theory $a_i$ are chosen. Furthermore, one could reasonably use the BG sensitivity of $M_Z$, rather than $M_Z^2$, as the indicator of fine-tuning. This would result in fine-tuning that is smaller by a factor of two. Such ambiguities in the way fine-tuning is calculated from the BG sensitivity suggest that it is not a precise way to quantify fine-tuning. A separate paper, \cite{Anderson:1994dz}, points out that the relationship between the proton mass and the strong coupling constant at a high scale exhibits high BG sensitivity, but is not actually finely tuned. They propose a more precise quantification of fine-tuning and show that the BG sensitivity actually overestimates the fine-tuning in some sample points in the MSSM.

\begin{figure}[!htbp]
	\centering
	\includegraphics[scale=1]{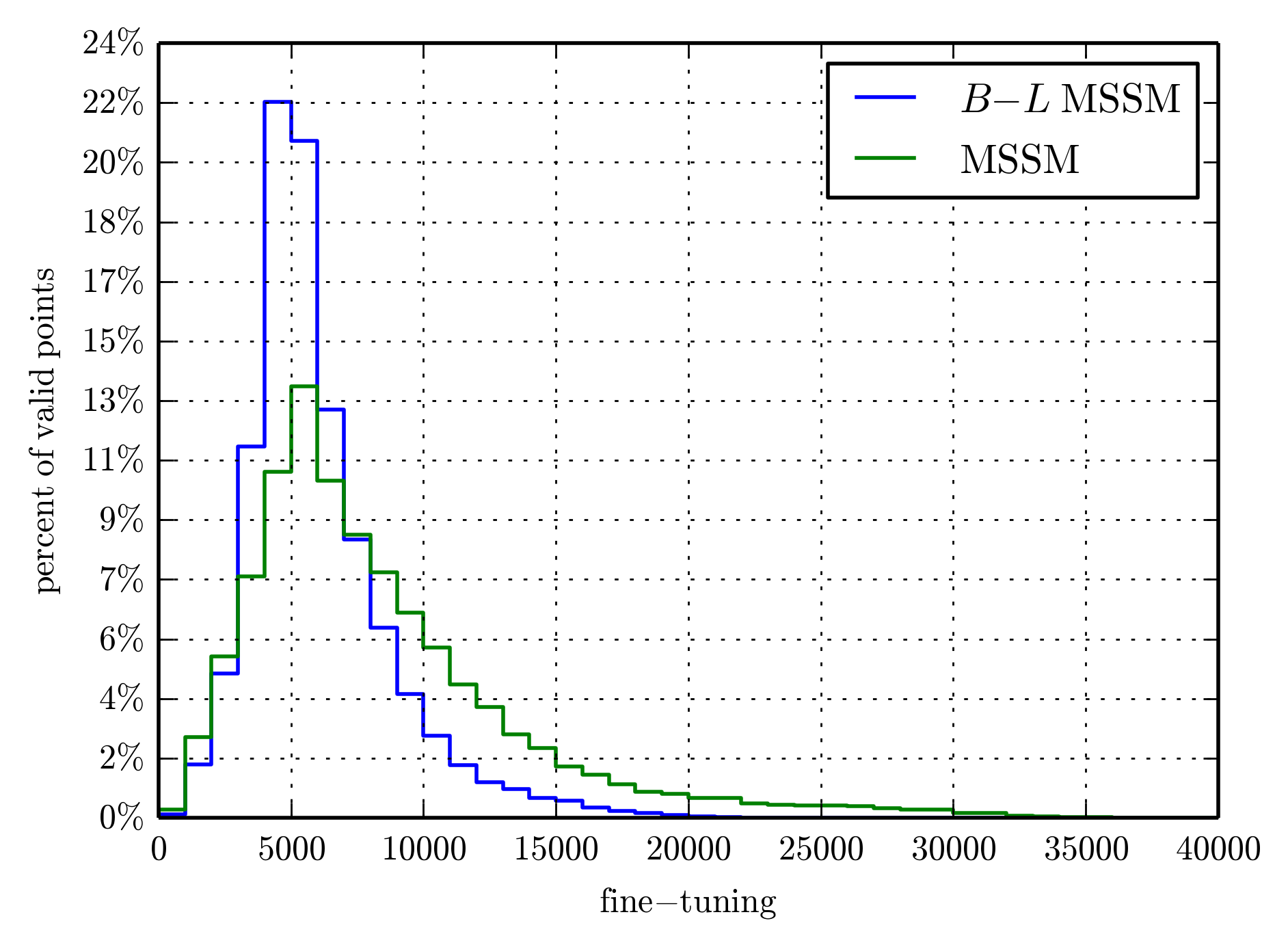}
	\caption{\small The blue line in the  histogram shows the amount of fine-tuning required for valid points in the main scan of the $B-L$ MSSM. Similarly, the green line specifies the amount of fine-tuning necessary for the valid points of the $R$-parity conserving MSSM--computed using the same statistical procedure as for the $B-L$ MSSM with $M=2700$ GeV and $f=3.3$. The $B-L$ MSSM shows slightly less fine-tuning, on average, than the  MSSM.}
	\label{fig:514}
\end{figure}

Despite the possible shortcomings, the BG sensitivity remains the most widely used tool for making rough quantitative analyses of fine-tuning in supersymmetric models. We, therefore, proceed using the BG sensitivity to quantify fine-tuning in the $B-L$ MSSM. For each of the valid points, we compute $F$. We allow $a_i$ to span all of the soft mass parameters of the theory, as well as $\mu$. In the case of scalar soft masses, we take $a_i$ to be the mass squared, while in the case of gaugino soft masses and $\mu$ we take $a_i$ to be the mass to the first power. This choice corresponds to how these parameters appear in the Lagrangian. We then create a histogram of $F$ for all of the valid points in our main scan. This data is shown as the blue line in Fig.~\ref{fig:514}. Note that the fine-tuning required by the highest percentage of valid points is $F \sim 5000$. Be that as it may, a reasonable number of valid points need significantly less fine-tuning--with about $ 2\%$ requiring $F\lesssim2000$. It is interesting to compare the amount of fine-tuning in the minimal $B-L$ extension of the MSSM model to the amount of fine-tuning required in an identical statistical scan of the $R$-parity invariant MSSM using $M=2700$ GeV and $f=3.3$. Due to the aforementioned ambiguities in how fine-tuning is quantified, it is critical that the fine-tuning be calculated the same way when two different models are being compared. Therefore, we use our own code, slightly modified, to produce a similar plot for the $R$-parity conserving MSSM. 
The results are shown as the green line in Fig.~\ref{fig:514}. Comparison of the blue and green lines in the figure show that the $B-L$ MSSM valid points tend to be slightly less finely tuned than valid points in the $R$-parity conserving MSSM. The difference is large enough to be apparent in the figure. However, due to the unresolved questions about how to properly quantify fine-tuning, we do not regard this difference between the $B-L$ MSSM and the MSSM to be significant.

\begin{figure}[!htbp]
	\centering
	\includegraphics[scale=1]{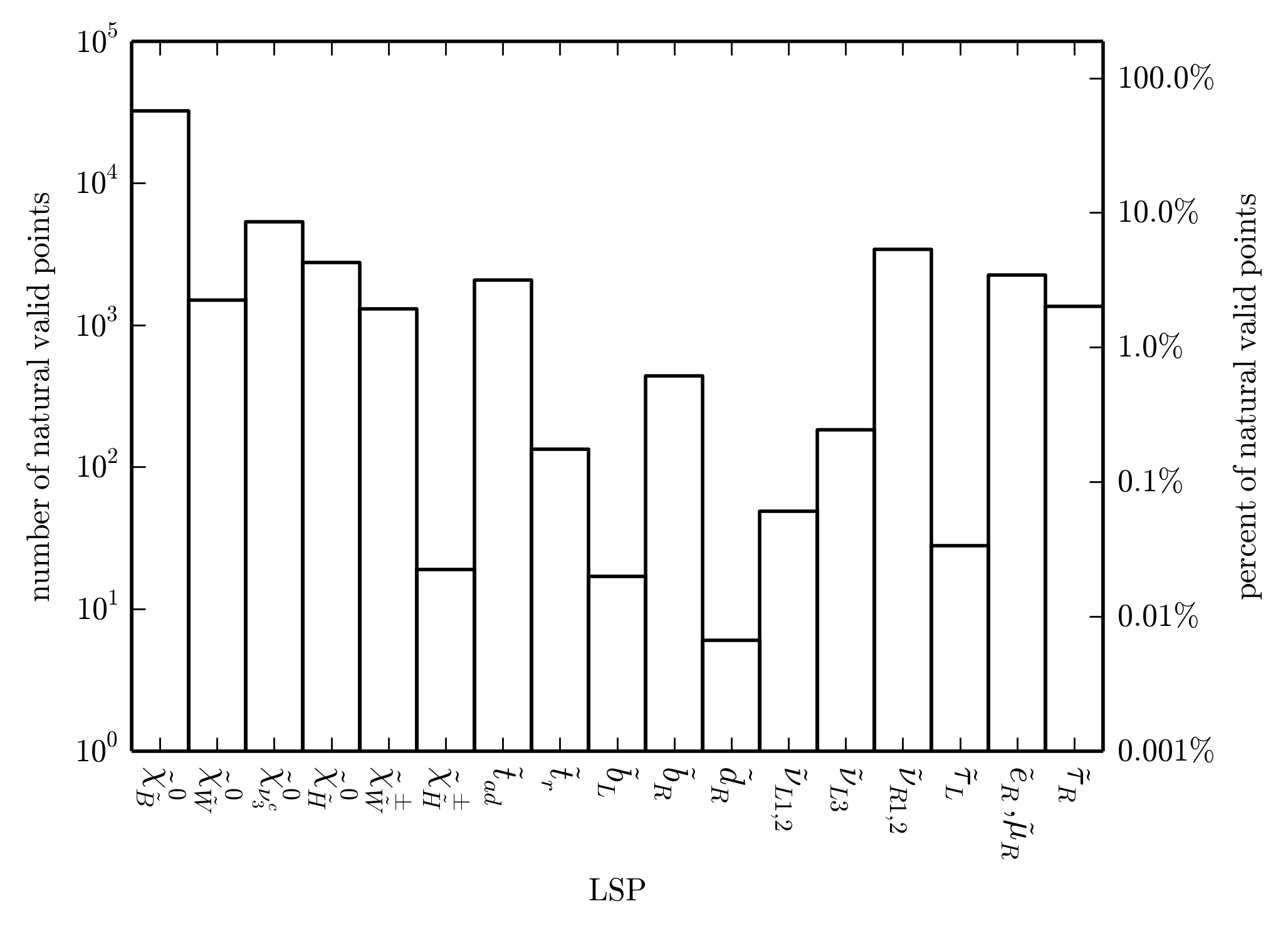}
	\caption{\small A histogram of the LSP's for the ``natural'' valid points with $F<1000.$ Sparticles which did not appear as LSP's are omitted. The y-axis has a log scale. The notation for the various states, as well as their most likely decay products, are given in Table~\ref{tbl:LSP}. Note that the natural valid points favor stop and sbottom LSP's more than the valid points presented in Fig.~\ref{fig:1039}. 
Note that we have combined left-handed first and second generation sneutrinos into one bin and each generation makes up about 50\% of the LSP's. The same is true for the first and second generation right-handed sleptons and sneutrinos. }
	\label{fig:417}
\end{figure}

With the fine-tuning of each randomly generated point in the $B-L$ MSSM now quantified, we are equipped to produce results for just the most natural points--that is, those requiring minimal fine-tuning. Figure~\ref{fig:417} shows a histogram of the LSP's for those points with $F<1000$, corresponding to the least fine-tuned $\sim0.1\%$ of points, from a larger scan of four hundred million points. We refer to these points as ``natural'' valid points. \comment{Note that the number of natural valid points in this scan is approximately equal to the number of valid points in our main scan.} There are three notable differences between Fig.~\ref{fig:417} and Fig.~\ref{fig:1039}. First, stop LSP's are more common. This includes both admixture and mostly right-handed stop LSP's. Stop LSP's are more common because heavy stops tend to cause fine-tuning, so low fine-tuning favors lighter stops and stop LSP's. Second, sbottom LSP's are more common. This is due to the fact that first, both the stop and sbottom masses depend on the soft mass $m_{Q_3}^2$ and second, because the right-handed stop and sbottom soft masses have similar terms in their RGE's. These two facts imply that favoring light stops tends to favor light sbottoms as well. Third, Fig.~\ref{fig:417} does not have the gluino LSP's shown in Fig.~\ref{fig:1039}, and it does have some $\tilde d_R$ LSP's not found in Fig.~\ref{fig:1039}. However, the disappearance and appearance of these states in the $F<1000$ histogram is not statistically significant and, hence, these states can be ignored.
The prevalence of stop and sbottom LSP's is the only significant difference between the natural valid points and the valid points. Note that the physical implications of having a stop or sbottom LSP in the $B-L$ MSSM have been studied in \cite{Marshall:2014kea,Marshall:2014cwa}.

%
\section{Conclusion}
\label{sec:conclusion}
%
In this paper, we presented a novel approach for relating UV physics to TeV scale physics and applied this analysis to the minimal SUSY $B-L$ model. This approach hypothesizes that all SUSY breaking parameters are about an order of magnitude away from a characteristic SUSY breaking mass scale. Practically, this translates into conducting an analysis where all relevant soft SUSY mass parameters are independently scanned over the same range at the UV scale,  and then RG evolved to the TeV scale. This program lends itself especially well to the string realization of the minimal $B-L$ MSSM model. However, our results are relevant for any high scale soft SUSY breaking minimal SUSY $B-L$ model with gauge coupling unification.

A central result of this work is the general region of initial parameter space that leads to radiative $B-L$ symmetry breaking.
While this depends on multiple parameters of the theory, it can be expressed in terms of the two $S$-parameters and is presented in this context in Fig.~\ref{fig:1204}. A subsequent figure, Fig~\ref{fig:1148}, shows how additional constraints, such as electroweak symmetry breaking and lower bounds on new sparticle masses, depend on the $S$-parameters. These two plots indicate that a significant amount of the initial parameter space leads to experimentally viable results. They are  followed by various spectrum graphs which show that acceptable spectra are relatively general and do not depend on a specific hierarchy of initial masses.

The phenomenology of a given point at the LHC strongly depends on the identity of the LSP. Therefore, another central result of this paper is the calculation of the probability that a given SUSY particle can be the LSP. This was addressed in Fig.~\ref{fig:1039}. As might be expected, a mostly bino neutralino is the most likely candidate. However, since binos cannot be directly produced at the LHC, signals associated with bino LSPs also depend on the rest of the SUSY spectrum. Therefore, an interesting future direction might be to investigate the phenomenology of mostly wino or Higgsino neutralinos. Mostly wino or Higgsino neutralinos can be directly produced at the LHC, independently of the rest of the SUSY spectrum, and have relatively large cross sections for colorless particles. The signals associated with different LSPs are summarized in Table~\ref{tbl:LSP}.

Finally, the fine-tuning associated with this statistical scan was investigated. While it is not drastically different than the fine-tuning in the MSSM with a similar UV completion, one might think that the new mass scale associated with $B-L$ breaking could introduce new contributions to fine-tuning. We showed that it does not. In fact, a given point in this model is typically less fine-tuned than a similar point in the MSSM. In addition, we explored possible LSPs for points with fine-tuning better than one part per thousand--in a way analogous to Fig.~\ref{fig:1039}. We found that stops and sbottoms become much more likely LSP candidates, as one might expect--see Fig.~\ref{fig:417}. The signals of stop and sbottom LSPs were discussed previously in~\cite{Marshall:2014kea, Marshall:2014cwa}.

\section{Acknowledgments}
S. Spinner would like to thanks to P. Fileviez Perez for long term collaboration on related topics. B.A. Ovrut, A. Purves and S. Spinner are supported in part by the DOE under contract No. DE-SC0007901 and by the NSF under grant No. 1001296.

\appendix

%
\section{Renormalization Group Equations}
\label{sec:719}
%

This Appendix lists the RGEs used in this study. Most RGEs are derived with the help of reference~\cite{Martin:1993zk}, unless otherwise stated.

The RGEs for gauge couplings were presented in Section~\ref{sec:unif}, but are repeated here for completeness. The RGE for a general gauge coupling is
\begin{equation}
	\frac{d}{dt}\alpha_a^{-1} = -\frac{b_a}{2\pi}.
\end{equation}
where $t$ is the logarithm of the renormalization scale and the index $a$ runs over the different gauge factors. The slope factors are different in each of the different scaling regimes:
\begin{itemize}
	\item Intermediate regime: $b_3 = 10,\ b_2 = 14,\ b_R=14,\ b_{B-L}=19$.
	\item $B-L$ MSSM: $b_3 = -3,\ b_2 = 1,\ b_{3R}=7,\ b_{B-L}=6$.
	\item MSSM: $b_3 = -3,\ b_2 = 1,\ b_1=\frac{33}{5}$.
	\item Non-SUSY $B-L$:  $b_3 = -7,\ b_2 = -\frac{19}{6},\ b_{3R}=\frac{53}{12},\ b_{B-L}=\frac{33}{8}$.
	\item SM: $b_3 = -7,\ b_2 = -\frac{19}{6},\ b_1=\frac{41}{10}$
\end{itemize}

The gaugino soft mass RGE is
\begin{equation}
	\frac{d}{dt} M_a = \frac{b_a \alpha_a M_a}{2 \pi},
\end{equation}
where the $b_a$ are the same slope factors given in Eqs.~(\ref{eq:646} -~\ref{eq:644}). It is helpful to observe that the gaugino mass renormalization group equation admits a rather compact analytic solution:
\begin{equation}
	M_a(t) = \frac{M_a(M_U)}{\alpha_U} \alpha_a(t) ,
\end{equation}
for all gaugino masses associated with $SO(10)$ and
\begin{equation}
	M_1(t) = \frac{M_1(M_{B-L})}{\alpha_1(M_{B-L})} \alpha_1(t) ,
\end{equation}
for the bino.

There are three significant Yukawa couplings for RGE analysis: $y_t$, $y_b$ and $y_\tau$.  In the SM scaling regime their RGEs can be found in~\cite{Machacek:1983fi}, for example, and are given by
\begin{eqnarray}
\label{eq:SM:yt}
	\frac{d}{dt}y_t&=&\frac{1}{16\pi^2} y_t\left(\frac32(y_t^2-y_b^2) + 3(y_t^2+y_b^2) + y_\tau^2-8g_3^2-\frac94g_2^2-\frac{17}{20}g_{1}^2\right)
	\\
	\label{eq:SM:yb}
	\frac{d}{dt}y_b&=&\frac{1}{16\pi^2} y_b\left(\frac32(y_b^2-y_t^2) + 3(y_t^2+y_b^2) + y_\tau^2-8g_3^2-\frac94g_2^2-\frac{1}{4}g_{1}^2\right)
	\\
	\label{eq:SM:ytau}
	\frac{d}{dt}y_\tau&=&\frac{1}{16\pi^2} y_\tau\left(\frac32y_\tau^2 + 3(y_t^2+y_b^2) + y_\tau^2-\frac94g_2^2-\frac{9}{4}g_{1}^2\right).
\end{eqnarray}
In the $U(1)$ extended SM regime of the upside-down case, the Yukawa coupling RGEs are
\begin{eqnarray}
	\label{eq:BLSM:yt}
	\frac{d}{dt}y_t&=&\frac{1}{16\pi^2} y_t\left(\frac32(y_t^2-y_b^2) + 3(y_t^2+y_b^2) + y_\tau^2\right.\nonumber\\
	&&\left.-8g_3^2-\frac94g_2^2-\frac{3}{4}g_R^2-\frac{1}{4}g_{BL}^2\right)\\
	\label{eq:BLSM:yb}
	\frac{d}{dt}y_b&=&\frac{1}{16\pi^2} y_b\left(\frac32(y_b^2-y_t^2) + 3(y_t^2+y_b^2) + y_\tau^2\right.\nonumber\\
	&&\left.-8g_3^2-\frac94g_2^2-\frac{3}{4}g_R^2-\frac{1}{4}g_{BL}^2\right)\\
	\label{eq:BLSM:ytau}
	\frac{d}{dt}y_\tau&=&\frac{1}{16\pi^2} y_\tau\left(\frac32y_\tau^2 + 3(y_t^2+y_b^2) + y_\tau^2\right.\nonumber\\
	&&\left.-\frac94g_2^2-\frac{3}{4}g_R^2-\frac{9}{4}g_{BL}^2\right).
\end{eqnarray}
The boundary condition at the $B-L$ scale is trivial. At the SUSY scale, however, the boundary condition is nontrivial:
\begin{eqnarray}
	y_{t}(M_\susy)&=&Y_{t}(M_\susy)\sin\beta\nonumber\\
	y_{b,\tau}(M_\susy)&=& Y_{b,\tau}(M_\susy)\cos\beta.
\end{eqnarray}
The Yukawa couplings above the SUSY scale will be denoted by $Y$ instead of $y$. This condition applies both in the upside-down case and in the right-side-up case. In the MSSM scaling regime of the right-side-up case the RGEs are 
\begin{eqnarray}
	\frac{d}{dt}Y_t&=&\frac{1}{16\pi^2} Y_t\left(6Y_t^2+Y_b^2-\frac{16}{3}g_3^2-3g_2^2-\frac{16}{15}g_{1}^2\right)
	\\
	\frac{d}{dt}Y_b&=&\frac{1}{16\pi^2} Y_b\left(6Y_b^2+Y_\tau^2+Y_t^2-\frac{16}{3}g_3^2-3g_2^2-\frac{4}{15}g_{1}^2\right)
	\\
	\frac{d}{dt}Y_\tau&=&\frac{1}{16\pi^2} Y_\tau\left(3Y_b^2+4Y_\tau^2-3g_2^2-\frac{12}{5}g_{1}^2\right)\nonumber.
\end{eqnarray}
In the $B-L$ MSSM scaling regime the RGEs are
\begin{eqnarray}
	\label{eq:$B-L$ MSSM.yt}
	\frac{d}{dt}Y_t&=&\frac{1}{16\pi^2} Y_t\left(6Y_t^2+Y_b^2-\frac{16}{3}g_3^2-3g_2^2-\frac{1}{6}g_{BL}^2-g_{I_3^R}^2\right)\\
	\label{eq:$B-L$ MSSM.yb}
	\frac{d}{dt}Y_b&=&\frac{1}{16\pi^2} Y_b\left(6Y_b^2+Y_\tau^2+Y_t^2-\frac{16}{3}g_3^2-3g_2^2-\frac{1}{6}g_{BL}^2-g_{I_3^R}^2\right)\\
	\label{eq:$B-L$ MSSM.ytau}
	\frac{d}{dt}Y_\tau&=&\frac{1}{16\pi^2} Y_\tau\left(3Y_b^2+4Y_\tau^2-3g_2^2-\frac{3}{2}g_{BL}^2-g_{I_3^R}^2\right).
\end{eqnarray}
The fact that these RGEs are non-linear means that the analytic solutions are much more cumbersome if they can be found at all. We use numerical integration techniques instead, yielding numerical values for the Yukawa couplings at any scale up to the intermediate scale, $M_\i$. These solutions will be subsequently used in the running of the soft tri-scalar couplings and some of the scalar soft masses because the RGEs of those parameters depend on the Yukawa couplings. The Yukawa couplings do not need to be evolved above the intermediate scale since the couplings that depend on them will not be evolved above the intermediate scale.

Tri-linear couplings are generated at the intermediate scale and evolved to the SUSY scale. Their RGEs in the $B-L$ MSSM scaling regime are
\begin{eqnarray}
	\label{eq:$B-L$ MSSM.at}
	\frac{d}{dt}a_t&=&\frac{1}{16\pi^2}a_t\left(8^2+Y_b^2-\frac{16}{3}g_3^2-3g_2^2-\frac{1}{6}g_{BL}^2-g_{I_3^R}^2\right)\nonumber\\
	&&+\frac{1}{16\pi^2}Y_t\left(10a_uY_t+2Y_b a_b+\frac{32}{3}g_3^2M_3\right.\nonumber\\
	&&\left.+6g_2^2 M_2+\frac{1}{3}g_{BL}^2M_{B-L}+2g_{I_3^R}^2M_{I_3^R}\right)\\
	\label{eq:$B-L$ MSSM.ab}
	\frac{d}{dt}a_b&=&\frac{1}{16\pi^2}a_b\left(8Y_b^2+Y_\tau^2+Y_t^2 - \frac{16}{3}g_3^2-3g_2^2-\frac{1}{6}g_{BL}^2-g_{I_3^R}^2\right)	\nonumber\\
	&&+\frac{1}{16\pi^2}Y_b\left(10a_bY_b+2a_\tau Y_\tau+2Y_ta_u+\frac{32}{3}g_3^2M_3\right.\nonumber\\
	&&\left.+6g_2^2 M_2+\frac{1}{3}g_{BL}^2M_{B-L}+2g_{I_3^R}^2M_{I_3^R}\right)\\
	\label{eq:$B-L$ MSSM.atau}
	\frac{d}{dt}a_\tau&=&\frac{1}{16\pi^2}a_\tau\left(3Y_b^2+6Y_\tau^2-3g_2^2-\frac{3}{2}g_{BL}^2-g_{I_3^R}^2\right)\nonumber\\
	&&+\frac{1}{16\pi^2}Y_\tau\left(6a_bY_b+6a_\tau Y_\tau+6g_2^2M_2+3g_{BL}^2M_{B-L}+2g_{I_3^R}^2M_{I_3^R}\right)
\end{eqnarray}
In the right-side-up case, the $B-L$ scale is above the SUSY scale so these parameters will also be run through the MSSM scaling regime from the $B-L$ scale to the SUSY scale. The RGEs in the MSSM scaling regime are
\begin{eqnarray}
	\label{eq:MSSM.at}
	\frac{d}{dt}a_t&=&\frac{1}{16\pi^2}a_t\left(8^2+Y_b^2-\frac{16}{3}g_3^2-3g_2^2-\frac{13}{15}g_{1}^2\right)\nonumber\\
	&&+\frac{1}{16\pi^2}Y_t\left(10a_uY_t+2Y_b a_b+\frac{32}{3}g_3^2M_3+6g_2^2 M_2+\frac{26}{15}g_{1}^2M_{1}\right)\\
	\label{eq:MSSM.ab}
	\frac{d}{dt}a_b&=&\frac{1}{16\pi^2}a_b\left(8Y_b^2+Y_\tau^2+Y_t^2 - \frac{16}{3}g_3^2-3g_2^2-\frac{7}{15}g_{1}^2\right)\nonumber\\
	&&+\frac{1}{16\pi^2}Y_b\left(10a_bY_b+2a_\tau Y_\tau+2Y_ta_u+\frac{32}{3}g_3^2M_3\right.\nonumber\\
	&&\left.+6g_2^2 M_2+\frac{14}{15}g_{1}^2M_{1}\right)\\
	\label{eq:MSSM.atau}
	\frac{d}{dt}a_\tau&=&\frac{1}{16\pi^2}a_\tau\left(3Y_b^2+6Y_\tau^2-3g_2^2-\frac{9}{5}g_{1}^2\right)\nonumber\\
	&&+\frac{1}{16\pi^2}Y_\tau\left(6a_bY_b+6a_\tau Y_\tau+6g_2^2M_2+\frac{18}{5}g_{1}^2M_{1}\right).
\end{eqnarray}
These equations are also do not yield tractable analytic solutions, of course.

Scalar soft mass squared parameters are also inputted at the intermediate scale and evolved down to the SUSY scale.  In the case of the right-side-up hierarchy, this will involve running through the $B-L$ scale and the brief MSSM scaling regime. The boundary condition at the $B-L$ scale is nontrivial because $D$-term interactions between the third-family right-handed sneutrino and the other scalars give rise to a new contribution to the soft masses when the third-family right-handed sneutrino acquires a VEV, Eq.~(\ref{eq:BC.BL}). As discussed in Section~\ref{sec:857}, we take the soft masses to be flavor diagonal in order to satisfy flavor constraints.

Before writing the scalar soft mass RGEs, it is useful to define the $S$-terms,
\begin{eqnarray}
	\label{eq:S.BL}
	S_{B-L}&=&\Tr(2m_{\tilde Q}^2-m_{\tilde u^c}^2-m_{\tilde d^c}^2-2m_{\tilde L}^2+m_{\tilde \nu^c}^2+m_{\tilde e^c}^2)\\
	\label{eq:S.R}
	S_{R}&=&m_{H_u}^2-m_{H_d}^2+\Tr\left(-\frac{3}{2}m_{\tilde u^c}^2+\frac{3}{2}m_{\tilde d^c}^2-\frac{1}{2} m_{\tilde \nu^c}^2+\frac{1}{2} m_{\tilde e^c}^2\right)\\
	\label{eq:S.Y}
	S_{Y}&=&m_{H_u}^2-m_{H_d}^2+\Tr\left(m_{\tilde Q}^2-2m_{\tilde u^c}^2+m_{\tilde d^c}^2+m_{\tilde L}^2-m_{\tilde e^c}^2\right),
\end{eqnarray}
where the traces are over generational indices. It can be shown, using the scalar soft mass RGEs, that the $S$-terms obey the RGEs:
\begin{equation}
	\frac{d}{dt} {S}_{a} = \frac{b_a \alpha_a {S}_a}{2 \pi},
\end{equation}
which admit the simple analytic solution
\begin{equation}
	S_a(t) = \frac{g^2_a(t)}{g^2_a(M_\i)}S_a(M_\i),
\end{equation}
for $S_R$ or $S_{B-L}$ and
\begin{eqnarray}
	S_Y(t) = \frac{g^2_Y(t)}{g^2_Y(M_\susy)}S_Y(M_\susy),
\end{eqnarray}

It is perhaps useful to separate the scalar mass RGEs into those that are analytically tractable and those that are not. In the $B-L$ MSSM scaling regime, the first- and second-family and sneutrino soft mass RGEs, analytically solvable, are
\begin{eqnarray}
	\label{eq:$B-L$ MSSM.mQ1}
	16\pi^2\frac{d}{dt}m_{\tilde Q_{1,2}}^2&=& - \frac{32}{3}g_3^2M_3^2-6g_2^2M_2^2 -\frac{1}{3}g_{BL}^2M_{B-L}^2+\frac{1}{4}g_{BL}	^2S_{B-L}\\
	16\pi^2\frac{d}{dt}m_{\tilde u^c_{1,2}}^2&=&-\frac{32}{3}g_3^2M_3^2-\frac{1}{3}g_{BL}^2M_{B-L}^2-2g_{R}^2M_{R}^2\nonumber\\
	&&-\frac{1}{4}g_{BL}^2S_{B-L}-g_{R}^2S_{R}\\
	16\pi^2\frac{d}{dt}m_{\tilde d^c_{1,2}}^2&=&-\frac{32}{3}g_3^2M_3^2-\frac{1}{3}g_{BL}^2M_{B-L}^2-2g_{R}^2M_{R}^2\nonumber\\
	&&-\frac{1}{4}g_{BL}^2S_{B-L}+g_{R}^2S_{R}\\
	16\pi^2\frac{d}{dt}m_{\tilde L_{1,2}}^2&=&-6g_2^2M_2^2-3g_{BL}^2M_{B-L}^2-\frac{3}{4}g_{BL}^2S_{B-L}\\
	16\pi^2\frac{d}{dt}m_{\tilde \nu^c_{1,2,3}}^2&=&-3g_{BL}^2M_{B-L}^2-2g_{R}^2M_{R}^2+\frac{3}{4}g_{BL}^2S_{B-L}-g_{R}^2S_{R}\\
	\label{eq:$B-L$ MSSM.mec}
	16\pi^2\frac{d}{dt}m_{\tilde e^c_{1,2}}^2&=&-3g_{BL}^2M_{B-L}^2-2g_{R}^2M_{R}^2+\frac{3}{4}g_{BL}^2S_{B-L}+g_{R}^2S_{R}.
\end{eqnarray}
In the MSSM scaling regime, which is only relevant to the case of the right-side-up hierarchy, the RGEs are 
\begin{eqnarray}
	\label{eq:MSSM.mQ1}
	16\pi^2\frac{d}{dt}m_{\tilde Q_{1,2}}^2&=&-\frac{32}{3}g_3^2M_3^2-6g_2^2M_2^2 -\frac{2}{15}g_1^2M_1^2+\frac{1}{5}g_1^2S_Y\\
	16\pi^2\frac{d}{dt}m_{\tilde u^c_{1,2}}^2&=&-\frac{32}{3}g_3^2M_3^2 -\frac{32}{15}g_1^2M_1^2-\frac{4}{5}Yg_1^2S_Y\\
	16\pi^2\frac{d}{dt}m_{\tilde d^c_{1,2}}^2&=&-\frac{32}{3}g_3^2M_3^2 -\frac{8}{15}g_1^2M_1^2+\frac{2}{5}Yg_1^2S_Y\\
	16\pi^2\frac{d}{dt}m_{\tilde L_{1,2}}^2&=&-6g_2^2M_2^2 -\frac{6}{5}g_1^2M_1^2-\frac{3}{5}Yg_1^2S_Y\\
	16\pi^2\frac{d}{dt}m_{\tilde \nu^c_{1,2}}^2&=&0\\
	\label{eq:MSSM.mec}
	16\pi^2\frac{d}{dt}m_{\tilde e^c_{1,2}}^2&=& -\frac{6}{5}Y^2g_1^2M_1^2+\frac{3}{5}Yg_1^2S_Y
\end{eqnarray}
The right-handed sneutrinos masses do not run in this regime because they are not charged under the MSSM gauge group. In the upside-down case the the right-handed sneutrinos are present in the brief scaling regime between $M_\susy$ and $M_{B-L}$. Their soft mass RGEs are
\begin{eqnarray}
	16\pi^2\frac{d}{dt} m_{\tilde \nu^c_{1,2,3}}^2&=&\frac34g_{BL}^2(m_{\tilde \nu^c_1}^2+m_{\tilde \nu^c_2}^2+m_{\tilde \nu^c_3}^2).
\end{eqnarray}

For the third family sfermions (excluding the sneutrinos) and for the MSSM Higgs, all of which are not analytically solvable, the RGEs In the $B-L$ MSSM scaling regime are
\begin{eqnarray}
	\label{eq:$B-L$ MSSM.Hu}
	16\pi^2\frac{d}{dt}m_{H_u}^2&=&6Y_t^2(m_{H_u}^2+m_{\tilde Q_3}^2+m_{\tilde t^c}^2)+6a_t^2\nonumber\\
	&& - 6g_2^2M_2^2-2g_R^2M_R^2+g_R^2S_R\\
	16\pi^2\frac{d}{dt}m_{H_d}^2&=&6Y_d^2(m_{H_d}^2+m_{\tilde Q_3}^2+m_{\tilde b^c}^2)+2Y_\tau^2(m_{H_d}^2+m_{\tilde L_3}	^2+m_{\tilde\tau^c})+6a^2_b+2a_\tau^2\nonumber\\
	&& - 6g_2^2M_2^2-2g_R^2M_R^2-g_R^2S_R\\
	\label{eq:$B-L$ MSSM.Q3}
	16\pi^2\frac{d}{dt}m_{\tilde Q_3}^2&=&2Y_t^2(m_{H_u}^2+m_{\tilde Q_3}^2+m_{\tilde t^c}^2) + 2Y_b^2(m_{H_d}^2+m_{\tilde Q_3}	^2+m_{\tilde b^c}) + 2a_t^2 + 2a_b^2\nonumber\\
	&& - \frac{32}{3}g_3^2M_3^2-6g_2^2M_2^2 -\frac{1}{3}g_{BL}^2M_{BL}^2+\frac{1}{4}g_{BL}^2S_{B-L}\\
	16\pi^2\frac{d}{dt}m_{\tilde L_3}^2&=&2Y_\tau^2(m_{H_d}^2+m_{\tilde L_3}^2+m_{\tilde\tau^c}^2)+2a_\tau^2\nonumber\\
	&&-6g_2^2M_2^2-3g_{BL}^2M_{BL}^2-\frac{3}{4}g_{BL}^2S_{B-L}\\
	\label{eq:$B-L$ MSSM.tc}
	16\pi^2\frac{d}{dt}m_{\tilde t^c}^2&=&4Y_t^2(m_{H_u}^2+m_{\tilde Q_3}^2+m_{\tilde t^c}) + 4a_t^2\nonumber\\
	&&-\frac{32}{3}g_3^2M_3^2-\frac{1}{3}g_{BL}^2M_{BL}^2-2g_R^2M_R^2-\frac{1}{4}g_{BL}^2S_{B-L}-g_R^2S_R\\
	16\pi^2\frac{d}{dt}m_{\tilde b^c}^2&=&4Y_b^2(m_{H_d}^2 + m_{\tilde Q_3}^2 + m_{\tilde b^c}^2) + 4a_b^2\nonumber\\
	&&-\frac{32}{3}g_3^2M_3^2-\frac{1}{3}g_{BL}^2M_{BL}^2-2g_R^2M_R^2-\frac{1}{4}g_{BL}^2S_{B-L}+g_R^2S_R\\
	\label{eq:$B-L$ MSSM.tauc}
	16\pi^2\frac{d}{dt}m_{\tilde \tau^c}^2&=&4Y_\tau^2(m_{H_d}^2 + m_{\tilde L_3}^2 + m_{\tilde\tau^c}^2)+4a_\tau^2\nonumber\\
	&&-3g_{BL}^2M_{BL}^2-2g_R^2M_R^2+\frac{3}{4}g_{BL}^2S_{B-L}+g_R^2S_R.
\end{eqnarray}
In the MSSM scaling regime they are
\begin{eqnarray}
	\label{eq:MSSM.Hu}
	16\pi^2\frac{d}{dt}m_{H_u}^2&=&6Y_t^2(m_{H_u}^2+m_{\tilde Q_3}^2+m_{\tilde t^c}^2)+6a_t^2\nonumber\\
	&& - 6g_2^2M_2^2-\frac65g_1^2M_1^2+\frac35g_1^2S_Y\\
	16\pi^2\frac{d}{dt}m_{H_d}^2&=&6Y_d^2(m_{H_d}^2+m_{\tilde Q_3}^2+m_{\tilde b^c}^2)+2Y_\tau^2(\tilde 	m_{H_d}^2+m_{\tilde L_3}^2+m_{\tilde\tau^c})+6a^2_b+2a_\tau^2\nonumber\\
	&& - 6g_2^2M_2^2-\frac65g_1^2M_1^2+\frac35g_1^2S_Y\\
	16\pi^2\frac{d}{dt}m_{\tilde Q_3}^2&=&2Y_t^2(m_{H_u}^2+m_{\tilde Q_3}^2+m_{\tilde t^c}^2) + 2Y_b^2(\tilde 	m_{H_d}^2+m_{\tilde Q_3}^2+m_{\tilde b^c}) + 2a_t^2 + 2a_b^2\nonumber\\
	&& - \frac{32}{3}g_3^2M_3^2-6g_2^2M_2^2-\frac{2}{15}g_1^2M_1^2+\frac15g_1^2S_Y
\end{eqnarray}
\begin{eqnarray}
	16\pi^2\frac{d}{dt}m_{\tilde L_3}^2&=&2Y_\tau^2(m_{H_d}^2+m_{\tilde L_3}^2+m_{\tilde\tau^c}^2)+2a_	\tau^2\nonumber\\
	&&-6g_2^2M_2^2-\frac{12}{5}g_1^2M_1^2-\frac35g_1^2S_Y\\
	16\pi^2\frac{d}{dt}m_{\tilde t^c}^2&=&4Y_t^2(m_{H_u}^2+m_{\tilde Q_3}^2+m_{\tilde t^c}) + 4a_t^2\nonumber\\
	&&-\frac{32}{3}g_3^2M_3^2-\frac{16}{5}g_1^2M_1^2-\frac45g_1^2S_Y\\
	16\pi^2\frac{d}{dt}m_{\tilde b^c}^2&=&4Y_b^2(m_{H_d}^2 + m_{\tilde Q_3}^2 + m_{\tilde b^c}^2) + 	4a_b^2\nonumber\\
	&&-\frac{32}{3}g_3^2M_3^2-\frac{8}{15}g_1^2M_1^2-\frac{2}{15}g_1^2S_Y\\
	\label{eq:MSSM.tauc}
	16\pi^2\frac{d}{dt}m_{\tilde \tau^c}^2&=&4Y_\tau^2(m_{H_d}^2 + m_{\tilde L_3}^2 + m_{\tilde\tau^c}^2)+4a_	\tau^2\nonumber\\
	&&-\frac{12}{5}g_1^2M_1^2+\frac{3}{5}g_1^2S_Y
\end{eqnarray}
The soft mass parameters are used in the calculation of the physical sparticle masses, discussed in the next appendix.

%
\section{Physical Masses}
\label{sec:131}
%

In this Appendix, we discuss how the physical masses of the sparticles and the Higgs are determined from the running parameters.

%
\subsection{Sparticle Masses}
\label{sec:546}
%

Because the first- and second-family Yukawa and tri-scalar couplings are negligible, mixing among the first- and second-family sfermions and the sneutrinos is negligible, greatly simplifying the relationship between physical masses and soft masses. However, there are electroweak $D$-term contributions associated with the electroweak scale. Although these are numerically small, they have the effect of splitting the masses of the otherwise degenerate $SU(2)_L$ doublets, which has implications for the lightest supersymmetric particle (see Section~\ref{sec:1058}):
\begin{eqnarray}
	\Delta_\phi= M_Z^2 \left(T_3-Q \sin^2 \theta_W \right) \cos 2 \beta ,
\end{eqnarray}
where $\theta_W$ is the weak mixing angle ($\sin^2 \theta_W \approx 0.23$) and $T_3$ and $Q$ are the left-handed isospin and electric charge of the scalar $\phi$. Here we lay out the physical masses with the electroweak $D$-term contributions, along with the notation for the physical masses.
\begin{eqnarray}
	m_{\tilde u_{L}}=m_{\tilde u} + \Delta_{\tilde Q_1}, & \ m_{\tilde u_{R}}=m_{\tilde u^c} + \Delta_{\tilde u^c},\nonumber\\
	m_{\tilde c_{L}}=m_{\tilde c} + \Delta_{\tilde Q_2}, & \ m_{\tilde c_{R}}=m_{\tilde c^c} + \Delta_{\tilde s^c},\nonumber\\
	m_{\tilde d_{L}}=m_{\tilde d} + \Delta_{\tilde Q_1}, & \ m_{\tilde d_{R}}=m_{\tilde d^c} + \Delta_{\tilde d^c},\nonumber\\
	m_{\tilde s_{L}}=m_{\tilde s} + \Delta_{\tilde Q_2}, & \ m_{\tilde s_{R}}=m_{\tilde s^c} + \Delta_{\tilde s^c},\nonumber\\
	m_{\tilde \nu_{L1}}=m_{\tilde \nu_1} + \Delta_{\tilde L_1}, & \ m_{\tilde \nu_{R1}}=m_{\tilde \nu^c_1} + \Delta_{\tilde 	\nu^c_1},\nonumber\\
	m_{\tilde \nu_{L2}}=m_{\tilde \nu_2} + \Delta_{\tilde L_2}, & \ m_{\tilde \nu_{R2}}=m_{\tilde \nu^c_2} + \Delta_{\tilde \nu^c_2},\nonumber\\
	m_{\tilde \nu_{L3}}=m_{\tilde \nu_3} + \Delta_{\tilde L_3}, & \ m_{\tilde \nu^c_R}=M_{Z_R},\nonumber\\
	m_{\tilde e_{L}}=m_{\tilde e} + \Delta_{\tilde L_1}, & \ m_{\tilde e_{R}}=m_{\tilde e^c} + \Delta_{\tilde e^c},\nonumber\\
	m_{\tilde \mu_{L}}=m_{\tilde \mu} + \Delta_{\tilde L_2}, & \ m_{\tilde \mu_{R}}=m_{\tilde \mu^c} + \Delta_{\tilde \mu^c}.
\end{eqnarray}
The third-family right-handed sneutrino physical state (referred to as $\tilde \nu^c_R$) mass is different because it acquires mass through the $B-L$ symmetry breaking mechanism and is degenerate with the $Z_R$ mass.

The Yukawa and tri-scalar couplings associated with third-family squarks and charged sleptons contribute non-negligible mixing terms among these scalars. These effects are captured in the stop, sbottom, and stau mixing matrices. Here we use the conventional notation $a_{t,b,\tau}=Y_{t,b,\tau}A_{t,b,\tau}$. The stop mixing matrix in the basis $(\tilde t, \tilde t^{c*})$ is\footnote{We present these matrices in terms of the fermion masses $M_{t,b,\tau}$ for simplicity. However, for numerical evaluation these fermion masses are replaced with the appropriate Higgs VEV times Yukawa coupling evaluated at the SUSY scale.}
\begin{eqnarray}
\mathcal{M}_{\tilde t}^2&=&\left(\begin{array}{cc}
		m_{\tilde Q_3}^2
		+ M_{t}^2
		+ \Delta_{\tilde Q_3}
		&
		M_t \left(A_t - \frac{\mu}{\tan \beta} \right)
	\\
		M_t \left(A_t - \frac{\mu}{\tan \beta} \right)
		&
		m_{\tilde t^c}^2
		+ M_{t}^2
		+ \Delta_{\tilde t^c}
	\end{array}\right).
\label{eq_stopmassmatrix}
\end{eqnarray}	
The eigenstates of this matrix will be referred to as $\tilde t_1$ and $\tilde t_2$ with mass eigenvalues defined such that $m_{\tilde t_1}<m_{\tilde t_2}$. The sbottom mixing matrix in the basis $(\tilde b, \tilde b^{c*})$ is
\begin{eqnarray}	
\mathcal{M}_{\tilde b}^2 &=& \left(\begin{array}{cc}
		m_{\tilde Q_3}^2
		+ M_{b}^2
		+\Delta_{\tilde Q_3}
		&
		M_b \left(A_b - \mu \tan \beta \right)
	\\
		M_b \left(A_b - \mu \tan \beta \right)
		&
		m_{\tilde b^c}^2
		+ M_{b}^2
		+\Delta_{\tilde b^c}
	\end{array}\right).
\label{eq_sbottommassmatrix}
\end{eqnarray}
The eigenstates of this mass matrix will be referred to similarly to the stops. The stau mixing matrix in the basis $(\tilde \tau, \tilde e^{c*})$ is
\begin{eqnarray}	
\mathcal{M}_{\tilde \tau}^2 &=& \left(\begin{array}{cc}
		m_{\tilde L_3}^2
		+ M_{\tau}^2
		+\Delta_{\tilde L_3}
		&
		M_\tau \left(A_\tau - \mu\tan\beta \right)
	\\
		M_\tau \left(A_\tau - \mu\tan\beta \right)
		&
		m_{\tilde \tau^c}^2
		+ M_{\tau}^2
		+ \Delta_{\tilde \tau^c}
	\end{array}\right).
\label{eq_staumassmatrix}
\end{eqnarray}
The eigenstates of this matrix will be referred to similarly to the stops and sbottoms. All of the running parameters in these matrices are evaluated at the SUSY scale.

For any of these matrices,
\begin{eqnarray}
	\trix{cc}L_{\tilde f}&X_{\tilde f}\\X_{\tilde f}&R_{\tilde f}\notrix,
\end{eqnarray}
the relevant mixing angle is given by 
\begin{eqnarray}
\tan 2\theta_{\tilde f}=\frac{-2|X_f|}{L_{\tilde f}-R_{\tilde f}},
\end{eqnarray}
where the angle $\theta_{\tilde f}$ may always be chosen to be between $0^\circ$ and $90^\circ$. Defined this way, a mixing angle close to zero means the lighter mass eigenstate consists of mostly the left-handed gauge eigenstate and a mixing angle close to $90^\circ$ means the lighter state is mostly right-handed.

The chargino content is identical to that of the MSSM in the approximation of vanishing $R$-parity violation. This is a good approximation for calculating masses but the mixing with the charged leptons need to be take into account when calculating decays, see~\cite{Marshall:2014kea} for example. Continuing with the approximation of vanishing $R$-parity violation, the results of \cite{Martin:1997ns} may be used. Those results, in our own notation, are
\begin{eqnarray}
	m_{\tilde \chi^\pm_{1}}^2=\frac12\left(M_2^2+\mu^2+2M_W^2-\sqrt{(M_2^2+\mu^2+2M_W^2)^2-4(\mu M_2-M_W^2\sin{2\beta})^2}\right)\\
	m_{\tilde \chi^\pm_{2}}^2=\frac12\left(M_2^2+\mu^2+2M_W^2+\sqrt{(M_2^2+\mu^2+2M_W^2)^2-4(\mu M_2-M_W^2\sin{2\beta})^2}\right).
\end{eqnarray}

In the basis $(\nu, \tilde W_R, \tilde B^\prime, \tilde W^0, \tilde H_u^0, \tilde H_d^0)$, the neutralino mass matrix is
\begin{eqnarray}
\trix{cccccc}
0			&-c_{\theta_R}M_{Z_R}		&s_{\theta_R}M_{Z_R}	&0				&0				&0				\\
-c_{\theta_R}M_{Z_R}	&M_R				&0			&0				&-c_\beta s_{\theta_W}M_Z	&s_\beta s_{\theta_W}M_Z		\\
s_{\theta_R}M_{Z_R}	&0				&M_{BL}		&0				&0				&0				\\
0			&0				&0			&M_2				&c_\beta c_{\theta_W}M_Z	&-s_\beta c_{\theta_W}M_Z	\\
0			&-c_\beta s_{\theta_W}M_Z	&0			&c_\beta c_{\theta_W}M_Z	&0				&-\mu				\\
0			&s_\beta s_{\theta_W}M_Z	&0			&-s_\beta c_{\theta_W}M_Z	&-\mu				&0				\\
\notrix,
\label{eq:1017}
\end{eqnarray}
where $c_\theta\equiv\cos\theta$ and $s_\theta\equiv\sin\theta$ etc. As with the charginos, we have assumed that mixing with the left-handed neutrinos, due to $R$-parity violation is 0. This is good approximation for calculating masses and will be used here, but cannot be used when calculating decay rates. As discussed in Section~\ref{sec:1115} some of the eigenstates of this matrix have masses associated with the $B-L$ scale while others have masses associated with the SUSY scale. A conventional approach to this situation would be to perturbatively diagonalize the matrix in the limit $M_\susy\gg M_{B-L}$ for the right-side-up case or $M_{B-L}\gg M_\susy$ for the upside-down case. However, these two scales may be comparable so the entire mass matrix must be diagonalized without the use of perturbative methods. This has the potential to introduce errors since it doesn't account for the fact that some states should be integrated out at different scales. However, the errors will always be small because the $B-L$ and SUSY scales are always of comparable size. We choose to evaluate all of the running parameters in this matrix at the SUSY scale. The error introduced by doing this should be smaller than the error introduced by associating the entire SUSY spectrum with a single scale, $M_\susy$. The mass eigenstates are referred to as $\tilde \chi^0_1\cdots\tilde \chi^0_6$ in a mass ordered basis with eigenvalues $m_{\tilde \chi^0_1}\cdots m_{\tilde \chi^0_6}$.

The physical gluino mass, $M_{\tilde g}$ is simply equal to the running gluino mass evaluated at the SUSY scale.
\begin{eqnarray}
	M_{\tilde g}=M_3(M_\susy).
\end{eqnarray}

\subsection{Higgs Masses}
\label{sec:915}

Supersymmetric models such as the MSSM and this $B-L$ MSSM contain five Higgs particles. The most important for the present discussion is the lightest neutral SM-like Higgs, $h^0$, which we refer to as ``the Higgs'' throughout this paper. This one is important because its mass is known and can be used to constrain some of the SUSY parameter space. The other four Higgses are the heavy Higgs, $H^0$, the Higgs pseudoscalar, $A^0$, and the charged Higgses, $H^\pm$.

The Higgs mass is calculated using methods discussed in \cite{ArkaniHamed:2004fb,Cabrera:2011bi,Giudice:2011cg}. The physical Higgs mass is
\begin{eqnarray}
	m_{h^0}=\sqrt{\lambda} v,
\end{eqnarray}
with the Higgs quartic coupling, $\lambda$, evaluated at the scale of the physical Higgs mass. Above the SUSY scale, $\lambda$ comes from the $D$-terms and is thereby fixed. Below the SUSY scale, RGE effects will cause $\lambda$ to deviate from its supersymmetric value. These effects come mainly from one-loop graphs involving the top quark. They are contained in the RGE for $\lambda$ in the SM scaling regime. We employ results from \cite{Giudice:2011cg}. Here we re-state the relevant equations in our own notation. The supersymmetric boundary condition on $\lambda$ is
\begin{eqnarray}
\lambda(M_\susy)=\frac14\left(g_L^2+\frac35g_1^2\right)\cos^22\beta+\delta\lambda.
\end{eqnarray}
The parameter $\delta\lambda$ contains threshold corrections applied at the SUSY scale. Including only the dominant stop contributions from \cite{Giudice:2011cg},
\begin{eqnarray}
16\pi^2\delta\lambda=3Y_t^4\left(2\frac{X_t^2}{m_{\tilde t_1}m_{\tilde t_2}}F\left(\frac{m_{\tilde t_1}}{m_{\tilde t_2}}\right)-\frac16\frac{X_t^4}{m_{\tilde t_1}^2m_{\tilde t_2}^2}G\left(\frac{m_{\tilde t_1}}{m_{\tilde t_2}}\right)\right),
\end{eqnarray}
where we define $X_t=A_t-\mu\cot\beta$ (note that this definition is different from that used in \cite{Giudice:2011cg}) and 
\begin{eqnarray}
F(x) &=& \frac{2x\ln x}{x^2-1}\\
G(x) &=& \frac{12x^2(1-x^2+(1+x^2)\ln x)}{(x^2-1)^3}.
\end{eqnarray}
The RGE for $\lambda$ in the SM regime is
\begin{eqnarray}
\frac{d}{dt}\lambda&=&4\lambda(3y_t^2+3y_b^2+y_\tau^2)-9\lambda(\frac15g_1^2+g_2^2)\nonumber\\
&&-4(3y_t^4+3y_b^4+y_\tau^4)+\frac{27}{100}g_1^4+\frac{9}{10}g_2^2g_1^2+\frac94g_2^4+12\lambda^2,
\end{eqnarray}
and in the upside-down case between $M_\susy$ and $M_{B-L}$ it is
\begin{eqnarray}
\frac{d}{dt}\lambda&=&4\lambda(3y_t^2+3y_b^2+y_\tau^2)-9\lambda(\frac13g_R^2+g_2^2)\nonumber\\
&&-4(3y_t^4+3y_b^4+y_\tau^4)+\frac{3}{4}g_R^4+\frac{3}{2}g_2^2g_R^2+\frac94g_2^4+12\lambda^2.
\end{eqnarray}
Since this depends on the Yukawa couplings, which are solved numerically, this must also be solved numerically. The dominant contributions come from the terms involving $y_t$. These terms are present because both stops are integrated out at $M_\susy$. This has the potential to introduce errors because the stops generally do not have the same mass. The errors introduced by this are minimized when the SUSY scale is chosen to be $M_\susy=\sqrt{m_{\tilde t_1}m_{\tilde t_2}}$. We find this method of calculating the Higgs mass is the best compromise between transparency and accuracy. \comment{We should point out that many other methods exist for the calculation of the Higgs mass in supersymmetric theories. See \todo{references} for a subset of the literature on the subject. There are also computer software packages for calculating the Higgs mass such as \todo{references}.}

Regarding the masses of the other four Higgses, the tree level results from \cite{Martin:1997ns} apply and are sufficient for the present purposes. We re-state them here.
\begin{eqnarray}
m_{A^0}^2&=&2b/\sin(2\beta)=2\mu^2+m_{H_u}^2+m_{H_d}^2\\
m_{H^0}^2&=&\frac12\left(m_{A^0}^2+M_Z^2+\sqrt{(m_{A^0}^2-M_Z^2)^2+4M_Z^2m_{A^0}^2\sin^2(2\beta)}\right)\\
m_{H^\pm}^2&=&m_{A^0}^2+M_W^2.
\end{eqnarray}

%
\section{Application of the Checks and Iterative Procedure}
\label{sec:536}
%
In this Appendix, we describe--for a single randomly generated initial point--two things: 1) the precise algorithm by which the checks described in Table~\ref{tab:320} are applied and 2) the iterative numerical method used to solve for the $B-L$ and SUSY scales. It is necessary to discuss these simultaneously since, as will become clear, they are interrelated. We include this Appendix to give the reader insight into the details of our statistical method and to elucidate technical comments made in the main text.

Before proceeding,  it is helpful to note several things. A ``point'' here refers to a randomly generated choice of the  parameters listed in Table~\ref{tbl:scan}. For each point, we make working ``guesses'' of the initial values of $M_\susy$ and $M_{B-L}$. These will be iteratively improved using a simple numerical method. For a fixed choice of randomly generated parameters and the two scales $M_\susy$ and $M_{B-L}$ specified, there is a unique solution for all of the RGEs and physical masses. That unique solution is found by our code using a combination of analytic solutions (discussed throughout this paper) and numerical methods (not discussed in this paper). For the purposes of this Appendix, it is sufficient to know that the solution can indeed be calculated.
It is also useful to note that, with the exception of the spill and convergence checks, the checks in Table~\ref{tab:320} are applied sequentially. For example, a point is subjected to the EW breaking check if and only if it passes the preceding $B-L$ breaking and $Z_R$ bound checks. This means that a point that fails a particular check a) has implicitly passed all previous checks and b) is immediately discarded and never subjected to subsequent checks. The sequential nature of these checks is what enables us to define the survival rates given in Table~\ref{tab:320}. The spill checks and the convergence check, however, are different because they are not necessarily applied in a particular order and may even be applied multiple times to a single point. Nevertheless, if any point fails a spill or convergence check, at any step in the iterative process, we count that point as having passed all spill checks that appear above the failed check in Table~\ref{tab:320}. This removes any ambiguity about how to define survival rates for the spill and convergence checks.

Now we are prepared to discuss the main goals of  this Appendix. For each randomly generated point, the initial guesses for $M_\susy$ and $M_{B-L}$ are always taken to be 1 TeV and 2.5 TeV respectively. If the point with these initial guesses does not satisfy $B-L$ breaking, then we count it as failing the $B-L$ breaking check. If the point does not satisfy the $Z_R$ lower bound, then it is so counted. If it does not satisfy EW breaking, then it is so counted. If it does not satisfy the non-tachyonic stops check, it is so counted.

If the guess for the $B-L$ scale satisfies its definition, that is, if the RG calculation of $M_{Z_{R}}(M_{B-L})=M_{B-L}$,
to within 1\%, and the guess for the SUSY scale satisfies its definition, that is, the RG calculation of $m_{\tilde t_1}$, $m_{\tilde t_2}$ satisfies $\sqrt{m_{\tilde t_1}(M_\susy)m_{\tilde t_2}(M_\susy)}=M_\susy$
to within 1\%, then ``convergence'' has occurred and the steps in the next two paragraphs are skipped.

If the guess for the $B-L$ scale satisfies its definition to within 1\%, then the rest of the steps in this paragraph are skipped. If not, the guess for the $B-L$ scale is changed to $M_{Z_R}$. Using the same value for $M_\susy$, and the new choice of $M_{B-L}$, we again run the RGEs for the same initial point. If $M_{Z_{R}}$ not within 1\% of $M_{B-L}$, then the process is repeated.  If the steps in this paragraph are repeated more than 300 times\footnote{a conveniently chosen number which provides adequate opportunity for the iteration to converge.} without success, then we count the point as having failed the convergence check.

If the guess for the SUSY scale satisfies its definition to within 1\%, the rest of the steps in this paragraph are skipped. If not, the guess for the SUSY scale is changed to $\sqrt{m_{\tilde t_1}(M_\susy)m_{\tilde t_2}(M_\susy)}$ and we rerun the RGEs. 
If the point now does not satisfy EW breaking, it is counted as failing the EW breaking spill check. If the point now does not satisfy the non-tachyonic stops check, it is counted as failing the non-tachyonic stops spill check. If it does pass these checks, but $M_\susy$ does not satisfy its definition to within 1\%,  then the steps in this paragraph are repeated. If they have been repeated more than 300 times without success, the point is counted as failing the convergence check. 

Having successfully passed all of the previous criterion, we now must check the remaining checks. If the point does not satisfy the $Z_R$ bound, it is counted as failing the $B-L$ bound spill check. If the point does not satisfy the sparticle bounds, it is so counted. If the point does not satisfy the Higgs mass check, it is so counted. If it does, however, satisfy all of these experimental checks, it is a valid point.

\begin{figure}[!b]
	\centering
	\includegraphics[scale=0.7]{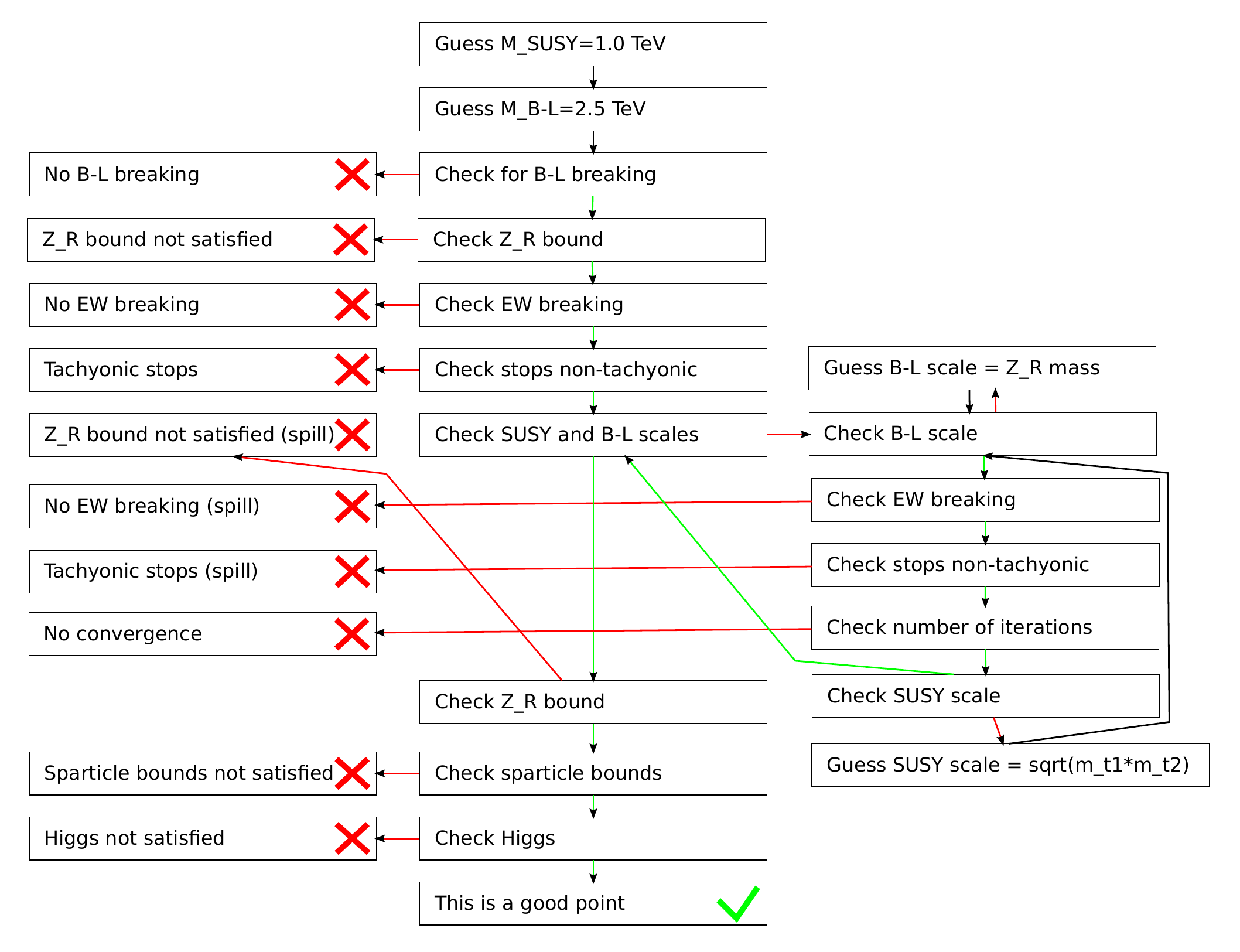}
	\caption{\small A ``flow chart'' showing how the checks are applied and how the iterative process of solving for the $B-L$ and SUSY scales works. Every block that begins with the word ``Check'' has an outgoing red and green arrow. The green arrow is followed if the check is satisfied and the red arrow is followed if the check is not satisfied.}
	\label{fig:650}
\end{figure}

The procedure described in the previous five paragraphs is represented pictorially by the ``flow chart'' in Fig.~\ref{fig:650}.

\newpage


\end{document}